\documentclass[a4paper,fleqn]{cas-dc}
\usepackage{amsmath,amssymb,amsfonts}
\usepackage[pdftex, svgnames, dvipsnames]{xcolor}
\usepackage{xspace}
\usepackage{subcaption}
\usepackage{xurl}
\usepackage{enumitem}
\usepackage{microtype}
\usepackage{pifont}
\usepackage{graphicx}

\definecolor{codegray}{rgb}{0.3,0.3,0.3}   
\definecolor{codegreen}{rgb}{0,0.7,0}      
\definecolor{codeblue}{rgb}{0,0.2,1}       
\definecolor{codered}{rgb}{0.85,0.1,0.1}   
\usepackage{tabularx,booktabs,makecell}

\usepackage[dvipsnames]{xcolor}
\usepackage[draft]{fixme}
\usepackage[most]{tcolorbox}
\usepackage{soul}   

\makeatletter
\newcommand*\fxauthor[4]{%
  \expandafter\def\csname fx@ac@#3\endcsname{#4}%
  \FXRegisterAuthor{#1}{#2}{#3}}
\newcommand*\fx@authorcolor{%
  \ifcsname fx@ac@\detokenize\expandafter{\cmdfx@note@author}\endcsname
    \csname fx@ac@\detokenize\expandafter{\cmdfx@note@author}\endcsname
  \else gray\fi}
\newcommand*\FXLayoutAuthorBox[3]{%
  \begin{tcolorbox}[enhanced, breakable, boxrule=0pt, leftrule=2.5pt, arc=1pt,
      colframe=\fx@authorcolor, colback=\fx@authorcolor!8,
      left=4pt, right=4pt, top=2pt, bottom=2pt,
      before skip=4pt, after skip=4pt, fontupper=\small\sffamily]
    \colorbox{\fx@authorcolor}{\color{white}\bfseries #3}\ #2%
  \end{tcolorbox}}
\FXRegisterLayout*{authorbox}{\FXLayoutAuthorBox}
\newcommand\FXTargetLayoutAuthorHL[2]{%
  {\color{\fx@authorcolor}#2}}
\FXRegisterTargetLayout{authorhl}{\FXTargetLayoutAuthorHL}
\makeatother

\fxsetup{status=final, layout=authorbox, targetlayout=authorhl}

\fxauthor{nd}{nde}{ND}{Purple}
\fxauthor{as}{asa}{AS}{Brown}
\fxauthor{lr}{lre}{LR}{Red}
\fxauthor{dm}{dma}{DM}{Violet}
\fxauthor{gg}{ggi}{GG}{Periwinkle}

\newcommand{\changed}[1]{#1}

\newcommand{\ie}{i.e.\@\xspace}
\newcommand{\eg}{e.g.\@\xspace}
\newcommand{\etal}{\textit{~et~al.\@}\xspace}
\newcommand{\wrt}{w.r.t.\@\xspace}

\usepackage{tikz}
\usepackage{circuitikz}
\usetikzlibrary{positioning,fit,arrows.meta,shapes.geometric,calc,graphs, backgrounds}
\usetikzlibrary{arrows.meta,shapes.multipart,positioning,calc}
\usetikzlibrary{arrows}

\usepackage[acronym,nomain]{glossaries}
\glsdisablehyper
\newacronym{stix}{STIX}{Structured Threat Information Expression}\newcommand{\stix}{\acrshort{stix}\xspace}
\newacronym{cti}{CTI}{Cyber Threat Intelligence}\newcommand{\cti}{\gls{cti}\xspace}
\newacronym{rag}{RAG}{Retrieval-Augmented Generation}\newcommand{\rag}{\gls{rag}\xspace}
\newacronym{llm}{LLM}{Large Language Model}\newcommand{\llm}{\gls{llm}\xspace}
\newacronym{ttps}{TTPs}{Tactics, Techniques, and Procedures}
\newacronym{iocs}{IoCs}{Indicators of Compromise}\newcommand{\iocs}{\gls{iocs}\xspace}
\newacronym{ml}{ML}{Machine Learning}
\newacronym{nlp}{NLP}{Natural Language Processing}\newcommand{\nlp}{\gls{nlp}\xspace}
\newacronym{objects}{Objects}{Domain Objects}
\newacronym{relations}{Relationships}{Relationship Objects}
\newacronym{taxii}{TAXII}{Trusted Automated Exchange of Intelligence Information}
\newacronym{ai}{AI}{Artificial Intelligence}
\newacronym{bm25}{BM25}{Best Matching 25}\newcommand{\bm}{\gls{bm25}\xspace}
\newacronym{ncsc}{NCSC}{National Cyber Security Centre}\newcommand{\ncsc}{\gls{ncsc}\xspace}
\newacronym{slm}{SLM}{Small Language Model}\newcommand{\slm}{\gls{slm}\xspace}

\newcommand{\as}{Attack Step\xspace}
\newcommand{\ms}{Milestone\xspace}

\newcommand{\precond}{\emph{pre-conditions}\xspace}
\newcommand{\postcond}{\emph{post-conditions}\xspace}

\newcommand{\mitre}{MITRE ATT\&CK\xspace}

\usepackage{tabularx,makecell,array,pifont,ragged2e}
\newcolumntype{C}{>{\Centering\arraybackslash}X}
\newcommand{\circlescale}[1]{%
\begin{tikzpicture}[baseline=-0.6ex]
  \def\r{2.1pt}
  \draw (0,0) circle (\r); 
  \ifnum#1=2
    \begin{scope}
      \clip (-\r,-\r) rectangle (0,\r); 
      \fill (0,0) circle (\r);
    \end{scope}
  \fi
  \ifnum#1=3
    \fill (0,0) circle (\r); 
  \fi
\end{tikzpicture}%
}

\newcommand{\legendBlueRed}[2]{%
  \begin{tabular}{@{}ll@{\qquad}ll@{}}
    \begin{tikzpicture}[baseline=-0.6ex]
      \draw[fill=red,draw=red] (0,0) rectangle (6pt,6pt);
    \end{tikzpicture} & #1 &
    \begin{tikzpicture}[baseline=-0.6ex]
      \draw[fill=blue,draw=blue] (0,0) circle (3pt);
    \end{tikzpicture} & #2
  \end{tabular}%
}

\usepackage{multicol}
\usepackage{float}
\usepackage{stfloats}

\usepackage[numbers]{natbib}

\def\tsc#1{\csdef{#1}{\textsc{\lowercase{#1}}\xspace}}
\tsc{WGM}
\tsc{QE}

\begin{document}
\let\WriteBookmarks\relax
\def\floatpagepagefraction{1}
\def\textpagefraction{.001}

\shorttitle{Merging CTI Through RAG and SLMs for Rich Threat Representation}    

\shortauthors{Deidda et al.} 

\title [mode = title]{Merging Cyber Threat Intelligence Through Retrieval-Augmented Generation and Small Language Models for Rich Threat Representation}  



%

\author[1,2]{Nicola Deidda}[orcid=0009-0005-7309-5703]
\cormark[1] 
\ead{nicola.deidda@unica.it}
\affiliation[1]{organization={Department of Electrical and Electronic Engineering (DIEE), University of Cagliari},
            addressline={Via Marengo 2},
            city={Cagliari},
            postcode={09123},
            country={Italy}}
\affiliation[2]{organization={IMT School for Advanced Studies Lucca},
            addressline={Piazza San Francesco 19},
            city={Lucca},
            postcode={55100},
            country={Italy}}
\credit{Conceptualization, Methodology, Software, Validation, Investigation, Data Curation, Visualization, Writing - Original Draft, Writing - Review \& Editing}

\author[1]{Leonardo Regano}[orcid=0000-0002-9259-5157]
\ead{leonardo.regano@unica.it}
\credit{Conceptualization, Methodology, Investigation, Writing - Review \& Editing}

\author[1]{Alessandro Sanna}[orcid=0000-0001-5663-4254]
\ead{alessandro.sanna96@unica.it}
\credit{Validation, Writing - Review \& Editing}

\author[1]{Davide Maiorca}[orcid=0000-0003-2640-4663]
\ead{davide.maiorca@unica.it}
\credit{Writing - Review \& Editing, Supervision}

\author[1]{Giorgio Giacinto}[orcid=0000-0002-5759-3017]
\ead{giorgio.giacinto@unica.it}
\credit{Resources, Writing - Review \& Editing, Funding acquisition, Supervision}

\cortext[1]{Corresponding author}


\begin{abstract}
Cybersecurity operations rely on \acrfull{cti} from heterogeneous sources, such as structured threat representations and narrative reports. They describe adversarial behaviors but rarely make explicit the conditions that enable each attack step, the consequences it produces, and the evidence supporting them, leaving analysts to correlate fragmented information manually. Existing approaches extract entities, map tactics and techniques, or reconstruct attack chains; enriching attack steps with explicit, evidence-linked semantics remains largely unexplored.

This paper presents a local-first pipeline for evidence-grounded attack-step enrichment that combines hybrid \acrfull{rag} with a locally deployed \acrfull{slm}. Starting from an analyst-provided attack representation and supporting \acrshort{cti} documents, the pipeline retrieves step-relevant evidence and generates enriched descriptions, \precond, \postcond, and traceable links to the supporting passages. The enriched steps are organized into a graph that groups \mitre-aligned steps into tactic-level milestones. Explicit conditions help defenders block attacks by removing their enabling conditions and support detection by checking for their expected consequences.

We evaluate the pipeline on ten public malware case studies derived from UK National Cyber Security Centre reports, using
automated measurements and a structured manual assessment with annotations from two annotators.
The results demonstrate that the pipeline is effective at generating conditions that are, for the most
part, evidence-supported and meaningful w.r.t. the analyzed attack step. Similar considerations hold
for the enriched descriptions. Finally, these findings demonstrate the feasibility of locally generating
structured, evidence-grounded enrichments that improve the interpretability and traceability of attack step
representations without requiring a manual reconstruction of attack execution.

\end{abstract}


\begin{highlights}
\item Evidence-linked enrichment of attack steps with pre- and post-conditions
\item Runs fully on a laptop, so sensitive CTI never leaves the organization
\item Enriched steps form a MITRE ATT\&CK-aligned graph with tactic milestones
\item On 10 NCSC cases, 89\% of conditions are atomic and 78\% meaningful
\item 70\% of conditions are directly supported by traceable CTI evidence
\end{highlights}

\begin{keywords}
 Cyber Threat Intelligence \sep Small Language Models \sep Attack Graphs \sep Attack Flow \sep Preconditions \sep Postconditions
\end{keywords}

\maketitle





\clearpage 

\section{Introduction}
\label{sec:introduction}

\cti is essential for understanding adversary behavior, prioritizing defensive actions, and supporting incident response. Modern \cti pipelines consume heterogeneous sources, including malware analysis reports, vendor blogs, structured indicators, \stix objects~\cite{krauss_analysis_2022,barnum_standardizing_nodate}, MISP events~\cite{10.1145/2994539.2994542}, and \mitre technique mappings~\cite{strom_mitre_nodate}. These sources provide complementary views of a threat: \iocs describe observable artifacts, \stix supports the standardized exchange of entities and relations, and \mitre provides a common vocabulary for adversary tactics and techniques. However, transforming these heterogeneous artifacts into an operational representation remains labor-intensive~\cite{jcp5030063}. Analysts must connect scattered evidence, reconcile inconsistent descriptions, and determine which assumptions, capabilities, and effects are associated with each adversary action~\cite{10.1145/3319535.3354239,thimmaraju_human_2025}.

\changed{Consider a \cti analyst} investigating \emph{Goofy Guineapig}, a Windows malware implementing a persistent backdoor with HTTPS command-and-control communications\footnote{\url{https://www.ncsc.gov.uk/sites/default/files/documents/NCSC-MAR-Goofy-Guineapig.pdf}}. In the setting considered in this work, the analysts receives from a \cti producer an initial structured representation of relevant attack steps (\eg, \stix objects or \mitre technique mappings), and has available a collection of supporting intelligence documents gathered from heterogeneous sources. This evidence collection may subsequently be extended with additional reports, indicators, or contextual information. The analyst will inspect the available material to determine what each step represents, which passages provide relevant evidence, and what conditions or consequences may be associated with the reported behavior. Importantly, this task extends beyond assigning technique labels, since it requires interpreting the operational context of individual actions while identifying information explicitly reported by a source.

A limitation of technique-oriented threat descriptions is that identifying an adversarial behavior does not necessarily capture the contextual information required to interpret its feasibility and effects. For instance, a \stix-derived description may state that a malware family maintains persistence as a Windows service and may associate this behavior with the \mitre technique \emph{Create or Modify System Process: Windows Service} (T1543.003). Although this mapping identifies the general behavior, it does not necessarily specify the relevant environmental requirements, the privileges involved, or the system changes resulting from the action. Depending on the reported implementation, these may include access to a Windows host, the ability to create or modify service-related resources, and the ability to establish a service configuration that supports subsequent execution. Such details may be distributed across several passages, remain implicit in the narrative, or be absent from the available evidence. Recovering and qualifying this context is relevant to threat interpretation, detection engineering, and incident investigation, but requires more than technique classification alone.

Recent work has explored the use of \llm{s}, knowledge graphs, and \nlp pipelines for \cti extraction, \mitre mapping, and threat representation~\cite{hu_llm_tikg_2024,10763668,10386611,gu2025surveyllmasajudge,zhang_attackg_2025}. These approaches demonstrate the potential of language models for structuring cyber threat information. Entity and relation extraction, technique mapping, and document summarization, however, leave a distinct step-level enrichment problem: connecting an already identified adversary action to explicit enabling conditions, direct effects, and inspectable source evidence. Preconditions and postconditions have been considered in prior attack representations~\cite{hou_reachable_2026}, but generating and assessing these fields from heterogeneous intelligence within an analyst-provided attack-step structure remains a useful applied research task. We address that task through a workflow that combines structured attack information, retrieval, locally executed generation, and field-level provenance.

This paper investigates evidence-linked enrichment of analyst-defined attack steps. Given an initial set of \cti-derived attack steps, \ie, the initial representation, and a corpus of supporting intelligence, the objective is to supplement each step with a concise contextual description and candidate \precond and \postcond associated with retrieved evidence. The task is narrower than autonomous attack reconstruction, \ie, the initial attack-step structure is provided rather than inferred from unstructured documents as a complete sequence. A \emph{pre-condition} denotes a state, requirement, or capability considered relevant before an attack step can be executed, such as an environmental dependency, a privilege requirement, an accessible resource, or a prior system state. A \emph{post-condition} denotes a state or effect associated with successful execution of the step, such as a created artifact, a modified configuration, an executed process, or an acquired capability. Conditions are expressed as atomic statements to facilitate inspection and subsequent analysis.

This explicit representation also makes the distinction between reported facts and plausible requirements visible to the analyst. A passage may describe an observed system change while only indirectly suggesting the permissions needed to produce it. By expressing such requirements and effects as separate candidate statements linked to their sources, the workflow exposes assumptions for inspection and helps identify which artifacts or system states warrant further investigation. Evidence support is therefore evaluated independently of the presence of a provenance link.

For instance, an initial \emph{Goofy Guineapig} attack step may identify persistence via a Windows service without explicitly specifying its operational context. The workflow retrieves passages relevant to the step and generates a contextual description together with candidate \precond and \postcond. A precondition might concern the permissions needed to establish the service, whereas a postcondition might describe the resulting persistence-related system state. By retaining the evidence associated with each field, the enriched representation allows the analyst to inspect the proposed requirements and consequences individually and revise those that are inadequately supported.

\changed{To investigate this task, we design a local-first pipeline that combines \rag, hybrid information retrieval, and a \slm. Figure~\ref{fig:simple_schema} provides an overview of the proposed workflow. The heterogeneous \cti sources available to the \cti analyst are processed into retrievable evidence chunks while retaining metadata identifying their origin. The initial representation serves as a guideline throughout the process. For each input attack step, lexical and semantic retrieval select potentially relevant passages, which are supplied to the local \slm through structured prompts. The model produces a contextual description and candidate \precond and \postcond associated with the retrieved material. The output representation retains the input step's tactic and technique context and records links to the evidence used during generation. These links support provenance inspection. The enriched steps are organized into a graph-based representation with tactic-level milestones. This preserves the analyst-provided attack steps organization while making descriptions, conditions, evidence links, and available ordering constraints accessible in a common structure. The graph is designed for inspection and further analysis; its ordering relations reflect the input representation and are not independently inferred causal dependencies.}

\begin{figure*}[t!]
\centering
\makebox[\textwidth][c]{%
\resizebox{1\textwidth}{!}{%
\begin{tikzpicture}[
  font=\large,
  every node/.style={align=center},
  box/.style={draw, minimum width=3.1cm, minimum height=1.1cm},
  light/.style={box, fill=gray!20},
  plain/.style={box, fill=white},
  arrow/.style={->, >=Stealth, line width=0.6pt},
  dashedbox/.style={draw, dashed, rounded corners, inner sep=6pt, thick}
]

\newcommand{\TrapH}{1.35cm}
\tikzset{
  trap/.style={
    draw, trapezium, trapezium left angle=70, trapezium right angle=110,
    minimum width=2.8cm, minimum height=\TrapH
  }
}
\node[plain] (documents) {HETEROGENEOUS\\ \cti SOURCES};
\node[plain, right=of documents] (chunks) {EVIDENCE\\CHUNKS};
\node[plain, right=of chunks] (rag) {RAG};
\node[plain, below=of documents] (int) {INITIAL\\REPRESENTATION\\(\eg, STIX, MISP)};
\node[plain, right=of int] (slm) {SLM};
\node[plain, right=of slm] (out) {OUTPUT\\REPRESENTATION};
\node[plain, right=of out] (graph) {GRAPH-LIKE\\REPRESENTATION};

\draw[arrow] (documents) -- (chunks);
\draw[arrow] (chunks) -- (rag);
\draw[arrow] (rag) -- (slm);
\draw[arrow] (int) -- (slm);
\draw[arrow] (slm) -- (out);
\draw[arrow] (out) -- (graph);

\end{tikzpicture}
}%
}%
\caption{Simplified schema of the proposed pipeline. An analyst provides the initial representation (\eg, STIX or MISP) and a heterogeneous set of documents. For each attack step in the initial representation, the pipeline retrieves potentially relevant evidence, and a local \slm generates a contextual description and candidate \precond and \postcond. The resulting graph retains the attack-step structure together with the enriched information.}
\label{fig:simple_schema}
\end{figure*}

The local-first design addresses practical constraints encountered when handling \cti. Intelligence collections may contain sensitive incident details, infrastructure information, or internal analytical material that organizations prefer to process within controlled environments. A locally executed \slm supports this deployment setting, while retrieval enables updating the evidence corpus without retraining the model. The study evaluates one feasible local configuration; deployment security ultimately depends on the surrounding data-handling practices.

Then, we specify how analyst-defined attack steps can be augmented with distinct, evidence-linked semantic fields, implement this transformation in a local workflow, and measure which properties of the generated outputs are consistently satisfied. This evaluation separates structural adequacy, condition semantics, and evidence alignment, providing a more informative account of enrichment quality than generation coverage alone.

To guide the empirical assessment, we formulate three research questions that address complementary aspects of the proposed enrichment workflow:
\begin{enumerate}
    \item[\textbf{RQ1.}] To what extent does the proposed workflow generate semantically well-formed attack-step conditions?
    \item[\textbf{RQ2.}] To what extent are the generated enrichments supported by the retrieved \cti evidence?
    \item[\textbf{RQ3.}] What are the observed computational characteristics of executing the enrichment workflow locally?
\end{enumerate}

The main contributions of this paper are as follows:
\begin{enumerate}
\item We formulate evidence-linked enrichment as a distinct step-level \cti task, specifying the initial attack representation, supporting document corpus, and generated semantic fields. This formulation focuses on making the operational context of known attack steps explicit and inspectable.

\item We define an enrichment representation that separates contextual step descriptions, candidate \precond, candidate \postcond, tactic and technique references, and source provenance. This field-level structure allows analysts to inspect requirements, effects, and their supporting material independently.

\item We implement a local-first enrichment workflow combining lexical and semantic retrieval with \rag-based generation using a \slm. The pipeline operates on analyst-provided, \mitre-aligned attack steps and retains the retrieval context and provenance associated with generated fields.

\item We organize the enriched attack steps into an evidence-linked graph structured around tactic-level milestones, retaining the input ordering constraints while exposing step-level context, conditions, and evidence in a common analytical representation.

\item We empirically characterize the workflow across 10 public malware case studies, assessing 860 generated conditions and 110 enriched descriptions with two independent annotators. The evaluation distinguishes meaningfulness and atomicity from condition-type correctness, and step specificity from evidence support and faithfulness. This reveals both the reliable structural properties of the output and the semantic errors that require analyst attention.
\end{enumerate}

The evaluation (Section~\ref{sec:evaluation}) identifies a substantive distinction between producing well-formed conditions and assigning them the correct operational semantics. Under strict joint-positive judgments, 88.95\% of conditions are atomic and 77.56\% are meaningful, while 70.00\% are directly supported by the cited evidence and 54.07\% have the correct precondition or postcondition type. For descriptions, 84.55\% are step-specific, although full faithfulness and the absence of unsupported claims remain more difficult to achieve. These findings characterize the current strengths and failure modes of local, evidence-linked enrichment and establish concrete priorities for improving its reliability.

The remainder of the paper is organized as follows. Section~\ref{sec:background} introduces the background on \cti representations, attack graphs, retrieval-augmented generation, hybrid retrieval, and local \slm{s}. Section~\ref{sec:related_work} compares the work with prior literature. Section~\ref{sec:methodology} defines the problem setting, threat and data model, formal concepts, and proposed pipeline. Section~\ref{sec:evaluation} presents the retrieval configuration study, the case study evaluation, the metrics, the manual annotation protocol, and the results. Section~\ref{sec:discussion} discusses the implications, utility, and limitations of the approach. Then, we analyze threats to validity. Finally, Section~\ref{sec:conclusion} concludes the paper.

\section{Background}
\label{sec:background}

In this section, we first review common \cti representations and the role of \mitre in describing adversary behavior. We then discuss attack-step semantics and attack-graph representations, which motivate the need for explicit \precond and \postcond. Finally, we introduce \rag, hybrid retrieval, and local \slm{s}, which provide the technical basis for the proposed enrichment pipeline.

\subsection{\acrlong{cti} Representations}
\label{sec:background_cti_representations}

\cti provides knowledge about adversaries, malware, campaigns, vulnerabilities, infrastructure, and observed behaviors. It is commonly used to support detection engineering, incident response, threat hunting, risk assessment, and strategic decision-making~\cite{AINSLIE2023103352,saeed_systematic_2023,WAGNER2019101589}. In practice, \cti is represented through a combination of unstructured reports, semi-structured descriptions, and standardized formats. This heterogeneity is useful because different sources capture different aspects of a threat, but it also creates integration challenges related to completeness, consistency, granularity, and data quality~\cite{sillaber_data_2016,qiang_quality_2018,yang_automated_2025}.

\mitre provides a structured knowledge base of adversary tactics, techniques, and procedures~\cite{strom_mitre_nodate}. Tactics describe adversary objectives, such as persistence, privilege escalation, credential access, lateral movement, and exfiltration, while techniques describe how adversaries may achieve those objectives. \mitre mappings are valuable because they provide a common behavioral vocabulary. Nevertheless, a technique label alone does not determine the concrete state of the target environment, the prerequisites of an observed action, or the direct consequences of that action in a specific campaign or report.

\stix is a widely used standard for representing and exchanging \cti~\cite{barnum_standardizing_nodate,krauss_analysis_2022}. It models threat intelligence through structured domain objects, such as malware, indicators, attack patterns, identities, tools, and reports, as well as relationship objects that connect them. This structure facilitates information exchange and automated processing. However, \stix objects may still contain textual fields that require interpretation, and their relations do not necessarily make explicit the operational assumptions and consequences associated with each adversary action.

MISP is a collaborative platform for sharing threat intelligence, especially events, attributes, indicators, and contextual information~\cite{10.1145/2994539.2994542}. It is used to collect and distribute \iocs, malware attributes, threat-actor information, and event-level metadata. MISP supports operational sharing, but the analyst must often combine these artifacts with external descriptions, reports, and taxonomies to interpret adversary behavior at the level of individual attack steps and their surrounding context.

\subsection{\mitre-Aligned Attack Steps}
\label{sec:background_attack_steps}

In this work, an attack step is understood as a unit of adversary behavior associated with one \mitre tactic and technique. An attack step may originate from a \stix attack-pattern object, a MISP event enriched with \mitre tags, a malware-analysis report, or another intermediate \cti representation. For instance, a report may state that a malware sample creates a Windows service for persistence and map this behavior to \emph{Create or Modify System Process: Windows Service} (T1543.003). The mapping identifies the type of behavior, but it does not necessarily specify the enabling conditions or resulting state changes.

This distinction is important. \mitre techniques describe classes of adversary behavior, whereas attack steps in a concrete \cti artifact refer to behavior observed, reported, or hypothesized in a specific context. The same \mitre technique may therefore correspond to different operational semantics depending on the malware family, compromised asset, privilege level, infrastructure, or report evidence. Consequently, enriching attack steps requires linking the step to evidence and making explicit the assumptions and effects that are otherwise implicit in textual descriptions.

\subsection{Attack Graphs, Attack Flows, and Step-Level Semantics}
\label{sec:background_attack_graphs}

Attack graphs are graph-based representations used to describe possible or observed adversary progressions~\cite{phillips_graph-based_1998,dantu_risk_2004,noel_correlating_2004,noel_measuring_nodate}. Classical attack graphs often model dependencies among vulnerabilities, privileges, network states, and attacker capabilities. In such settings, nodes and edges may represent conditions, exploits, privileges, reachability, or transitions between security states. These representations are useful for risk analysis, attack-path discovery, and reasoning about how one compromise may enable another.

More recent \cti-oriented graph representations organize information extracted from reports, structured artifacts, or \mitre mappings~\cite{li_attackg_2022,zhang_attackg_2025,hu_llm_tikg_2024}. These graphs can help analysts inspect relationships among threat actors, malware, techniques, infrastructure, and campaign-level behaviors. However, graph construction alone does not guarantee that each node has sufficient step-level semantics. A graph may group techniques into phases or connect related behaviors, but the analyst may still need to infer what must have been true before a step and what changed afterward.

Explicit \precond and \postcond provide a way to expose this missing semantics. A \emph{pre-condition} captures a required or enabling state before a step, such as access to a host, possession of credentials, reachability of a service, or availability of a vulnerable component. A \emph{post-condition} captures a state that plausibly follows from the step, such as a new persistence mechanism, an acquired credential, an executed payload, or an observable artifact. These conditions do not by themselves prove a complete causal chain, but they make the assumptions and consequences of attack steps inspectable and reusable.

\subsection{Local Small Language Models for Sensitive \acrshort{cti}}
\label{sec:background_local_slms}

Language models have increasingly been used for cyber threat analysis, including extraction, summarization, classification, question answering, and graph construction~\cite{arazzi_nlp-based_2025,naveed_comprehensive_2024,hu_llm_tikg_2024,zhang_attackg_2025}. Large models can provide strong general-language and reasoning capabilities, but their deployment may be constrained by privacy, confidentiality, cost, latency, and reproducibility requirements. These constraints are particularly relevant for \cti because the input may include sensitive incident details, internal infrastructure, defensive capabilities, or non-public intelligence.

Local \slm{s}~\cite{belcak_small_2025} offer an alternative deployment model. They can be executed within an organization's infrastructure, reducing the need to transmit sensitive intelligence to external services. Their smaller size may also make them easier to deploy in constrained environments. However, local \slm{s} may have weaker reasoning ability, smaller context windows, or lower instruction-following reliability than larger models. For this reason, they benefit from explicit task constraints, structured prompts, retrieval grounding, and post-generation validation.

\changed{Providing data to language models requires considering their context window, which defines the maximum sequence length of tokens that these models can simultaneously process and retain in their working memory during a single inference cycle.
Data can be appended to the model's textual input, \ie, the prompt, along with other instructions. This text is bounded by the context window limit.}

\changed{In this work, we use a \slm as a generator conditioned on retrieved \cti evidence. The \slm receives the evidence retrieved from the \rag architecture in its context; this evidence is used to generate the conditions and the enriched description. Thus, the \slm does not act as an unconstrained source of cyber-domain knowledge. As output, depending on the provided prompt, it produces a set of \precond, a set of \postcond, or an enriched description. For each generated content, we attach the evidence used to trace the results.}

\subsection{\acrlong{rag}}
\label{sec:background_rag}
\rag combines information retrieval and text generation by conditioning a language model on external context retrieved from a document store~\cite{gao_retrieval-augmented_2024,tural_retrieval-augmented_2024}. A typical \rag system comprises a retriever that selects relevant text units from a knowledge base and a generator, such as an \llm or \slm, that produces an output conditioned on the retrieved context. The central motivation is to reduce reliance on parametric model knowledge and to ground generated outputs in an explicit, updateable evidence corpus.

In \cti, \rag is useful because evidence about an attack step is often dispersed across semi-structured threat representations, technical reports, malware descriptions, indicators, and auxiliary artifacts. By retrieving and consolidating these fragments, a \rag system can provide the generator with local context relevant to the specific step being enriched. This is especially important when the desired output must remain auditable: the analyst should be able to inspect which evidence units supported a generated description, \precond, or \postcond.

The effectiveness of \rag depends on the quality of the evidence units made available for retrieval. Chunking splits documents into smaller units to support efficient retrieval of semantically coherent evidence~\cite{kucecka_selective_2013,smith2024evaluating}. In \rag pipelines, chunking affects both recall and precision. Small chunks may separate related facts, while large chunks may mix relevant and irrelevant information. Common approaches include fixed-size chunking, which is simple but may fragment meaning, and semantic chunking, which attempts to preserve coherent boundaries. Chunk overlap can mitigate boundary effects but may also introduce redundancy.

\changed{In this work, the \rag architecture is the module providing the \slm evidence about the considered \as. We query the \rag with all details useful for identifying relevant chunks of evidence; thus, it returns the top-$k$ chunks according to the retriever's parameters.}

\subsection{Hybrid Retrieval}
\label{sec:background_hybrid_retrieval}

Hybrid retrieval combines lexical retrieval with embedding-based semantic retrieval~\cite{bhagdev_hybrid_2008}. Lexical retrieval ranks documents by explicit term matching, \eg, using models such as \bm~\cite{bm25,chen_keyword-based_2011}. Embedding-based retrieval represents queries and documents in a dense vector space and ranks documents by semantic similarity~\cite{huang_embedding-based_2020}. The two retrieval families provide complementary strengths.

This combination is well-suited to \cti. Some elements should be matched exactly, including \mitre identifiers, malware names, file paths, registry keys, domains, IP addresses, service names, and command-line fragments. Other elements may be described with variable terminology across reports, such as persistence mechanisms, credential theft, command-and-control behavior, or lateral movement. Lexical retrieval can achieve high precision for exact artifacts, whereas semantic retrieval can improve robustness to paraphrasing and vocabulary mismatches. In practice, lexical and semantic scores can be computed independently and combined, for instance, through weighted scoring, to select candidate evidence units for generation.

\section{Related Work}
\label{sec:related_work}
This work intersects structured \cti extraction, attack-graph and knowledge-graph construction, retrieval-augmented \cti analysis, and evidence-grounded generation. Several of these capabilities have been studied individually, including explicit \emph{pre-} and \emph{post-condition} extraction. We therefore position our contribution not as a new mechanism in isolation, but as a distinct \emph{evidence-linked attack-step enrichment task}: given an analyst-guided representation, our pipeline augments existing steps with case-specific descriptions, \precond, \postcond, and traceable evidence references using hybrid retrieval and a locally deployed \slm. The output is an inspectable enrichment of the supplied scaffold, not a reconstructed or formally verified causal attack trace. Table~\ref{tab:related_comparison} summarizes this positioning.

\subsection{\cti Extraction and \mitre Mapping}
\label{sec:related_cti_extraction}
A substantial body of research extracts indicators, entities, relations, adversarial behaviors, and ATT\&CK techniques from unstructured \cti reports. Alam\etal~\cite{alam2023lookingiocsautomaticallyextracting} investigate the extraction of attack patterns beyond indicators of compromise, while Orbinato\etal~\cite{orbinato_automatic_2022} study the automatic mapping of unstructured \cti to attack techniques. aCTIon~\cite{siracusano_action_2023} provides an analyst-curated benchmark and an automated approach for extracting structured STIX intelligence from real-world reports. Related evaluation resources include CTI-HAL~\cite{ctihal}, a human-annotated ATT\&CK-structured \cti dataset, and CTISum~\cite{ctisum}, a benchmark for \cti summarization including attack-process summaries.

More recent work makes grounding an explicit part of technique extraction. CTI2TTP~\cite{ttprint2026} decomposes reports into atomic behaviors, anchors candidate ATT\&CK techniques to localized source spans, and verifies candidates against both those spans and the corresponding technique definitions. Such methods address \emph{which} behaviors or techniques are present and supported by a report. Our setting begins after this identification stage: candidate steps and ATT\&CK metadata are provided in an analyst-guided intermediate representation, and the goal is to enrich each step with its case-specific enabling assumptions, direct consequences, and supporting evidence.

\subsection{Attack Graphs, Knowledge Graphs, and Step Semantics}
\label{sec:related_attack_graphs}
Classical attack graphs represent possible adversarial paths through vulnerabilities, privileges, and security states for attack-path and risk analysis~\cite{phillips_graph-based_1998,dantu_risk_2004,noel_correlating_2004,noel_measuring_nodate}. In contrast, \cti-oriented systems increasingly construct graph representations from incident narratives. AttacKG~\cite{li_attackg_2022} extracts behavior graphs and aggregates technique-level knowledge across reports. AttacKG+~\cite{zhang_attackg_2025} uses language models to build temporally organized attack knowledge graphs, including behavior, ATT\&CK, and state information. CRUcialG~\cite{cheng2024crucialgreconstructintegratedattack} reconstructs attack-scenario graphs and incorporates tactical-phase checks and graph-repair procedures.

Other approaches focus on broader entity--relation knowledge graphs. LLM-TIKG~\cite{hu_llm_tikg_2024} uses language models for topic classification and entity--relationship extraction; Liu and Zhan~\cite{10386611} likewise investigate language-model-based \cti knowledge-graph construction. CTINexus~\cite{ctinexus} employs optimized in-context learning, demonstration retrieval, and entity/relation alignment to construct cybersecurity knowledge graphs from limited annotated data. These systems provide structured threat knowledge, whereas our pipeline enriches the semantics of \emph{existing} ATT\&CK-aligned attack steps without claiming to infer all missing events or dependencies.

The closest work on state-transition semantics is the framework of Hou\etal~\cite {hou_reachable_2026}. It explicitly extracts attack units comprising preconditions, attack behavior, and postconditions, normalizes these elements into predicates, repairs dependencies, and compiles rules for reachability analysis. Thus, pre-/post-condition generation is \emph{not} novel by itself. The methodological distinction lies in the objective and starting point: Hou\etal automate the extraction and repair of reasoning-ready reachable chains, whereas we retain an analyst-provided ATT\&CK/STIX scaffold, attach natural-language conditions and descriptions to its existing steps, and preserve references to supporting report evidence. Our representation is intended for analyst inspection rather than formal attack-goal reachability.

\subsection{Retrieval-Augmented \cti Analysis}
\label{sec:related_llm_rag_cti}
Retrieval-augmented language-model systems increasingly incorporate external security knowledge into \cti analysis. IntelEX~\cite{xu_intelex_2024} retrieves ATT\&CK knowledge using a vector database, applies language-model judging to mitigate incorrect technique assignments, and uses GraphRAG to generate attack-level procedure intelligence. CTIConnect~\cite{cticonnect} benchmarks retrieval-augmented reasoning across heterogeneous \cti sources and demonstrates that domain-specific retrieval strategies are important when moving between analyst narratives and structured security knowledge. CTIFoundry~\cite{ctifoundry2026} provides an agent-oriented substrate with typed ontology links, provenance-carrying report chunks, and hybrid dense--lexical retrieval.

The mixture of exact artifacts (e.g., malware names, ATT\&CK identifiers, file paths, and commands) and paraphrased behaviors makes both lexical and semantic signals relevant to \cti retrieval. General hybrid-retrieval research such as Blended RAG~\cite{sawarkar_blended_2024} also motivates combining sparse and dense retrieval. Our pipeline uses a fixed, weighted combination of \bm and dense-vector retrieval, but differs in its downstream purpose: rather than open-ended question answering, autonomous investigation, or new technique extraction, the retrieved context is bounded to each supplied attack step and used to generate structured, evidence-linked enrichment fields. The configuration is pragmatic; we do not claim \cti-specific retrieval optimality.

\subsection{Evidence Grounding, Provenance, and Verification}
\label{sec:related_grounding_evaluation}

Retrieving relevant documents does not, by itself, establish that generated claims are supported. Provenance~\cite{sankararaman2024provenance} uses compact natural-language-inference models to detect nonfactual retrieval-augmented outputs and associate them with context chunks. In the \cti domain, KGV~\cite{kgv2024} combines language-model processing with knowledge graphs for claim credibility assessment, while LRCTI~\cite{tang2025lrctilargelanguagemodelbased} iteratively retrieves evidence and applies credibility verification with interpretable justifications. TTPrint~\cite{ttprint2026} applies evidence-based verification to ATT\&CK candidate labels.

TRACE-CTI~\cite{trace_cti_2026} addresses a complementary governance task by retaining evidence, extraction configuration, model provenance, and validation history for ATT\&CK claims. Our pipeline instead integrates evidence references into the generated \emph{step-level representation}: descriptions and generated conditions remain traceable to retrieved evidence units.

\subsection{Positioning of This Work}
\label{sec:related_positioning}

The literature provides precedents for structured extraction~\cite{siracusano_action_2023}, attack-graph and knowledge-graph construction~\cite{li_attackg_2022,zhang_attackg_2025,ctinexus}, explicit pre-/post-condition semantics~\cite{hou_reachable_2026}, hybrid \cti retrieval and retrieval-augmented analysis~\cite{cticonnect,ctifoundry2026}, and evidence provenance~\cite{sankararaman2024provenance,trace_cti_2026}. Our contribution lies in their \emph{task-specific integration}: a locally deployable pipeline that preserves an analyst-guided, tactic-aligned ATT\&CK/STIX step scaffold while enriching each step with descriptions, explicit \precond and \postcond, and field-level evidence links. This provides an inspectable intermediate representation for subsequent analyst reasoning without asserting complete temporal ordering, causal reconstruction, or automatically verified attack feasibility.

\begin{table*}[t]
\centering
\caption{Comparison with representative approaches. \(\checkmark\): explicitly supported; \(\sim\): partial or ancillary; \(\times\): outside the stated design; ---: not established in the source.}
\label{tab:related_comparison}
\small
\setlength{\tabcolsep}{5pt}
\renewcommand{\arraystretch}{1.12}
\begin{tabular}{@{}lcccccccc@{}}
\toprule
\textbf{Work} & \textbf{Graph} & \textbf{Retr.} & \textbf{Prov.} & \textbf{Pre/Post} & \textbf{Local} & \textbf{Hybrid} & \textbf{Scaff.} \\
\midrule
AttacKG~\cite{li_attackg_2022}  & \(\checkmark\) & \(\times\) & \(\sim\) & \(\times\) & --- & \(\times\) & \(\times\) \\
AttacKG+~\cite{zhang_attackg_2025} & \(\checkmark\) & \(\times\) & \(\sim\) & \(\times\) & --- & \(\times\) & \(\times\) \\
CRUcialG~\cite{cheng2024crucialgreconstructintegratedattack} & \(\checkmark\) & --- & \(\sim\) & \(\times\) & --- & --- & \(\times\) \\
CTINexus~\cite{ctinexus} & \(\checkmark\) & \(\sim\) & --- & \(\times\) & --- & --- & \(\times\) \\
IntelEX~\cite{xu_intelex_2024} & \(\sim\) & \(\checkmark\) & \(\sim\) & \(\times\) & --- & --- & \(\times\) \\
TTPrint~\cite{ttprint2026} & \(\times\) & \(\sim\) & \(\checkmark\) & \(\times\) & --- & --- & \(\times\) \\
CTIConnect~\cite{cticonnect} & \(\times\) & \(\checkmark\) & \(\sim\) & \(\times\) & --- & \(\sim\) & \(\times\) \\
CTIFoundry~\cite{ctifoundry2026} & \(\checkmark\) & \(\checkmark\) & \(\checkmark\) & \(\times\) & \(\sim\) & \(\checkmark\) & \(\times\) \\
TRACE-CTI~\cite{trace_cti_2026} & \(\checkmark\) & --- & \(\checkmark\) & \(\times\) & --- & --- & \(\times\) \\
Hou\etal~\cite{hou_reachable_2026} & \(\checkmark\) & --- & \(\sim\) & \(\checkmark\) & --- & --- & \(\times\) \\
\midrule
\textbf{This work} & \(\checkmark\) & \(\checkmark\) & \(\checkmark\) & \(\checkmark\) & \(\checkmark\) & \(\checkmark\) & \(\checkmark\) \\
\bottomrule
\end{tabular}
\vspace{2pt}
\parbox{\textwidth}{\footnotesize\emph{Columns:} Graph = structured graph output; Retr. = evidence/knowledge retrieval; Prov. = evidence provenance; Pre/Post = explicit step-level pre-/post-conditions; Local = local/small-model deployment focus; Hybrid = combined lexical and dense retrieval; Scaff. = enrichment of a pre-existing analyst-guided attack-step scaffold.}
\end{table*}

\section{Methodology}
\label{sec:methodology}

This section defines evidence-linked enrichment of analyst-defined attack steps as a transformation from a structured threat representation and heterogeneous \cti evidence into a richer, inspectable representation. It specifies the input assumptions, output semantics, retrieval and generation stages, and construction of the tactic-aligned graph. The pipeline preserves the initial analytical scaffold while adding contextual descriptions, candidate \precond and \postcond, and provenance links. Its explicit separation of input structure, generated assertions, and supporting evidence enables the empirical assessment of each enrichment field.

\subsection{Problem Formulation}
\label{sec:problem_formulation}

Let $D=\{d_1,\ldots,d_n\}$ be a local corpus of heterogeneous \cti sources. The corpus may include malware analysis reports, vendor blogs, structured indicators, and other analyst-provided documents. Let $T$ be an analyst-guided intermediate representation, such as a \stix bundle or an equivalent structure derived from MISP or another \cti format. We assume that $T$ contains, or can be mapped to, a set of attack-step candidates with associated \mitre tactic and technique information. The mapping from a particular external format to the internal representation is an input-adaptation task; it is not an autonomous inference of attack steps from unstructured evidence.

The task is to produce an evidence-linked enriched attack-step graph \(G=(M,A,E)\),
where $M$ is a set of tactic-level milestone nodes, $A$ is a set of enriched attack-step nodes, and $E$ is a set of relations supported by the intermediate representation or analyst input. The graph retains the organizational information available in $T$ while incorporating additional semantic annotations. It does not introduce independently verified temporal or causal relationships between attack steps. Each enriched attack step is associated with: (\emph{i}) a normalized step description $\delta$, (\emph{ii}) a set of \precond $Pre(\cdot)$, (\emph{iii}) a set of \postcond $Post(\cdot)$, and (\emph{iv}) evidence links to the chunks used during generation.

Formally, the enrichment process can be viewed as a transformation $\mathcal{F}$ that takes an initial attack representation and a supporting evidence corpus as input and produces an augmented representation:
\(
\mathcal{F}(T,D) = G.
\)
The transformation preserves the distinction between the input attack-step structure, the evidence retrieved from $D$, and the semantic content generated by the model. This distinction is important because the input structure, retrieved passages, and generated statements may have different levels of reliability and may require different forms of analyst validation.

The transformation makes candidate requirements and consequences explicit while preserving the evidence context needed to inspect them. The attack-step structure and any ordering constraints remain those supplied by the analyst; the enrichment stage does not independently infer occurrence, causality, or temporal completeness.

The evaluation of this transformation distinguishes three properties. First, the semantic quality of generated conditions concerns whether the statements are meaningful, atomic, and assigned to the appropriate precondition or postcondition category (RQ1). Second, evidence alignment concerns whether generated statements are adequately supported by the retrieved intelligence and whether descriptions remain faithful to the represented attack step (RQ2). Third, local execution characteristics concern the processing cost of performing retrieval and generation on the evaluated hardware (RQ3). These properties are assessed separately because a well-formed condition may lack evidential support, an evidence-associated description may contain unsupported details, and successful local execution does not by itself establish operational efficiency.

\subsection{Threat and Data Model and Assumptions}
\label{sec:assumptions}
The proposed methodology operates on analyst-selected \cti and assumes an analyst-guided workflow. Two principal inputs are provided: an intermediate threat representation $T$ and a heterogeneous evidence corpus $D$. The representation $T$ supplies the initial attack-step candidates and any available tactic, technique, milestone, source, or ordering information. The corpus $D$ supplies the material from which contextual enrichments are generated. The framework does not assume that the documents provide a complete account of the intrusion or that every condition relevant to an attack step is explicitly documented.

We make the following assumptions:
\begin{enumerate}
\item \textbf{Analyst-guided structure.} The input representation $T$ is syntactically valid and contains attack-pattern, malware, or equivalent objects from which candidate attack steps can be derived. The quality and completeness of this initial structure depend on the upstream analyst or \cti production process.

\item \textbf{Evidence availability.} The document corpus $D$ contains at least some material relevant to the attack steps represented in $T$. It may nevertheless be incomplete, redundant, inconsistent, or insufficient to support particular generated statements. Relevance of the corpus does not imply that all step-level requirements and effects can be established from it.

\item \textbf{Analyst-controlled corpus.} Documents are collected, selected, and sanitized by an analyst or trusted \cti workflow before ingestion. The proposed implementation assumes this controlled ingestion process; it does not implement a general-purpose mechanism for securely processing arbitrary, potentially adversarial external content.

\item \textbf{Bounded generation.} The \slm receives the attack-step metadata and a bounded set of retrieved evidence units. Prompt instructions constrain the requested output to descriptions, \precond, and \postcond associated with the step. These constraints are intended to encourage evidence-based generation, but they do not guarantee that the model will avoid unsupported inferences or the use of knowledge not established by the supplied passages.

\item \textbf{No adversarial poisoning.} Knowledge-base poisoning, prompt injection, and malicious manipulation of the evidence corpus are outside the scope of this work. These threats remain relevant even in a local deployment and would require additional safeguards before processing untrusted intelligence at scale.

\item \textbf{Analyst-facing output.} Generated descriptions, \precond, and \postcond are analytic artifacts intended to support inspection. They should not be interpreted as verified forensic facts unless independently validated.
\end{enumerate}

These assumptions define a bounded analyst-support task: the pipeline exposes candidate step-level semantics and their evidence context for review. Its effect on analyst effort and decisions is a separate operational question, beyond the output-quality evaluation considered here.

\subsection{Formal Definitions}
\label{sec:formal_definitions}

We define the main objects used by the pipeline.

\paragraph{Evidence unit.}
An evidence unit is a retrievable fragment extracted from the corpus $D$. We denote the set of evidence units as $\mathcal{E}=\{e_1,\ldots,e_m\}$. Each evidence unit $e_j$ is represented as:
\[
e_j = \langle id_j, text_j, source_j, type_j, meta_j \rangle,
\]
where $id_j$ is a unique identifier, $text_j$ is the textual content, $source_j$ identifies the originating document or object, $type_j$ describes the source type, and $meta_j$ stores provenance metadata such as token count and chunking strategy.

An evidence unit is a retrieval and provenance object, not a verified factual assertion. Its inclusion in the retrieved context indicates that the retrieval procedure selected it as potentially relevant to the attack step. The evidential relationship between a passage and a particular generated statement must be assessed separately.

\paragraph{Attack step.}
An attack step is a unit of adversary behavior derived from $T$ and aligned with one or more \mitre tactics and techniques. We represent an attack step as:
\[
a_i = \langle id_i, label_i, desc_i, tactic_i, technique_i \rangle,
\]
where $id_i$ is the step identifier, $label_i$ is the step name, $desc_i$ is the initial textual description, and $tactic_i$ and $technique_i$ are \mitre references.

The attack step is treated as part of the input analytical scaffold. Its presence in $T$ does not independently establish that the corresponding behavior was observed during the incident, and its initial description may omit implementation details that are available elsewhere in the evidence corpus. The enrichment process supplements this initial representation without replacing its provenance or asserting that the input mapping has been independently validated.

\paragraph{Pre-condition.}
A \emph{pre-condition} for an attack step $a_i$ is an atomic statement describing what must plausibly hold before $a_i$ can occur. It may express an environmental requirement, an attacker capability, a privilege level, a configuration state, a dependency, or a prior system state. We denote the set of \precond for $a_i$ as:
\[
Pre(a_i)=\{pre^{i,1},\ldots,pre^{i,r}\}.
\]

In this work, a generated pre-condition is a candidate analytical requirement associated with the step. It is not necessarily an observed property of the compromised environment. For example, a technique may generally require a particular capability, but the available report may not explicitly establish that capability or describe how it was obtained. The distinction between a plausible requirement and a source-supported incident fact is therefore relevant when interpreting the output.

\paragraph{Post-condition.}
A \emph{post-condition} for an attack step $a_i$ is an atomic statement describing what becomes true, changed, observable, or enabled after successful execution of $a_i$. It may represent a created artifact, a modified configuration, a process execution, a telemetry event, a credential exposure, a persistence mechanism, or an attacker capability. We denote the set of \postcond for $a_i$ as:
\[
Post(a_i)=\{post^{i,1},\ldots,post^{i,s}\}.
\]
A \emph{post-condition} should describe a direct consequence of the current step rather than a generic downstream possibility.

Post-conditions are interpreted relative to the successful execution of the represented action. Their generation does not establish that the action succeeded in a particular incident. In addition, an effect that is technically plausible for a \mitre technique is not necessarily supported for the specific malware implementation under analysis. The intended representation therefore favors step-specific effects that can be related to the available evidence.

\paragraph{Evidence support.}
For each generated field $x$, including a description, \precond, or \postcond, the pipeline stores a support set:
\[
Supp(x)\subseteq \mathcal{E}.
\]
The support set contains the evidence units cited for $x$. The term \emph{support set} denotes the evidence associated with a generated field, rather than a guarantee of logical entailment or factual correctness. The associated evidence units provide a basis for inspecting the generation context. During manual evaluation, annotators assess whether the generated statements are actually supported by the available evidence.

Where the same retrieved context is associated with multiple generated fields, the provenance links identify the shared evidence context rather than proving that every passage supports every field. Consequently, evidence association and evidence-support correctness are treated as distinct concepts throughout the evaluation.

\paragraph{Enriched attack step.}
An enriched attack step is defined as:
\[
\hat{a}_i = \langle a_i, \delta_i, Pre(a_i), Post(a_i), Supp_i \rangle,
\]
where $a_i$ is the original attack step, $\delta_i$ is the enriched description, $Pre(a_i)$ and $Post(a_i)$ are the generated condition sets, and $Supp_i$ contains provenance links between generated fields and evidence units.

The enriched step retains the original attack-step information while adding model-generated contextual annotations. This separation allows the output to be inspected as an augmentation of the input representation rather than as a replacement for the analyst-provided account. The provenance component identifies the evidence context used for enrichment, while the quality and validity of the generated fields remain to be evaluated.

\subsection{Pipeline Overview}
\label{sec:pipeline_overview}

\changed{In the following, we illustrate the blocks and flows in Figure~\ref{fig:high_level_schema}, which summarizes the proposed pipeline}. The first blocks perform \emph{ingestion and preparation}. Starting from the analyst-provided inputs, namely the initial intermediate threat representation, the heterogeneous documents, and the pipeline configuration, the pipeline (1) loads the attack representation, (2) normalizes the input documents, segments them into evidence units (chunks), and (3) builds the corresponding lexical and vector indexes. Each evidence unit is stored with stable metadata, including document identifiers, chunk identifiers, source references, and indexing information. This creates a traceable evidence base that can be reused by the downstream components.

\begin{figure*}[t!]
\centering
\makebox[\textwidth][c]{%
\resizebox{1\textwidth}{!}{%
\begin{tikzpicture}[
  font=\large,
  every node/.style={align=center},
  box/.style={draw, minimum width=3.1cm, minimum height=1.1cm},
  light/.style={box, fill=gray!20},
  plain/.style={box, fill=white},
  arrow/.style={->, >=Stealth, line width=0.6pt},
  dashedbox/.style={draw, dashed, rounded corners, inner sep=6pt, thick}
]

\newcommand{\TrapH}{1.35cm}
\tikzset{
  trap/.style={
    draw, trapezium, trapezium left angle=70, trapezium right angle=110,
    minimum width=2.8cm, minimum height=\TrapH
  }
}
\node[trap] (pre) {PRE\\CONDITIONS};
\node[trap, right=0.25cm of pre] (post) {POST\\CONDITIONS};
\node[trap, right=0.25cm of post, minimum width=3.6cm] (attack) {ENRICHED\\DESCRIPTION};
\node[dashedbox, fit=(pre)(post)(attack), inner sep=10pt,
      label={[font=\large, anchor=north east]below left:Evidence-Linked Attack-Step Enrichment}] (attackbox) {};

\coordinate (ragcenter) at ($(attackbox.north west)!0.74!(attackbox.north east)+(0,2.9cm)$);

\node[plain, anchor=center] (vs)
  at ($(ragcenter)+(-1.55cm,0.55cm)$) {\normalsize VECTOR STORE};

\node[plain, anchor=center] (bm25)
  at ($(ragcenter)+(1.55cm,0.55cm)$) {\normalsize LEXICAL\\RETRIEVER};

\node[plain, anchor=center, minimum width=6.2cm] (retriever)
  at ($(ragcenter)+(0,-0.55cm)$) {\normalsize ENSEMBLE RETRIEVER};

\node[dashedbox, fit=(vs)(bm25)(retriever)] (ragbox) {};
\node[font=\large, above=0.1cm of ragbox.north] {RAG Architecture};

\def\ChunksGap{3.2cm}
\node[plain, anchor=west] (chunks)
  at ($ (ragbox.east |- ragbox.center) + (\ChunksGap,0) $) {CHUNKS};

\node[light, anchor=west] (files)
  at ($ (chunks.north) + (-1.85,2.0cm) $) {HETEROGENEOUS\\DOCUMENTS};

\draw[arrow] (files.south) -- (chunks.north)
  node[midway, fill=white, inner sep=1pt]{chunking}
  node[midway, right=10mm, circle, draw, inner sep=1pt, font=\small, fill=white] {2};

\draw[arrow] (chunks.west) -- (ragbox.east)
  node[midway, above, fill=white, inner sep=1pt]{embedding}
  node[midway, above=5mm, circle, draw, inner sep=1pt, font=\small, fill=white] {3};

\def\BelowGap{0.6cm}
\def\UserThreatGap{0cm}

\node[plain, anchor=north] (interm)
  at ($(attackbox.south)+(0,-\BelowGap)$) {THREAT\\REPRESENTATION};

\node[light, anchor=north] (ioc)
  at ($(interm.south)+(-\UserThreatGap,-1.5cm)$) {USER};

\draw[arrow] (ioc.north) -- (interm.south)
  node[midway, fill=white, inner sep=1pt]{preliminary\\modelization}
  node[midway, left=15mm, circle, draw, inner sep=1pt, font=\small, fill=white] {1};

\coordinate (intermTop) at ($(interm.north |- attackbox.south)$);
\draw[arrow] (interm.north) -- (intermTop)
  node[midway, right=2mm, fill=white, inner sep=1pt] {};

\node[box, left=2.2cm of retriever, minimum width=4.3cm] (llm) {\large SLM};

\coordinate (ctxX)   at ($(attackbox.north west)!0.12!(attackbox.north east)$);
\coordinate (ctxTop) at ($(ctxX |- llm.south)$);
\draw[arrow] (ctxX) -- (ctxTop)
  node[midway, anchor=center, fill=white, inner sep=1pt]{context \&\\prompt}
  node[midway, left=13mm, circle, draw, inner sep=1pt, font=\small, fill=white] {6};

\coordinate (respX)   at ($(attackbox.north west)!0.30!(attackbox.north east)$);
\coordinate (respTop) at ($(respX |- llm.south)$);
\draw[arrow] (respTop) -- (respX)
  node[midway, anchor=center, fill=white, inner sep=1pt]{response}
  node[midway, right=10mm, circle, draw, inner sep=1pt, font=\small, fill=white] {7};

\coordinate (ragDropA) at ($(ragbox.south west)!0.35!(ragbox.south east)$);
\coordinate (ragDropB) at ($(ragbox.south west)!0.72!(ragbox.south east)$);
\coordinate (flowTopA) at ($(ragDropA |- attackbox.north)$);
\coordinate (flowTopB) at ($(ragDropB |- attackbox.north)$);

\draw[arrow] (flowTopA) -- (ragDropA)
  node[midway, anchor=center, fill=white, inner sep=1pt]{fetch relevant\\chunks}
  node[midway, left=15mm, circle, draw, inner sep=1pt, font=\small, fill=white] {4};

\draw[arrow] (ragDropB) -- (flowTopB)
  node[midway, anchor=center, fill=white, inner sep=1pt]{chunks}
  node[midway, right=10mm, circle, draw, inner sep=1pt, font=\small, fill=white] {5};

\node[plain, right=1cm of attackbox.east, minimum width=3.3cm] (refine) {SEMANTIC DEDUPLICATION\\AND GRAPH UPDATE};
\node[light, right=1cm of refine] (out) {ENRICHED\\REPRESENTATION};

\draw[arrow] (attackbox.east) -- (refine)
  node[midway, below=5mm, circle, draw, inner sep=1pt, font=\small, fill=white] {8};

\draw[arrow] (refine) -- (out)
  node[midway, below=5mm, circle, draw, inner sep=1pt, font=\small, fill=white] {9};

\end{tikzpicture}
}%
}%
\caption{High-level overview of the proposed analyst-guided enrichment workflow. Heterogeneous \cti documents are segmented into evidence units and indexed for hybrid retrieval. An analyst-provided threat representation supplies the initial attack-step scaffold. For each step, the retriever selects potentially relevant evidence, and a local \slm generates a contextual description and candidate \precond and \postcond. Semantic deduplication removes sufficiently similar generated conditions before enriching the output representation. The resulting graph retains the attack-step structure, tactic-level organization, and links to the evidence used during generation; these links do not constitute independent verification of the generated statements.}
\label{fig:high_level_schema}
\end{figure*}

For each attack step, the pipeline then constructs a retrieval query (4) using the available step information, including the malware name, \mitre technique identifier, technique name, tactic, original description, and relevant structured fields. The hybrid retriever combines lexical and semantic retrieval and returns the top-ranked evidence units (5) according to the configured retrieval parameters. The retrieval record retains the query, selected chunk identifiers, source metadata, ranks, and latency. These records characterize what the system retrieved; they do not independently establish the relevance or correctness of the retrieved passages.

The selected evidence units are formatted as source-labeled context blocks (6) together with the attack-step metadata. Separate local \slm calls generate an enriched description, candidate \precond, and candidate \postcond (7). Context size is limited by the retrieval cut-off and the configured \slm context window; the formatting function does not truncate individual evidence chunks. The output fields are intended to be concise and step-specific. Their provenance is attached deterministically: each generated description and condition records all retrieved evidence-chunk identifiers, rather than a model-selected subset of passages. This provides the generation context for inspection, while the actual support of each claim is assessed separately.

The pipeline subsequently applies semantic deduplication (8) to the generated condition sets. \changed{This operation is intended to reduce repeated or semantically overlapping statements by computing the semantic similarities between the generated conditions. Overlapping conditions are removed}. 
\changed{We highlight that neither the embedding model nor the \slm is fine-tuned for this task.}
Finally, the enriched descriptions, retained conditions, and provenance information are incorporated into the initial attack representation (9). 
The output preserves the analyst-guided structure and organizes enriched steps into tactic-level milestones. It should be interpreted as an evidence-linked analytical representation rather than a verified reconstruction of adversarial execution.

\subsection{Document Ingestion and Evidence Creation}
\label{sec:ingestion_creation}
\changed{To illustrate the methodology, consider the \emph{Goofy Guineapig} example introduced in Section~\ref{sec:introduction}. Given an initial representation (\eg, a STIX representation) that models the attack, we assume that heterogeneous documents are already available. In this case, we gathered 7 documents, spanning PDFs and unstructured formats, \eg, YAML files. These documents may have been gathered using different strategies, \eg, by consulting specialized platforms and blogs or by ingesting feeds. We document our selection process in Section~\ref{sec:datasets_case_studies}.
Once these documents are available and loaded (1), the pipeline produces the evidence units to be stored in the \rag architecture (2, 3). Producing the evidence means applying the chunking process detailed in Section~\ref{sec:preprocessing_evaluation} to each available heterogeneous document. In the case of this running example, 175 evidence were created. Nevertheless, the initial representation is parsed to extract the \as{s} and the other useful information (\eg, the malware name). In this case, 16 attack steps are available in the STIX file.}

\subsection{Retrieval and Evidence Selection}
\label{sec:retrieval_evidence_selection}
\changed{Consider that the input representation (the STIX objects) contains an attack step describing persistence through a Windows service, mapped to \mitre technique T1543.003. The initial step identifies the relevant behavior but may omit details about the environment, required capabilities, service configuration, and consequences of execution.}
The retrieval stage (4,5) selects a bounded set of evidence units for each attack step. \changed{The pipeline constructs a retrieval query from the malware name, the original step description, and the tactic and technique information. In our example, the retrieval query contains \emph{Goofy Guineapig}, the description provided by the STIX file, and the information concerning TA0003 (Persistence) and T1543.003 (Create or Modify System Process: Windows Service).}

\changed{The hybrid retriever combines the most relevant chunks retrieved using different strategies (Section~\ref{sec:background_hybrid_retrieval}).} In this setup, we combine the results of a semantic and lexical retriever. 
For this purpose, we use Rank fusion~\cite{cormack_reciprocal_2009}, which combines complementary lexical and semantic retrieval signals without requiring their numerical similarity scales to be comparable. Lexical matching can favor exact identifiers and technical terminology, whereas dense retrieval can identify passages expressed using different vocabulary. The combined ranked list provides a bounded, inspectable evidence context for generation.

Retrieval relevance and generation faithfulness are distinct properties. A passage may be indirectly relevant to the attack step without supporting a particular precondition, postcondition, or description detail. Conversely, a generated statement may be technically plausible but unsupported by the selected evidence. The methodology therefore records retrieval provenance and evaluates the resulting content separately, rather than treating retrieval scores or the presence of evidence links as indicators of factual correctness.

At the end of this stage, the pipeline has available a set of evidence considered relevant to the \as representing the technique T1543.003. The same process is repeated for all the \as{s} in the initial representation.

\subsection{Evidence-Grounded Enrichment and Semantic Deduplication}
\label{sec:evidence_grounded_enrichment}

\changed{The set of retrieved evidence and attack information (\eg, the malware name) are used to construct the instruction prompt of the \slm (6,7). Using the selected context, the local \slm generates a description that characterizes the reported \as, \eg, the persistence mechanism (T1543.003), and proposes separate \precond and \postcond.} For instance, a candidate \emph{pre-condition} might concern execution on a Windows host, while another might concern the permissions required to create or modify the relevant service configuration. A candidate \emph{post-condition} might describe the existence of a configured service entry supporting persistence. The precise statements generated for a particular case depend on the retrieved context and model output, and should not be interpreted as verified findings merely because they are technically plausible. \changed{Interestingly, different failure modes may be observed during the generation process; we documented them in Section~\ref{sec:limitations_failure_modes}.}

\changed{Subsequently, the pipeline removes redundant statements within each attack step, comparing normalized text separately for the two declared condition types and retaining the first occurrence. Moreover, the pipeline performs semantic deduplication of \precond and \postcond separately, using a user-defined similarity threshold. In our experiments, we used a threshold of 0.7. We discuss the effects of such choice, along with
the validity of the generated descriptions and conditions is assessed through the evaluation procedure described in Section~\ref{sec:evaluation}.}

\subsection{Attack Graph Construction and Output Schema}
\label{sec:attack_step_graph_schema}

\changed{The resulting enriched step retains its original \mitre mapping and is associated with the evidence units supplied during generation. An analyst can inspect these passages to determine whether each generated statement is supported, whether a purported pre-condition is actually a prerequisite, and whether a post-condition describes a direct effect of the represented action. The enriched representation does not establish that the complete attack chain occurred, do not contain all possible consequences of the persistence mechanism, nor can be used to validate causal dependencies with other steps. Instead, it exposes candidate step-level semantics and their evidence context for subsequent manual assessment.}

\changed{Finally, the enriched attack steps are organized} into an evidence-linked graph-based representation. The graph retains the structural information supplied by the intermediate representation and adds the generated descriptions, condition sets, and evidence provenance. It contains two principal types of nodes:
\begin{itemize}
\item \textbf{Milestone nodes}, representing \mitre tactics or other high-level phases available in the intermediate representation;
\item \textbf{Attack-step nodes}, representing enriched \mitre-aligned behaviors derived from the threat representation.
\end{itemize}

Milestone nodes provide a coarse organizational structure, while attack-step nodes retain the contextual information associated with individual behaviors. The relations in the output reflect the available input structure and any ordering constraints represented by that structure. They should not be interpreted as causal dependencies established by the language model or verified through independent forensic evidence.

Let $M_i$ and $M_{i+1}$ denote two consecutive milestones in the input representation, and let $A_i$ be the set of attack steps associated with the corresponding milestone interval. When ordering constraints are available, they may be represented by a partial-order relation $\preceq_i$ over $A_i$. The resulting structure is denoted by $(A_i,\preceq_i)$, where the relation expresses only the ordering information available from the input representation. In the absence of an ordering constraint between two distinct steps, their relative execution order remains unspecified. Thus, the representation does not impose a total sequence on the steps within a milestone interval.

The partial-order interpretation is organizational rather than evidential: it does not demonstrate that the encoded ordering occurred in the analyzed incident, nor does it establish that one step causally enables another. Similarly, a \emph{post-condition} associated with one step and a \emph{pre-condition} associated with another do not, by themselves, constitute a validated dependency edge. Establishing such relationships would require additional reasoning rules and evidence beyond the enrichment procedure considered here.

The resulting graph therefore provides a structured view of tactic-level organization, attack-step context, candidate enabling conditions, candidate effects, and evidence provenance. Its principal purpose is to support the inspection and interpretation of an existing threat representation while leaving unsupported ordering and causal relationships unresolved.

\changed{Concerning our running example, we provide the resulting graph in Figure~\ref{fig:case1_graph}. Table~\ref{tab:goofy_example_enrichment} presents one representative enrichment.}

\begin{table*}[t!]
\centering
\caption{Illustrative enrichment for attack step $a_1$ in the Goofy Guineapig case study. The values are generated outputs, not independently verified ground truth.}
\label{tab:goofy_example_enrichment}
\small
\renewcommand{\arraystretch}{1.15}
\begin{tabular}{p{0.20\textwidth}p{0.70\textwidth}}
\toprule
\textbf{Field} & \textbf{Value} \\
\midrule
Attack step &
Persistence through Windows service creation or modification. \\
ATT\&CK mapping &
T1543.003 -- Create or Modify System Process: Windows Service. \\
Description &
The malware creates persistence by installing a Windows service named ``GoogleUpdate'' with display name ``GoogleUpdate'', configured to Auto Start, [...] The service creation mechanism utilizes hardcoded strings for communication targeting an HTTPS server at static.tcplog.com:4443, and is embedded within a UPX packed NSIS installer masquerading as legitimate Firefox installation files. \\
\emph{\precond} &
\begin{minipage}[t]{0.69\textwidth}
Goofy Guineapig creates a Windows service named GoogleUpdate.

The persistence mechanism requires execution of Goopdate.dll from within the \texttt{C:\textbackslash{}ProgramData\textbackslash{}GoogleUpdate} directory.
\end{minipage}
\\
\emph{\postcond} &
\begin{minipage}[t]{0.69\textwidth}
A Windows service named GoogleUpdate is created.

The GoogleUpdate service has an Auto start type.
\end{minipage}
\\
\bottomrule
\end{tabular}
\end{table*}

\tikzstyle{myarrow} = [->, thin, >=latex', rounded corners]
\tikzset{
as/.style={
  align=center,
  fill=white,
  inner sep=1pt,
  minimum width=0.7cm,
  shape=rectangle,
  draw=red
  }
}

\tikzset{
ms/.style={
  shape=circle,
  draw=blue,
  align=center,
  inner sep=1pt,
  fill=white
  }
}

\begin{figure*}[t!]
\scriptsize
\centering
\begin{minipage}{0.23\linewidth}
    \begin{tabular}{cc}
        \toprule
        \makecell{Attack\\Step} & \makecell{MITRE\\Technique}\\
        \midrule
        a1 & T1543.003 \\
        \cmidrule(l){1-2}
        a2 & T1036.005 \\
        a3 & T1497.003 \\
        a4 & T1497.001 \\
        a5 & T1497.002 \\
        a6 & T1027 \\
        a7 & T1140 \\
        a8 & T1564.003 \\
        a9 & T1070.010 \\
        a10 & T1574.001 \\
        a11 & T1055.012 \\
        a12 & T1218.011 \\
        \cmidrule(l){1-2}
        a13 & T1049 \\
        \cmidrule(l){1-2}
        a14 & T1071.001 \\
        a15 & T1008 \\
        a16 & T1571 \\
        \bottomrule
    \end{tabular}
\end{minipage}
\begin{minipage}{0.70\linewidth}
    \begin{tikzpicture}
        \node[shape=rectangle,draw=black,fill=white] (start) at (0,0) {Start};
        \node[ms][right=0.2cm of start] (m1) {$m_{1}$};
        \node[as][right=0.2cm of m1] (a1) {$a_{1}$};
        \node[ms][right=0.2cm of a1,align=center] (m2) {$m_{2}$};
        \node[as][right=-0.7cm of m2] (a2) at (4,2.5) {$a_{2}$};
        \node[as][right=-0.7cm of m2] (a3) at (4,2) {$a_{3}$};
        \node[as][right=-0.7cm of m2] (a4) at (4,1.5) {$a_{4}$};
        \node[as][right=-0.7cm of m2] (a5) at (4,1) {$a_{5}$};
        \node[as][right=-0.7cm of m2] (a6) at (4,0.5) {$a_{6}$};
        \node[as][right=-0.7cm of m2] (a7) at (4,0) {$a_{7}$};
        \node[as][right=-0.7cm of m2] (a8) at (4,-0.5) {$a_{8}$};
        \node[as][right=-0.7cm of m2] (a9) at (4,-1) {$a_{9}$};
        \node[as][right=-0.7cm of m2] (a10) at (4,-1.5) {$a_{10}$};
        \node[as][right=-0.7cm of m2] (a11) at (4,-2) {$a_{11}$};
        \node[as][right=-0.7cm of m2] (a12) at (4,-2.5) {$a_{12}$};
        \node[ms][right=1.8cm of m2] (m3) {$m_{3}$};
        \node[as][right=0.2cm of m3] (a13) {$a_{13}$};
        \node[ms][right=0.2cm of a13,fill=white] (m4) {$m_{4}$};
        \node[as][right=-1cm of m4] (a14) at (8,0.5) {$a_{14}$};
        \node[as][right=-1cm of m4] (a15) at (8,0) {$a_{15}$};
        \node[as][right=-1cm of m4] (a16) at (8,-0.5) {$a_{16}$};
        \node[shape=rectangle,draw=black,right=1.4cm of m4,fill=white] (end) {End};

        \begin{scope}[on background layer]
        \draw[myarrow] (start.east) -- (m1.west);
        \draw[myarrow] (m1.east) -- (a1.west);
        \draw[myarrow] (a1.east) -- (m2.west);
        \draw[myarrow] (m2.east) -- ++(0.4mm,0mm) |- (a2.west);
        \draw[myarrow] (m2.east) -- ++(0.8mm,0mm) |- (a3.west);
        \draw[myarrow] (m2.east) -- ++(1.2mm,0mm) |- (a4.west);
        \draw[myarrow] (m2.east) -- ++(1.6mm,0mm) |- (a5.west);
        \draw[myarrow] (m2.east) -- ++(2mm,0mm) |- (a6.west);
        \draw[myarrow] (m2.east) -- (a7.west);
        \draw[myarrow] (m2.east) -- ++(2mm,0mm) |- (a8.west);
        \draw[myarrow] (m2.east) -- ++(1.6mm,0mm) |- (a9.west);
        \draw[myarrow] (m2.east) -- ++(1.2mm,0mm) |- (a10.west);
        \draw[myarrow] (m2.east) -- ++(0.8mm,0mm) |- (a11.west);
        \draw[myarrow] (m2.east) -- ++(0.4mm,0mm) |- (a12.west);
        \draw[myarrow] (a2.east) -- ++(3.4mm,0mm) |- (m3.west);
        \draw[myarrow] (a3.east) -- ++(3mm,0mm) |- (m3.west);
        \draw[myarrow] (a4.east) -- ++(2.6mm,0mm) |- (m3.west);
        \draw[myarrow] (a5.east) -- ++(2.2mm,0mm) |- (m3.west);
        \draw[myarrow] (a6.east) -- ++(1.8mm,0mm) |- (m3.west);
        \draw[myarrow] (a7.east) -- (m3.west);
        \draw[myarrow] (a8.east) -- ++(1.8mm,0mm) |- (m3.west);
        \draw[myarrow] (a9.east) -- ++(2.2mm,0mm) |- (m3.west);
        \draw[myarrow] (a10.east) -- ++(2.6mm,0mm) |- (m3.west);
        \draw[myarrow] (a11.east) -- ++(3mm,0mm) |- (m3.west);
        \draw[myarrow] (a12.east) -- ++(3.4mm,0mm) |- (m3.west);
        \draw[myarrow] (m3.east) -- (a13.west);
        \draw[myarrow] (a13.east) -- (m4.west);
        \draw[myarrow] (m4.east) -- ++(0.3mm,0mm) |- (a14.west);
        \draw[myarrow] (m4.east) -- (a15.west);
        \draw[myarrow] (m4.east) -- ++(0.3mm,0mm) |- (a16.west);
        \draw[myarrow] (a14.east) -- ++(1.8mm,0mm) |- (end.west);
        \draw[myarrow] (a15.east) -- (end.west);
        \draw[myarrow] (a16.east) -- ++(1.8mm,0mm) |- (end.west);
        \end{scope}
    \end{tikzpicture}
\end{minipage}

\caption{Tactic-aligned enriched attack-step representation for the Goofy Guineapig case study. Milestone nodes provide coarse tactic-level grouping. Attack steps within the same milestone interval are not interpreted as a fully ordered execution trace. The table on the left lists the \mitre technique identifiers associated with each attack step.}
\label{fig:case1_graph}
\vspace{4pt}
\footnotesize
\textbf{Legend:} \legendBlueRed{\as}{\ms}
\end{figure*}

\section{Empirical Evaluation of the Quality of \acrshort{cti} Reports}
\label{sec:evaluation}

This section empirically characterizes the proposed workflow for enriching analyst-defined \cti attack steps with contextual descriptions, candidate \precond, candidate \postcond, and evidence provenance. The evaluation comprises two complementary parts. First, a generic retrieval benchmark is used to select practical parameters for the hybrid retrieval layer. Second, ten public malware case studies are used to assess the generated enrichments and characterize the computational cost of executing the workflow locally.

The case-study evaluation addresses three research questions introduced in Section~\ref{sec:introduction}. RQ1 examines the semantic quality of generated conditions, focusing on meaningfulness, atomicity, and precondition/postcondition type correctness. RQ2 examines evidence alignment, focusing on the support of generated conditions and the faithfulness, specificity, and unsupported content of enriched descriptions. RQ3 characterizes the observed computational cost of local execution using recorded processing times and the experimental hardware configuration. Table~\ref{tab:research_questions_mapping} summarizes the correspondence between these questions and the available measurements.

The retrieval-configuration study is treated as a preliminary experimental step rather than an independent demonstration of \cti retrieval effectiveness. Its benchmark provides answer-span-based relevance judgments for general-domain question answering, not task-specific relevance judgments for the malware case-study corpus. Consequently, the selected parameters are regarded as practical operating settings, and the case-study annotations characterize the evidence alignment of the resulting enrichments without establishing domain-optimal retrieval performance.

\begin{table*}[t!]
\centering
\caption{Mapping between research questions, evaluation dimensions, and available empirical evidence.}
\label{tab:research_questions_mapping}
\small
\begin{tabular}{p{0.09\textwidth}p{0.27\textwidth}p{0.54\textwidth}}
\hline
\textbf{RQ} & \textbf{Evaluation dimension} & \textbf{Assessment criteria and measurements} \\
\hline
RQ1 &
Semantic quality of conditions &
Expert judgments of condition meaningfulness, atomicity, and precondition/postcondition type correctness. \\
RQ2 &
Evidence alignment and description quality &
Expert judgments of condition evidence support, description faithfulness, description evidence support, step specificity, and unsupported claims. \\
RQ3 &
Local execution characteristics &
Recorded pipeline processing time, generation time, retrieval latency, and experimental hardware configuration. \\
\hline
\end{tabular}
\end{table*}

\subsection{Evaluation Design}
\label{sec:evaluation_design}

The evaluation follows a two-stage design. The first stage configures the retrieval component using a \changed{general-purpose} dataset as a generic, reproducible benchmark. 
This stage examines retrieval depth, ranking effectiveness, and the trade-off between retrieved context size and retrieval metrics. Because \changed{this dataset} differs from the target \cti enrichment task in both domain and relevance criteria, its results support parameter selection but do not demonstrate that the resulting configuration is optimal for retrieving \cti-specific evidence. \changed{The results of this preliminary stage are presented in Appendix~\ref{app:rag}.}

The second stage evaluates the integrated workflow on public malware case studies and supporting intelligence documents. For each case, the pipeline processes an analyst-guided intermediate representation and a local evidence corpus, retrieves potentially relevant passages, and generates enriched descriptions and candidate \precond and \postcond. The generated content is then assessed through a structured manual annotation protocol. Recorded processing times are used to characterize local execution.

The primary annotation units are individual generated conditions and enriched attack-step descriptions. For RQ1, condition-level judgments assess whether generated statements are meaningful, atomic, and assigned to the appropriate precondition or postcondition category. For RQ2, condition-level judgments assess support from cited evidence, while description-level judgments assess faithfulness, evidence support, specificity, and the presence of unsupported claims. Non-redundancy is reported as an additional output-quality property because the pipeline includes semantic deduplication.

RQ3 is assessed using the recorded end-to-end processing time, generation time, and retrieval latency under the specified local hardware configuration. These measurements characterize the implementation's observed execution cost.

The case-study evaluation also documents the number of processed documents, evidence units, input attack steps, and generated fields. These descriptive measurements characterize the scope of the experiments and the output representation.

\subsection{Datasets and Case Studies}
\label{sec:datasets_case_studies}

\changed{
To select the evaluation corpus, we examined recent and well-known \cti datasets to assess whether an existing benchmark could serve as ground truth for generating \precond and \postcond. CTISum~\cite{ctisum} introduces a benchmark for \cti summarization and includes an attack-process summarization subtask over 1,345 documents from diverse web sources. This makes it relevant to our setting because it explicitly considers attack processes in \cti reports. However, its annotation target is a document- or process-level summary, not a set of validated attack-step \precond and \postcond. CTINexus~\cite{ctinexus} targets automatic cybersecurity knowledge-graph construction from \cti reports, using ground-truth annotations for triplet extraction, entity alignment, and relation prediction. Although this is close to our objective of producing structured threat representations, its labels concern entities and relations rather than the operational conditions associated with individual attack steps. CTI-HAL~\cite{ctihal} provides a human-annotated dataset for mapping statements in real \cti reports to \mitre techniques. It is highly relevant for evaluating TTP extraction and evidence-to-technique mapping, but it does not annotate the prerequisites, effects, or state changes associated with each technique instance. Finally, CTIConnect~\cite{cticonnect} proposes a benchmark for retrieval-augmented \llm{s} over heterogeneous \cti sources, integrating structured knowledge bases and unstructured reports into expert-verified question--answer pairs. This benchmark is particularly relevant to our use of \rag, but its ground truth is better suited to QA-style retrieval and reasoning tasks than to evaluating generated \precond{} and \postcond{}.}

\changed{These datasets provide relevant evaluation resources for summarization, \mitre mapping, knowledge-graph construction, and retrieval-augmented question answering. However, the benchmarks considered here do not directly provide validated, case-specific, attack-step-level \precond and \postcond for the selected malware reports.}

\changed{We therefore selected as case studies for the evaluation of the proposed pipeline a set of malware analysis reports published by the UK \ncsc\footnote{\url{https://www.ncsc.gov.uk/section/keep-up-to-date/malware-analysis-reports}}. These reports provide technical descriptions of malware behavior and accompanying structured artifacts (\ie, \stix representations). We select them because they offer a consistent and publicly accessible source of intelligence, contain technical details relevant to \mitre-aligned behavior, and can be redistributed or referenced more easily than private operational \cti.}

\changed{We chose ten case studies among all the reports published by \ncsc according to multiple selection criteria. The primary selection criterion was the availability of both a technical malware analysis report and a corresponding structured \stix representation, enabling the construction of an analyst-guided attack-step representation from publicly available material. For each candidate case, we examined public repositories, security research blogs, and additional technical reports to identify supplementary intelligence and verify its correspondence with the malware or campaign described in the \ncsc material. We also sought to include different malware families and behavioral profiles, rather than concentrating on a single family or attack pattern. The resulting collection is therefore intended to provide diversity across the examined cases, but it is not a statistically representative sample of malware threats. Candidate reports were excluded when a corresponding \stix representation was unavailable or when the relationship between the structured artifacts and the published technical analysis could not be established with sufficient confidence.}

\changed{The corpus is ingested by the pipeline and normalized into evidence units. The analyst-guided intermediate representation provides candidate attack steps, tactic-level milestones, \mitre-aligned metadata, and any available ordering constraints. We evaluate the generated enrichments through a structured manual protocol, described in Section~\ref{sec:manual_evaluation_protocol}. Table~\ref{tab:summary_case_studies} summarizes the case-study set.}

\begin{table*}[t!]
\centering
\caption{Summary of the case studies used in the evaluation. AS denotes attack steps.}
\label{tab:summary_case_studies}
\small
\renewcommand{\arraystretch}{1.2}
\setlength{\tabcolsep}{5pt}
\begin{tabular*}{\textwidth}{
@{\extracolsep{\fill}}
>{\centering\arraybackslash}p{0.04\textwidth}
p{0.34\textwidth}
>{\centering\arraybackslash}p{0.10\textwidth}
>{\centering\arraybackslash}p{0.09\textwidth}
>{\centering\arraybackslash}p{0.11\textwidth}
>{\centering\arraybackslash}p{0.11\textwidth}
>{\centering\arraybackslash}p{0.11\textwidth}
}
\toprule
\textbf{Case} & \textbf{Description} & \textbf{\# Docs} & \textbf{\# AS} & \textbf{\# Evidence Units} & \textbf{\# Precond} & \textbf{\# Postcond} \\
\midrule
1 & Persistent Windows backdoor with HTTPS C2 communications. & 7 & 16 & 175 & 49 & 60 \\
2 & Highly targeted credential and OAuth 2.0 token-stealing malware targeting Outlook. & 10 & 8 & 233 & 42 & 26 \\
3 & Windows malware implemented as an RPC server. & 6 & 14 & 275 & 46 & 65 \\
4 & A fully featured Windows remote access tool. & 5 & 12 & 238 & 33 & 40 \\
5 & A lightweight, staged downloader targeting Windows, delivered via spear-phishing. & 6 & 9 & 303 & 32 & 25 \\
6 & Cisco IOS malware that collects device information and enables backdoor access. & 9 & 11 & 201 & 43 & 48 \\
7 & Network device backdoor. & 13 & 12 & 526 & 47 & 77 \\
8 & Telegram Bot API-based Python backdoor with file download and execution capability. & 9 & 7 & 158 & 23 & 35 \\
9 & macOS supply-chain malware that exfiltrates victim data using a custom data-encoding algorithm over HTTPS. & 7 & 9 & 365 & 34 & 43 \\
10 & Malware targeting Fortinet devices. & 12 & 12 & 242 & 42 & 50 \\
\midrule
& \textbf{Total} & \textbf{84} & \textbf{110} & \textbf{2716} & \textbf{391} & \textbf{469} \\
\bottomrule
\end{tabular*}
\end{table*}

\changed{The first case study, Goofy Guineapig,} is also used as a running example throughout the paper. The remaining cases are used to assess whether the proposed methodology can be applied across malware families, platforms, and behavioral profiles. Since all cases are based on public reports, the evaluation should be interpreted as a controlled case-study evaluation rather than a large-scale operational deployment study.

\subsection{Experimental Setup}
\label{sec:experimental_setup}

The case-study pipeline is implemented as a modular local-first architecture. The case-study experiments are conducted on a commercial laptop with 32~GB of RAM and an Intel Core i7-14650HX (2.20~GHz) processor, equipped with an NVIDIA GeForce RTX 4060. The pipeline is implemented in Python and uses a local \changed{Ollama service, a lightweight platform to execute language models locally.}

The local \slm used for enrichment is \texttt{DeepSeek-R1}~\cite{deepseek-ai_deepseek-r1_2025}, distilled with 8 billion parameters from Qwen3~\cite{yang2025qwen3technicalreport}, \changed{part of a family of language models developed by the Alibaba Group. }
We select this model as a practical local \slm because it provides a feasible compromise between reasoning capability, instruction-following behavior, and deployability on commodity hardware.
\changed{Ollama's default context is set to 4,096 tokens; however, this was not sufficient to fit all the information we fed to the local models. Hence, we increased Ollama's default context to 16384 tokens.}

This choice is a design decision rather than a conclusion established by a comparative model experiment. The model is kept fixed across the case studies, and no model ablation is performed. The evaluation therefore characterizes the outputs of this particular local configuration, without establishing that the selected model offers a better quality--cost trade-off than other available models.

\changed{Similarity computations described in Section \ref{sec:methodology}} rely on \texttt{all-MiniLM-L6-v2}, a compact sentence-transformer model selected for its efficiency, mature library support, and suitability for local semantic retrieval on constrained computational resources.

The pipeline combines retrieval, structured prompting, explicit evidence links, and local generation over an analyst-controlled evidence corpus. The experiments assess the resulting enrichments, but do not isolate the contribution of each component.

\subsection{Preprocessing and Evidence-Unit Construction}
\label{sec:preprocessing_evaluation}

To configure the evidence base used by the \rag architecture, documents are normalized and segmented into evidence units. The semantic chunking configuration uses a maximum size of 200 tokens and zero overlap. The supplied chunker relies on semantic segmentation, with a fixed-splitting fallback if semantic chunking cannot be executed. This setting follows the motivation of keeping locally related content together while bounding retrieval units~\cite{smith2024evaluating}. Each evidence unit retains provenance metadata, including a source identifier and chunk identifier, together with any available section, page, object, or \mitre reference metadata.

For the case-study evaluation, we report the number of documents, evidence units, attack steps, generated \precond, and generated \postcond for each case study in Table~\ref{tab:summary_case_studies}. These values characterize the size of the evidence base and the enrichment output.

\subsection{\rag Retrieval Configuration}
\label{sec:rag_hyperparameters}
\changed{In Section~\ref{sec:methodology} we described the \rag architecture}. It contains three retrieval components: a lexical retriever, a vector-database retriever, and a hybrid ensemble retriever. 
We implemented the lexical retriever with the \bm~\cite{bm25} ranking algorithm. The top-$k_{bm25}$ results of the \changed{lexical} retriever and the top-$k_{vdb}$ results of the vector retriever are merged through weighted reciprocal-rank fusion (RRF)~\cite{cormack_reciprocal_2009}. For an evidence unit $e$ retrieved for query $q$, its fused ranking score can be expressed as:
\[
s_{\mathrm{RRF}}(q,e)=\sum_{j\in\{bm25,vdb\}:\,e\in L_j(q)}\frac{w_j}{c+r_j(q,e)}
\]
{where $L_j(q)$ is the ranked list of retriever $j$, $r_j(q,e)$ is the one-based rank of $e$ in that list, $c$ is the rank-fusion constant, and $w_{bm25}+w_{vdb}=1$. A candidate absent from one component contributes only through the component in which it appears. Fusion therefore combines ranks rather than raw \changed{lexical} and dense similarity values. The implementation uses the library's default fusion constant and document-identity behavior; the retrieved results are subsequently deduplicated by chunk identifier when available and truncated to top-$k_h$.

The lexical retriever uses $k_{bm25}\in\mathbb{N}$, the number of evidence units returned by \bm. The vector retriever uses $k_{vdb}\in\mathbb{N}$, the number of evidence units retrieved by embedding similarity. The ensemble retriever is parameterized by $(w_{bm25},w_{vdb})$, where $w_{bm25},w_{vdb}\in[0,1]$ and $w_{bm25}+w_{vdb}=1$. These weights govern the contributions of the two ranked lists of evidence units to reciprocal-rank fusion. Retrieval query text, selected chunk identifiers, ranks, and available source metadata are retained for provenance inspection.
\changed{To evaluate these parameters, we relied on the SQuAD dataset~\cite{rajpurkar-etal-2016-squad}, which consists of questions posed on a set of general-purpose text articles, where the answer to every question is a segment of text. We selected this dataset since it is designed for question-answer tasks, similar to querying the \rag to retrieve relevant evidence. The experiments to tune these parameters and the results are provided in Appendix~\ref{app:rag}. }

\subsection{Manual Evaluation Protocol}
\label{sec:manual_evaluation_protocol}
\changed{We need to assess the quality of the generated descriptions for \precond and \postcond. With a ground truth available, \ie, the exact expected sentence, we could perform semantic similarity.} However, the variability of \llm{s}~\cite{wagner_towards_2024} and \slm{s} generation makes the automatic evaluation of enriched descriptions, \precond, and \postcond difficult. Each generated piece of content may vary in form while preserving its meaning; therefore, comparison with a fixed gold sentence may bias the evaluation. In addition, multiple valid conditions may exist for the same attack step, and their quality depends on validity, evidence support, meaningfulness, directness, and non-redundancy. For this reason, we adopt a structured manual evaluation protocol centered on generated fields and the evidence they cite. Prior works relied on manual validation to assess the performance of \llm{s} in \cti-related tasks~\cite{jcp5030063,ctihal}.

For each condition, annotators are shown:
\begin{itemize}
\item the case-study identifier;
\item the attack-step label, tactic, technique, and original description;
\item the generated enriched description for context;
\item the generated condition;
\item whether the condition is labeled as a \emph{pre-condition} or \emph{post-condition};
\item the cited evidence units and their source metadata.
\end{itemize}

Two annotators independently evaluate the generated outputs. The first annotator is a second-year Ph.D. student in cybersecurity with a focus on \cti and a Master's degree in the same field. The second annotator is a postdoctoral researcher in cybersecurity, with a particular focus on malware analysis. Agreement statistics are computed on the independent annotations.

\changed{The annotation process aims to set, for each generated content (\ie, the enriched description and the conditions), the scores for the metrics defined in the remainder of this section.} 

\changed{Annotators evaluate each generated condition (\ie, \emph{pre-condition} or \emph{post-condition}) according to the criteria in Table~\ref{tab:annotation_rubric}. \emph{Evidence support} measures whether the cited chunks support the condition. \emph{Meaningfulness} captures whether the condition is specific enough to be useful rather than generic. \emph{Atomicity} checks whether the condition expresses a single state, requirement, artifact, capability, or consequence. \emph{Type validity} checks whether the condition is correctly classified as a \emph{pre-condition} or a \emph{post-condition}. \emph{Non-redundancy} checks whether the condition duplicates another condition for the same attack step.}

\changed{In addition to condition-level labels, annotators assess enriched descriptions using the rubric in Table~\ref{tab:description_rubric}. This separates description quality from condition quality.}

The annotation material is provided in a tabular form. \changed{To each row corresponds a generated content, paired with generated field, its declared type, cited chunk identifiers, cited chunk text, source document metadata, and the labels to be assigned by the annotators.} The annotators do not discuss labels during the independent annotation phase.

The annotation dataset contains 860 generated conditions, comprising 391 \precond and 469 \postcond, and 110 enriched descriptions. Each item is independently evaluated by both annotators. Thus, the manual evaluation covers 970 distinct generated items, with two sets of judgments for each item. The condition-level and description-level denominators are kept separate throughout the analysis. Both annotators received written guidelines on annotation metrics, which we reported in Table~\ref{tab:annotation_rubric} and Table~\ref{tab:description_rubric}.

\begin{table*}[t!]
\centering
\caption{Manual annotation rubric for generated conditions.}
\label{tab:annotation_rubric}
\small
\renewcommand{\arraystretch}{1.15}
\begin{tabular}{p{0.16\textwidth}p{0.35\textwidth}p{0.44\textwidth}}
\toprule
\textbf{Criterion} & \textbf{Question} & \textbf{Labels} \\
\midrule
Evidence support &
Does the cited evidence support the condition? &
2 = Directly supported; 1 = Inferentially supported; 0 = Unsupported \\
Meaningfulness &
Does the condition capture a non-trivial, step-specific requirement or consequence rather than a generic cybersecurity statement? &
1 = Meaningful; 0 = Generic \\
Atomicity &
Does the condition express one state, requirement, artifact, capability, or consequence? &
1 = Atomic; 0 = Compound \\
Type validity &
Is the condition correctly classified as a \precond{} or \postcond{}? &
2 = Correct type; 1 = Ambiguous type; 0 = Wrong type \\
Non-redundancy &
Does the condition avoid duplicating another condition for the same step? &
2 = Non-redundant; 1 = Partially overlapping; 0 = Redundant \\
\bottomrule
\end{tabular}
\end{table*}

\begin{table*}[t!]
\centering
\caption{Manual annotation rubric for enriched descriptions.}
\label{tab:description_rubric}
\small
\renewcommand{\arraystretch}{1.15}
\begin{tabular}{p{0.20\textwidth}p{0.38\textwidth}p{0.36\textwidth}}
\toprule
\textbf{Criterion} & \textbf{Question} & \textbf{Labels} \\
\midrule
Faithfulness &
Is the description faithful to the cited evidence and original attack step? &
2 = Faithful; 1 = Partially faithful; 0 = Not faithful \\
Evidence support &
Does the cited evidence support the description? &
2 = Supported; 1 = Partially supported or unclear; 0 = Unsupported \\
Specificity &
Does the description contain step-specific details rather than only restating the \mitre technique? &
2 = Specific; 1 = Partially specific; 0 = Generic \\
Unsupported claims &
Does the description introduce claims not supported by the evidence? &
2 = No unsupported claims; 1 = Minor unsupported claims; 0 = Major unsupported claims \\
\bottomrule
\end{tabular}
\end{table*}

At the representation level, the evaluation checks the presence of milestone nodes, attack-step nodes, evidence links, enriched descriptions, \precond, and \postcond, and whether attack steps are grouped under the expected tactic-aligned milestones. These checks concern output structure and alignment with the analyst-defined \changed{input representation}. They do not establish the causal validity of graph edges or a verified execution order.

We report Cohen's $\kappa$~\cite{cohen} and Gwet's AC1~\cite{gwet_computing_2008} for binary criteria, and quadratic Cohen's $\kappa$ and Gwet's AC2 for ordinal criteria. Raw agreement is also reported.

\subsection{Metrics}
\label{sec:evaluation_metrics}

We report automatic and manual metrics. Automatic metrics characterize coverage, structure, provenance, and cost \changed{and are defined as measurements that can be run automatically, \eg, during the pipeline execution or by analyzing the results}. Manual metrics assess semantic quality \changed{and cannot be run automatically, \ie, a manual assessment is required due to the issues presented in Section~\ref{sec:manual_evaluation_protocol}.}

\paragraph{Automatic metrics.}
We present the automatic metrics computed on the generated descriptions and conditions. The automatic metrics are:
\begin{itemize}
\item \textbf{Field coverage:} percentage of attack steps for which the pipeline generates an enriched description, at least one \emph{pre-condition}, or at least one \emph{post-condition}, reported separately by field type.
\item \textbf{Average field length:} average number of tokens per generated enriched description or condition, reported by field type.
\item \textbf{Duplicate-condition removal rate:} percentage of condition candidates removed by the reported deduplication procedure, separately for \precond and \postcond.
\item \textbf{Retrieval latency:} time required to retrieve evidence for an attack step.
\item \textbf{Generation latency:} time spent generating enrichments.
\item \textbf{Pipeline total latency:} end-to-end execution time for a case study.
\end{itemize}

\paragraph{Manual metrics.}

\changed{In a real-world setting, an analyst must be supported by a pipeline capable of generating content that is supported by evidence and valid \wrt to the documents and the considered \as. The former is mandatory to provide the analyst with factual data, the latter to provide properly contextualized information. 
Measuring the quality of the generated content with a manual approach allowed us to consider this requirements. Moreover, the quality of the generation outcomes cannot always be represented with a binary set of values, \eg, conditions may be partially valid or may partially overlap with other conditions.}

Therefore, the annotation rubric includes binary and ordinal criteria. For binary criteria, labels distinguish satisfaction from non-satisfaction of the corresponding property. For ordinal criteria, label 2 denotes full satisfaction, label 1 denotes partial or inferential satisfaction, and label 0 denotes failure.

Different forms of summary statistics are distinguished. 
\emph{Inter-annotator agreement} measures consistency between annotators and is reported using raw agreement and chance-corrected coefficients. \changed{Concerning agreement, we define a \emph{minor disagreement} when the annotators' scores differ by one, whereas a \emph{major disagreement} occurs when they differ by two. The corresponding counts, respectively $m$ and $M$, are defined as:}
\[
m = \left|\left\{i : |a_i-b_i|=1\right\}\right|,
\qquad
M = \left|\left\{i : |a_i-b_i|=2\right\}\right|,
\]
\changed{where $a_i$ and $b_i$ are the labels assigned by the two annotators to item $i$. For binary criteria, disagreements are reported without this ordinal distinction.}
\emph{Joint-positive rates} count items for which both annotators assign a label indicating satisfaction of a criterion. For ordinal criteria, a strict joint-positive rate requires both annotators to assign label 2. A relaxed joint-positive rate requires both annotators to assign either label 1 or label 2. For binary criteria, a joint-positive rate requires both annotators to assign label 1.

For clarity, let $y_{i1}$ and $y_{i2}$ denote the two annotations for item $i$, and let $N$ be the number of items evaluated for the criterion. For an ordinal criterion, the strict and relaxed joint-positive rates are defined as:
\[
J_{\mathrm{strict}} =
\frac{1}{N}
\sum_{i=1}^{N}
\mathbb{I}[y_{i1}=2 \land y_{i2}=2],
\]
\[
J_{\mathrm{relaxed}} =
\frac{1}{N}
\sum_{i=1}^{N}
\mathbb{I}[y_{i1}\geq 1 \land y_{i2}\geq 1].
\]

For generated conditions, we report:
\begin{itemize}
\item \textbf{Evidence-support rate:} fraction of conditions jointly judged directly supported under the strict setting, or directly/inferentially supported under the relaxed setting.
\item \textbf{Meaningfulness rate:} fraction of conditions jointly labeled meaningful.
\item \textbf{Atomicity rate:} fraction of conditions jointly labeled atomic.
\item \textbf{Type-validity rate:} fraction of conditions jointly assigned the correct type under the strict setting, or correct/ambiguous type under the relaxed setting.
\item \textbf{Non-redundancy rate:} fraction of conditions jointly labeled non-redundant under the strict setting, or non-redundant/partially overlapping under the relaxed setting.
\end{itemize}

For enriched descriptions, we report:
\begin{itemize}
\item \textbf{Description-faithfulness rate:} fraction of descriptions jointly labeled fully faithful under the strict setting, or faithful/partially faithful under the relaxed setting.
\item \textbf{Description-support rate:} fraction of descriptions jointly labeled supported under the strict setting, or supported/partially supported under the relaxed setting.
\item \textbf{Description-specificity rate:} fraction of descriptions jointly labeled specific under the strict setting, or specific/partially specific under the relaxed setting.
\item \textbf{Absence-of-unsupported-claims rate:} fraction of descriptions for which both annotators report no unsupported claims under the strict setting, or no/minor unsupported claims under the relaxed setting.
\end{itemize}

The last metric is stated in positive form to match the annotation scale, where label 2 means that no unsupported claims were identified.

\subsection{Results}
\label{sec:evaluation_results}

We report the case-study results in four parts. First, automatic measurements characterize the generated outputs and the cost of local execution, providing evidence for RQ3. Second, inter-annotator agreement statistics characterize the consistency of manual judgments. Third, condition-level and description-level annotations provide evidence for RQ1 and RQ2. Finally, a qualitative case study and error analysis illustrate the types of enrichment produced by the workflow and the limitations observed during annotation.

\begin{table*}[t!]
\centering
\caption{Automatic output metrics for the complete pipeline, reported as averages over the ten case studies.}
\label{tab:automatic_results}
\small
\renewcommand{\arraystretch}{1.15}
\begin{tabular}{
p{0.32\textwidth}
>{\centering\arraybackslash}p{0.17\textwidth}
>{\centering\arraybackslash}p{0.17\textwidth}
>{\centering\arraybackslash}p{0.17\textwidth}
}
\toprule
\textbf{Metric} &
\textbf{Descriptions} &
\textbf{\emph{Pre-conditions}} &
\textbf{\emph{Post-conditions}} \\
\midrule
Field coverage (\%) & 100 & 96.78 & 97.50 \\
Average field length (tokens) & 69.86 & 15.48 & 13.39 \\
Duplicate-condition removal rate (\%) & -- & 15.85 & 55.65 \\
\bottomrule
\end{tabular}
\end{table*}

\subsubsection{Automatic Metrics}
Table~\ref{tab:automatic_results} summarizes automatically recorded output characteristics. The pipeline produced 110 enriched descriptions, 391 retained \precond, and 469 retained \postcond, corresponding to 970 generated items included in the manual assessment. Before annotation, conditions were filtered using a semantic similarity threshold of 0.7, as discussed in Section~\ref{sec:evidence_grounded_enrichment}. Deduplication has been applied to \precond and \postcond separately. The reported duplicate-condition removal rates were 15.85\% for \precond and 55.65\% for \postcond. These percentages describe the behavior of the deduplication procedure and do not establish that all removed conditions were semantically redundant or that all retained conditions were distinct.

With respect to RQ3, the reported average end-to-end execution time was 36.10 minutes per case study, while generation averaged 35.72 minutes per case study. Generation therefore represented approximately 98.9\% of the reported end-to-end processing time. The average recorded retrieval latency was 34.80 milliseconds. These measurements indicate that language-model generation dominates the observed execution cost under the evaluated configuration.

\subsubsection{Inter-Annotator Agreement}
\label{sec:inter_annotator_agreement}

\begin{table*}[t!]
\centering
\caption{Inter-annotator agreement for condition- and description-level criteria. Triplets are reported as original-scale / optimistic binary recoding / pessimistic binary recoding. For binary criteria, a single value is reported.}
\label{tab:iaa_results}
\small
\renewcommand{\arraystretch}{1.15}
\begin{tabular}{
p{0.32\textwidth}
>{\centering\arraybackslash}p{0.19\textwidth}
>{\centering\arraybackslash}p{0.19\textwidth}
>{\centering\arraybackslash}p{0.19\textwidth}
}
\toprule
\textbf{Annotation unit / criterion} &
\textbf{Raw agreement} &
\textbf{Cohen's $\kappa$} &
\textbf{Gwet's AC} \\
\midrule
Evidence support &
0.72/0.95/0.74 &
0.29/0.37/0.17 &
0.87/0.95/0.64 \\
Meaningfulness &
0.79 &
0.12 &
0.74 \\
Atomicity &
0.91 &
0.35 &
0.90 \\
Type validity &
0.68/0.74/0.70 &
0.36/0.36/0.32 &
0.53/0.57/0.48 \\
Non-redundancy &
0.74/0.87/0.75 &
0.28/0.31/0.21 &
0.78/0.85/0.65 \\
Description faithfulness &
0.44/0.93/0.45 &
0.10/0.20/0.05 &
0.67/0.93/0.04 \\
Description evidence support &
0.57/1.00/0.57 &
0.08/Undefined/0.08 &
0.82/1.00/0.30 \\
Description specificity &
0.84/0.95/0.85 &
0.11/0.00/0.09 &
0.92/0.95/0.82 \\
Unsupported claims &
0.40/0.74/0.43 &
0.05/0.04/0.04 &
0.35/0.66/$-0.02$ \\
\bottomrule
\end{tabular}
\end{table*}

For binary criteria, agreement is summarized using raw agreement, Cohen's $\kappa$~\cite{cohen}, and Gwet's AC1~\cite{gwet_computing_2008}. For ordinal criteria, the original three-level labels are assessed using raw agreement, weighted Cohen's $\kappa$, and Gwet's AC2. The binary recodings are assessed using the corresponding binary agreement coefficients.

For ordinal criteria, the original-scale agreement uses the three labels 0, 1, and 2. The optimistic binary recoding groups labels 1 and 2, while the pessimistic binary recoding groups labels 0 and 1. These transformations measure the sensitivity of inter-annotator agreement to the treatment of intermediate judgments. They are distinct from strict and relaxed joint-positive quality rates, which count items that satisfy a criterion in both annotators' ratings.

Table~\ref{tab:iaa_results} reports inter-annotator agreement for the manual evaluation criteria.  The agreement results indicate that some criteria are more difficult to assess consistently than others. In particular, several description-level criteria show substantial differences between raw agreement and chance-corrected agreement. These discrepancies can arise from the distribution of labels as well as from genuine disagreement. \changed{As the two annotators, also two analysts may diverge in evaluating a generated content. Nevertheless, evidence citations can support in-depth analysis for critical disagreements. The additional effort required for this assessment may be limited by having the evidence already correlated to the generated content and the considered \as.}

\subsubsection{Condition-Level Results}
\label{sec:condition_level_results}

\begin{table*}[t!]
\centering
\caption{Raw inter-annotator agreement for generated conditions, separated by declared condition type. Triplets denote original-scale / optimistic recoding / pessimistic recoding.}
\label{tab:condition_manual_results}
\small
\renewcommand{\arraystretch}{1.15}
\begin{tabular}{
p{0.19\textwidth}
>{\centering\arraybackslash}p{0.17\textwidth}
>{\centering\arraybackslash}p{0.13\textwidth}
>{\centering\arraybackslash}p{0.12\textwidth}
>{\centering\arraybackslash}p{0.17\textwidth}
>{\centering\arraybackslash}p{0.17\textwidth}
}
\toprule
\textbf{Condition type} &
\textbf{Evidence support} &
\textbf{Meaningfulness} &
\textbf{Atomicity} &
\textbf{Type validity} &
\textbf{Non-redundancy} \\
\midrule
\emph{\precond{}} &
0.73/0.95/0.75 &
0.73 &
0.89 &
0.66/0.71/0.70 &
0.84/0.93/0.85 \\
\emph{\postcond{}} &
0.73/0.95/0.73 &
0.85 &
0.93 &
0.69/0.76/0.70 &
0.65/0.83/0.67 \\
All conditions &
0.72/0.95/0.74 &
0.79 &
0.91 &
0.68/0.74/0.70 &
0.74/0.87/0.75 \\
\bottomrule
\end{tabular}
\end{table*}

Table~\ref{tab:condition_manual_results} reports agreement rather than condition quality. These values indicate how consistently the annotators applied the rubric, including agreement on negative labels. They should not be interpreted as the proportion of conditions that are meaningful, atomic, supported, or correctly typed.

Figure~\ref{fig:agreement_conditions} distinguishes exact agreement from disagreements of different magnitudes, \changed{as defined in Section~\ref{sec:manual_evaluation_protocol}. }

\begin{figure*}[h!]
    \centering
    \includegraphics[width=\linewidth]{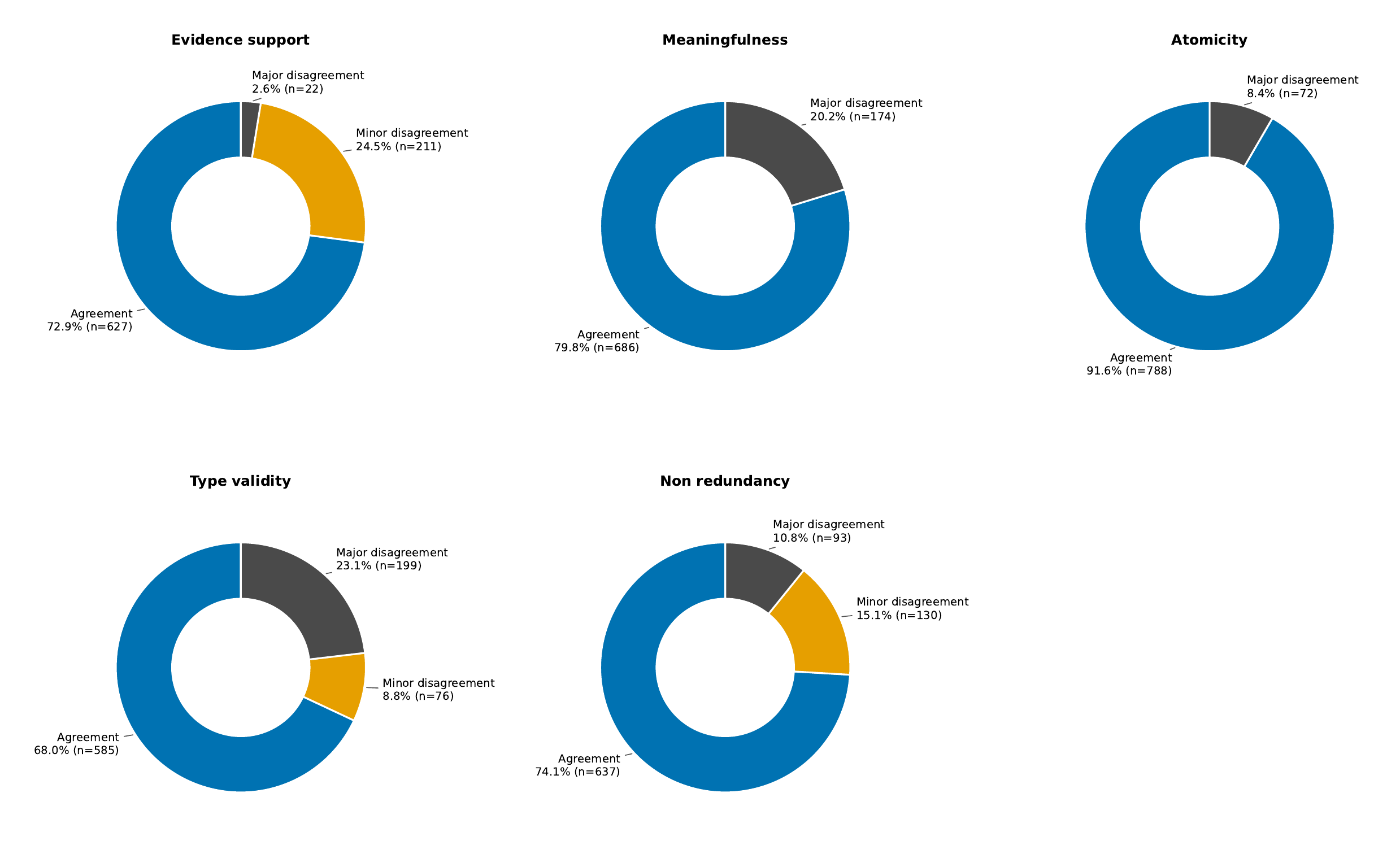}
    \caption{Proportions of agreement, minor disagreement, and major disagreement for each condition criterion.}
    \label{fig:agreement_conditions}
\end{figure*}

To answer RQ1, we examine joint-positive judgments for condition meaningfulness, atomicity, and type validity. \changed{These metrics assess the \slm capability of producing semantically well-formed conditions, \ie, they assess to which extent the conditions considered by the analyst can be assumed correctly typed and relevant for the considered \as}. Requiring both annotators to assign the highest applicable label, 667 of 860 conditions (77.56\%) were judged meaningful and 765 of 860 (88.95\%) atomic. In contrast, 465 of 860 conditions (54.07\%) were jointly judged to have the correct precondition or postcondition type. These findings indicate that the workflow more consistently produces statements judged meaningful and atomic than statements jointly judged to have the correct semantic type.

This distinction matters for the intended representation. A condition can be concise and relevant while still being assigned to the wrong category. Consequently, generated condition sets should be regarded as candidate enrichments requiring analyst inspection, particularly where the difference between an enabling state and a resulting state is operationally significant. Table~\ref{tab:condition_joint_positive_strict} and Table~\ref{tab:condition_joint_positive_relaxed} report, respectively, the strict and relaxed joint-positive judgments for generated conditions. \changed{From these tables emerge that both annotators agreed that most of conditions are evidence-grounded, meaningfull \wrt the considered \as, and non-redundant. This metric is particularly interesting in terms of produced noise, \ie, an analyst using the pipeline will obtain conditions that are, for the majority, novel and semantically well-formed. The most critical metric is \emph{Type Validity}; thus, additional efforts will be required to analysts to evaluate the generated conditions in terms of prerequisites and consequences.}

\begin{table}[t!]
\centering
\caption{Strict joint-positive judgments for generated conditions. Both annotators must assign the highest applicable label.}
\label{tab:condition_joint_positive_strict}
\small
\renewcommand{\arraystretch}{1.15}
\begin{tabular}{p{0.44\linewidth}rr}
\toprule
\textbf{Criterion} & \textbf{Count} & \textbf{Rate (\%)} \\
\midrule
Evidence support & 602/860 & 70.00\\
Meaningfulness & 667/860 & 77.56\\
Atomicity & 765/860 & 88.95\\
Type validity & 465/860 & 54.07\\
Non-redundancy & 599/860 & 69.65\\
\bottomrule
\end{tabular}
\end{table}

\begin{table}[t!]
\centering
\caption{Relaxed joint-positive judgments for generated conditions. Binary metrics have the same values reported in Table~\ref{tab:condition_joint_positive_strict}. }
\label{tab:condition_joint_positive_relaxed}
\small
\renewcommand{\arraystretch}{1.15}
\begin{tabular}{p{0.44\linewidth}rr}
\toprule
\textbf{Criterion} & \textbf{Count} & \textbf{Rate (\%)} \\
\midrule
Evidence support & 812/860 & 94.41\\
Meaningfulness & 667/860 & 77.56\\
Atomicity & 765/860 & 88.95\\
Type validity & 519/860 & 60.34\\
Non-redundancy & 724/860 & 84.18\\
\bottomrule
\end{tabular}
\end{table}

\subsubsection{Evidence Alignment and Description-Level Results}
\label{sec:description_level_results}

RQ2 concerns the relationship between generated enrichments and the retrieved \cti evidence. At the condition level, 602 of 860 conditions (70.00\%) were judged directly supported by both annotators under the strict joint-positive criterion. This indicates that a substantial proportion of conditions received positive evidence-support judgments, while also showing that provenance links alone are insufficient to establish the support of every generated statement.

Description-level annotations provide a complementary perspective. Faithfulness assesses whether the generated description preserves the meaning of the cited evidence and original attack step. Evidence support assesses whether the retrieved passages substantiate the description. Specificity evaluates whether the description adds details relevant to the particular attack step rather than merely repeating the general \mitre technique definition. The unsupported-claims criterion assesses whether the description introduces information not established by the available evidence. Table~\ref{tab:description_joint_positive_strict} and Table~\ref{tab:description_joint_positive_relaxed} provide both strict and relaxed joint-positive judgments for enriched descriptions. \changed{These results highlight that the presence of unsupported claims is critical in the generation of descriptions, \ie, an analyst may pay additional attention to the provided content. However, the presence of linked evidence allows for in-depth assessment of the description faithfulness. \emph{Specificity} demonstrates the capability of the pipeline of generating descriptions specific to the considered \as. \emph{Faithfulness} and \emph{evidence support} suffer from the subjectivity of the annotators' judgments. Indeed, there is an increment of almost 50\% between strict and relaxed judgments. Nevertheless, under a relaxed setting all the description have been considered evidence supported.}

\begin{table}[t!]
\centering
\caption{Strict joint-positive judgments for enriched descriptions. Both annotators must assign label 2.}
\label{tab:description_joint_positive_strict}
\small
\renewcommand{\arraystretch}{1.15}
\begin{tabular}{p{0.45\linewidth}rr}
\toprule
\textbf{Criterion} & \textbf{Count} & \textbf{Rate (\%)} \\
\midrule
Faithfulness & 46/110 & 41.82 \\
Evidence support & 58/110 & 52.72 \\
Specificity & 93/110 & 84.55\\
No unsupported claims & 41/110 & 37.27\\
\bottomrule
\end{tabular}
\end{table}

\begin{table}[t!]
\centering
\caption{Relaxed joint-positive judgments for enriched descriptions.}
\label{tab:description_joint_positive_relaxed}
\small
\renewcommand{\arraystretch}{1.15}
\begin{tabular}{p{0.45\linewidth}rr}
\toprule
\textbf{Criterion} & \textbf{Count} & \textbf{Rate (\%)} \\
\midrule
Faithfulness & 102/110 & 92.72\\
Evidence support & 110/110 & 100.00 \\
Specificity & 105/110 & 95.45\\
No unsupported claims & 81/110 & 73.63\\
\bottomrule
\end{tabular}
\end{table}

Under the strict joint-positive criterion, 93 of 110 descriptions (84.55\%) were judged specific to their corresponding attack steps. However, only 46 descriptions (41.82\%) were jointly judged fully faithful, and only 41 (37.27\%) received the highest label from both annotators for the absence of unsupported claims. These findings reveal a distinction between producing descriptions that contain step-specific details and producing descriptions whose content is fully aligned with the retrieved evidence. Specificity is therefore not a substitute for factual faithfulness.

\begin{figure*}[h!]
    \centering
    \includegraphics[width=\linewidth]{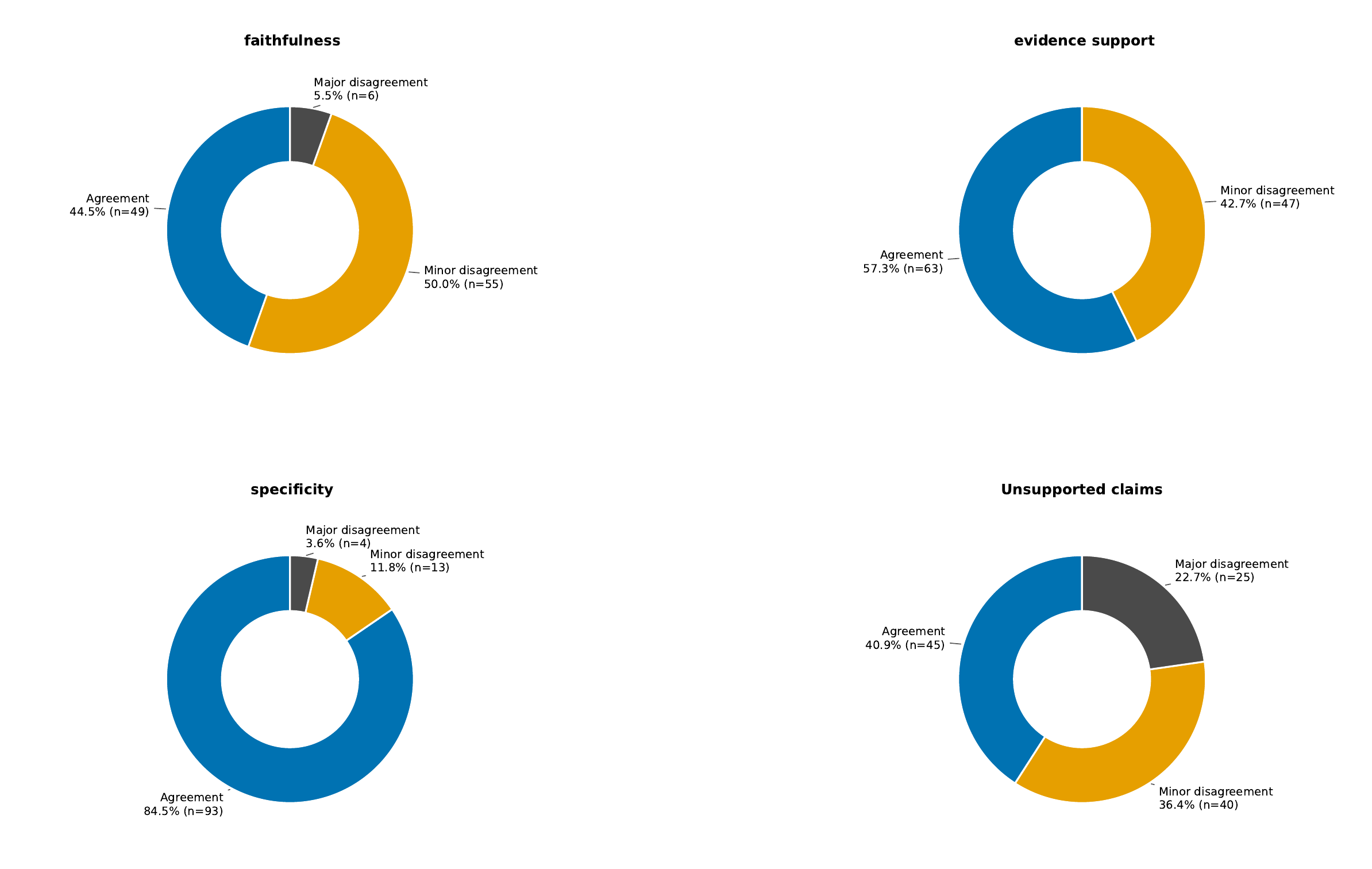}
    \caption{Proportions of agreement, minor disagreement, and major disagreement for each description criterion.}
    \label{fig:agreement_descriptions}
\end{figure*}

Taken together, the condition- and description-level results provide a qualified answer to RQ2. The workflow generates content that annotators judge specific and, in many cases, supported by retrieved evidence. Nevertheless, description faithfulness and unsupported content remain substantial concerns. Evidence links improve inspectability by identifying retrieval context, but they should not be interpreted as independent verification of generated statements.

The results provide distinct answers to the three research questions. For RQ1, condition annotations indicate relatively high atomicity and meaningfulness but substantially lower joint-positive type validity. For RQ2, evidence-support and description annotations show that generated content can be specific to an attack step while still containing statements that are insufficiently supported or not fully faithful to the retrieved intelligence. For RQ3, execution measurements characterize the cost of running the workflow locally and show that generation dominates processing time in the evaluated configuration.

These findings support the feasibility of producing structured, evidence-linked attack-step enrichments from an analyst-provided representation. They also identify limitations directly relevant to analyst use: condition types may require correction, generated descriptions may introduce unsupported details, and processing time may constrain operational applicability. The evaluation does not establish that the resulting graph is causally correct or that the workflow improves analyst performance relative to existing practices.

\subsection{Error Analysis}
\label{sec:error_analysis}

To complement the aggregate annotation results, we examine error categories identified through the manual assessment. Table~\ref{tab:error_taxonomy_conditions} reports the number of generated conditions for which both annotators assigned the lowest label for a given criterion. These counts identify jointly recognized failures and should not be interpreted as the total number of problematic conditions, because items receiving partial judgments or disagreement are not included.

The error categories include conditions judged unsupported by the cited evidence, statements considered too generic to characterize the attack step, non-atomic conditions, incorrectly classified \precond or \postcond, and redundant statements. In particular, the 120 conditions jointly assigned the lowest type-validity label are consistent with the weakness identified for RQ1. The evidence-support failures are relevant to RQ2, although the strict failure count does not capture all instances of partial support, ambiguity, or annotator disagreement.

\begin{table}[h!]
\centering
\caption{Condition-level failure analysis. Counts indicate items for which both annotators assigned label 0 for the corresponding criterion. Categories are not mutually exclusive.}
\label{tab:error_taxonomy_conditions}
\small
\renewcommand{\arraystretch}{1.15}
\begin{tabular}{p{0.56\linewidth}r}
\toprule
\textbf{Error type} & \textbf{Count} \\
\midrule
Unsupported condition & 12 \\
Generic condition & 19 \\
Non-atomic condition & 23 \\
Wrong condition type & 120 \\
Redundant condition & 32 \\
\bottomrule
\end{tabular}
\end{table}

The error counts should not be subtracted from the strict joint-positive counts to infer the number of acceptable conditions. For ordinal criteria, intermediate labels and disagreements produce additional outcomes. Similarly, the 41 descriptions jointly judged to contain no unsupported claims are not the complement of the number of descriptions jointly assigned the most severe unsupported-claims label. These measures capture different portions of the annotation distribution.

The observed errors may arise from several stages of the workflow. Retrieval may fail to select passages containing the necessary details; generation may misinterpret relevant evidence, introduce unsupported assumptions, or assign a condition to the wrong semantic category; and the available intelligence may itself be incomplete or ambiguous. The present evaluation does not experimentally isolate these causes, so individual errors cannot be conclusively attributed to a specific component. Nevertheless, the taxonomy motivates task-specific retrieval evaluation, stronger evidence-support checking, more explicit condition-type constraints, and analyst review when source material is insufficient.

\changed{Concerning a real-world setting, the criticality of these errors may weight differently. Retrieval failures are of greatest concern, since neither the generated content nor the evidence will contain the proper \cti. Indeed, such errors are unrelated to the \slm but stem from the \rag architecture's insufficient capability to identify relevant evidence.
Instead, a generation error will lead to inaccuracies that can be assessed by using the linked evidence. These errors are related to the \slm's ability to properly ingest the provided data and may be overcome with a more capable model or prompting techniques. Incomplete, ambiguous, or compromised documents are out of the scope of this work (Section~\ref{sec:assumptions}). }

\subsection{Reproducibility}
\label{sec:evaluation_reproducibility}

To support reproducibility, we publicly release the pipeline source code, configuration files, prompts, case-study inputs, and annotation files~\cite{deidda_paper-slm-attackgraphs_2026}, together with the generated outputs and the evaluation results~\cite{deidda_paper-slm-attackgraphs_data_2026}.
\section{Discussion}
\label{sec:discussion}
Figure~\ref{fig:case1_graph} illustrates the evidence-linked representation produced by enriching an analyst-provided attack-step scaffold. Unlike infrastructure-centric attack graphs whose nodes model hosts, services, or reachability relations, this representation retains tactic-aligned milestones (\ms{s}) and attack steps (\as{s}) as an organizational structure for examining adversary behavior. The enrichment adds distinct candidate execution requirements (\precond), potential direct effects (\postcond), contextual descriptions, and evidence references to each step. This separation makes it possible to inspect what a step describes, what must plausibly hold beforehand, what state may follow, and which source passages informed each statement. The analysis below examines the quality and support of these generated fields; it does not evaluate the correctness of the analyst-provided graph edges.

\subsection{Results Discussion}
The pipeline output was assessed using the automatic and manual metrics presented in Section~\ref{sec:evaluation}. 
Although both annotators received guidelines defining the evaluation criteria, subjective interpretation may still affect the assigned scores, particularly for criteria that admit intermediate judgments. To account for this uncertainty, we analyze the results using strict and relaxed scoring criteria. These strategies provide alternative interpretations of partially satisfactory or ambiguous assessments. They should not be confused with inter-annotator agreement coefficients, which quantify the consistency of the original annotations and are reported in Table~\ref{tab:iaa_results}.

\paragraph{Evidence support and retrieval transferability.}
Figure~\ref{fig:agreement_conditions} reports the distribution of agreements, minor disagreements, and major disagreements across the condition-level metrics. Particular attention is given to \emph{evidence support}, which provides an indirect indication of whether the retrieval configuration selected on SQuAD supplies adequate evidence for downstream \cti enrichment. Specifically, we examine whether the evidence retrieved using the selected hybrid configuration ($k_{bm25}=100$, $k_{vdb}=50$, $k_h=50$, $w_{bm25}=0.25$, and $w_{vdb}=0.75$) supports the \precond and \postcond generated by the \slm.

Under the strict criterion, both annotators classified 602 out of 860 conditions (70.00\%) as supported by the retrieved evidence. When intermediate judgments are also accepted under the optimistic approach, the number of conditions increases to 812 (94.42\%). Conversely, 12 conditions (1.40\%) were explicitly classified as unsupported by both annotators, while 39 (4.53\%) fail to satisfy the evidence-support criterion under the pessimistic approach. The difference between the strict and optimistic estimates indicates that a non-negligible proportion of conditions received intermediate assessments, highlighting the uncertainty involved in determining whether narrative \cti evidence sufficiently supports generated statements.

The strict support rate shows that the selected configuration provides evidence judged directly supportive of many generated conditions in the examined cases. The increase under the relaxed criterion is also informative: much of the remaining assessment concerns inferential rather than wholly absent support. This matters for \cti because narrative reports frequently describe actions and outcomes more explicitly than their prerequisites. It motivates separating a condition's provenance from the strength of the inference it expresses. These results assess the evidence alignment of the complete retrieval-and-generation workflow; they are not a measurement of retrieval recall or proof that its SQuAD-selected parameters are optimal for \cti.

\paragraph{Quality of generated conditions.}
Beyond evidence support, the quality of the generated conditions is assessed through \emph{meaningfulness}, \emph{atomicity}, \emph{type validity}, and \emph{non-redundancy}. Meaningfulness and atomicity are evaluated using binary labels ($m\in\{0,1\}$), whereas type validity and non-redundancy use three-level ordinal labels ($m\in\{0,1,2\}$).

Regarding \emph{meaningfulness}, both annotators classified 667 conditions (77.56\%) as meaningful and 19 (2.21\%) as generic, while disagreeing on the remaining 174 conditions (20.23\%). For \emph{atomicity}, both annotators agreed that 765 conditions (88.95\%) were atomic and 23 (2.67\%) were compound, with disagreements affecting 72 conditions (8.37\%). These results indicate that most generated conditions are perceived as meaningful and atomic by both annotators. In particular, the high atomicity rate suggests that the pipeline generally produces conditions expressing individual requirements or consequences rather than compound statements. Nevertheless, the disagreement rate observed for meaningfulness indicates that distinguishing step-specific information from generic statements remains partially dependent on annotator judgment.

\emph{Type validity} presents a more challenging result. Under the strict criterion, both annotators classified 465 conditions (54.07\%) as correctly typed, increasing to 519 (60.35\%) under the optimistic approach. Conversely, 120 conditions (13.95\%) were classified as incorrectly typed by both annotators under the strict criterion, increasing to 142 (16.51\%) under the pessimistic approach. The substantial disagreement rate, including 23.1\% major disagreements under the original ordinal labels and 31.9\% disagreements after pessimistic recoding, highlights the difficulty of distinguishing \precond from \postcond in certain contexts. Some conditions may describe states that can be interpreted either as requirements for the current step or as consequences of preceding behavior, depending on the available context. These findings suggest that condition-type classification remains a limitation of the enrichment process and may require further refinement of the generation constraints and annotation guidelines.

Regarding \emph{non-redundancy}, both annotators classified 599 conditions (69.65\%) as non-redundant under the strict criterion, increasing to 724 (84.19\%) under the optimistic approach. Conversely, 32 conditions (3.72\%) were classified as redundant by both annotators under the strict criterion, increasing to 49 (5.70\%) under the pessimistic approach. These results should be considered alongside the automatic deduplication mechanism applied during pipeline execution. As reported in Table~\ref{tab:automatic_results}, 15.85\% of generated \precond and 55.65\% of generated \postcond were identified as duplicates and removed before producing the final output. These removal rates show that the \slm frequently generates overlapping conditions, particularly \postcond, and motivate the inclusion of similarity-based deduplication. Although the residual redundancy observed during manual evaluation suggests that some overlapping statements persist, the automatic filtering stage provides a practical mechanism to limit duplication in the final representation.

\paragraph{Quality of enriched descriptions.}
Figure~\ref{fig:agreement_descriptions} reports the distributions of agreements and disagreements for the 110 enriched descriptions. These outputs are assessed according to four three-level criteria: \emph{faithfulness}, \emph{evidence support}, \emph{specificity}, and \emph{unsupported claims}.

Concerning \emph{evidence support}, both annotators classified 58 descriptions (52.73\%) as supported under the strict criterion. No description was classified as unsupported by both annotators under the strict criterion, whereas five descriptions (4.55\%) fail to satisfy the support criterion under the pessimistic approach. This difference indicates that intermediate evidence-support judgments also affect the interpretation of description quality, and that the strict positive rate should not be equated with the proportion of descriptions that contain no supporting evidence.

For \emph{faithfulness}, 46 descriptions (41.82\%) were classified as fully faithful by both annotators under the strict criterion. Under the optimistic approach, which also accepts partially faithful descriptions, this number increases to 102 (92.73\%). Conversely, one description (0.91\%) was classified as not faithful by both annotators under the strict criterion, increasing to four (3.64\%) under the pessimistic approach. The substantial difference between strict and optimistic outcomes indicates that partial faithfulness is an important component of the assessment. Therefore, while most descriptions are at least partially consistent with the original attack steps and evidence, fewer are consistently judged to be fully faithful.

The results for \emph{specificity} are more favorable. Under the strict criterion, both annotators classified 93 descriptions (84.55\%) as specific, increasing to 105 (95.45\%) under the optimistic approach. No description was classified as generic by both annotators under the strict criterion, while one description failed to satisfy the specificity criterion under the pessimistic approach. These findings suggest that the enrichment process generally produces descriptions containing information specific to the considered attack step, rather than merely reformulating the corresponding \mitre technique.

Finally, the \emph{unsupported claims} metric evaluates whether enriched descriptions introduce statements that cannot be justified by the available evidence. Under the strict criterion, 41 descriptions (37.27\%) were classified by both annotators as containing no unsupported claims. Under the optimistic approach, which also accepts descriptions containing minor unsupported claims, this number increases to 81 (73.64\%). One description was classified by both annotators as containing major unsupported claims under the strict criterion, while seven descriptions (6.36\%) were identified as containing at least minor unsupported claims under the pessimistic approach. These results indicate that unsupported or insufficiently justified details remain a relevant concern, even when the generated descriptions are generally considered specific and at least partially faithful.

\paragraph{Synthesis: statement form, operational semantics, and evidential adequacy.}
The evaluation reveals three distinguishable dimensions of enrichment quality. First, \emph{statement form} is comparatively strong: most retained conditions are judged atomic, and most descriptions contain step-specific details. Second, \emph{operational semantics} are more difficult: assigning an enabling state or direct consequence to the correct side of an attack step remains uncertain for a substantial part of the output. Third, \emph{evidential adequacy} requires a stricter standard than relevance: descriptions can be specific and broadly consistent with their sources without being judged fully faithful or entirely free of unsupported details. These findings explain the value of separating descriptions, conditions, declared types, and evidence references in the representation. They allow an analyst to inspect the aspects that the pipeline produces consistently while focusing validation on semantic typing and claims requiring stronger support. This is a design implication of the observed annotations, rather than a measured improvement in analyst efficiency.

\subsection{Limitations and Failure Modes}
\label{sec:limitations_failure_modes}
Although the evaluation provides encouraging evidence of the pipeline's ability to produce meaningful, evidence-linked attack-step enrichments, several limitations and failure modes remain. These limitations arise from the quality and completeness of the available \cti, the retrieval and generation processes, and the inherent ambiguity of the semantics associated with individual attack steps.

\paragraph{Evidence availability and retrieval limitations.}
The quality of the generated enrichments depends on the information available in the analyst-provided documents and on the \rag architecture's ability to retrieve relevant evidence. \cti reports may omit implementation details, describe behaviors at different levels of abstraction, or provide incomplete information about the conditions enabling an attack step and its consequences. Consequently, some \precond and \postcond may remain only partially supported or cannot be derived from the available evidence. Moreover, the hybrid-retrieval configuration was selected using a general-domain dataset rather than a dedicated \cti benchmark. Although the manual evaluation provides preliminary evidence of its applicability to the examined case studies, it does not establish retrieval completeness or domain-specific optimality. Relevant evidence may therefore remain outside the retrieved context, potentially affecting the quality and completeness of the generated enrichments.

\paragraph{Generation errors and semantic ambiguity.}
Despite the use of bounded retrieved context and structured prompting, the \slm may generate inaccurate, overly generic, or insufficiently supported statements. In particular, the manual evaluation highlights difficulties in distinguishing between \precond and \postcond. This distinction is not always explicit in narrative \cti: a system state may represent a consequence of a preceding step while simultaneously constituting a requirement for a subsequent one. Without sufficient contextual information, the model may incorrectly classify a condition or formulate a statement that is plausible but not directly attributable to the attack step under consideration. Similar ambiguity may affect human annotation, as reflected in disagreements over condition-type validity and other subjective metrics. These results suggest that structured generation does not entirely eliminate semantic ambiguity or the need for analyst interpretation.

\paragraph{Redundancy limitations.}
Similarity-based deduplication introduces a trade-off between reducing redundant information and preserving distinct conditions with similar wording. The automatic results show that duplicate generation is particularly relevant for \postcond, while the manual evaluation indicates that some redundancy remains in the final output. Conversely, overly aggressive similarity thresholds could remove semantically distinct conditions. These limitations motivate retaining evidence provenance and enabling analysts to inspect the resulting enrichments.



\subsection{Answers to the Research Questions}
\label{sec:rq_answers}

\paragraph{RQ1: Semantic quality of generated conditions.}
The manual assessment indicates that the generated conditions frequently satisfy basic structural and semantic criteria, but that their classification remains substantially less reliable. Under the strict agreement criterion requiring both annotators to assign the highest score, 765 of 860 conditions (88.95\%) were judged atomic and 667 (77.56\%) meaningful. In contrast, only 465 conditions (54.07\%) satisfied the corresponding criterion for condition-type correctness. These results suggest that producing concise and interpretable statements is more consistently achieved than distinguishing enabling requirements from direct consequences. Therefore, the workflow provides a mechanism for generating candidate step-level conditions, but the resulting \precond and \postcond cannot be assumed to be correctly typed without further inspection.\changed{Nevertheless, the pipeline capability of linking evidence used to generate the \precond and \postcond supports the human analyst in validating the generated content. This capability is further assessed by RQ2.}

\paragraph{RQ2: Evidence alignment of generated enrichments.}
The evidence-alignment results reveal a substantial difference between generating step-specific content and ensuring that this content is fully supported. Under the strict agreement criterion, 602 of 860 conditions (70.00\%) were judged supported by the retrieved evidence. For enriched descriptions, 93 of 110 (84.55\%) were judged specific to the corresponding attack step, whereas only 46 (41.82\%) satisfied the criterion for full faithfulness and 41 (37.27\%) satisfied the criterion for the absence of unsupported claims. These findings indicate that the workflow can produce descriptions that are relevant to the intended behavior while still introducing details that are not adequately established by the retrieved material. Consequently, provenance links should be interpreted as references to the generation context rather than as proof that each generated statement is correct.

\paragraph{RQ3: Local execution characteristics.}
The workflow was executed locally on the hardware described in the experimental setup. Across the evaluated malware cases, the reported average processing time was approximately 36.1 minutes per case, of which approximately 35.72 minutes were attributable to generation. These measurements indicate that language-model generation dominates the observed processing time in the evaluated configuration. They establish that the workflow can be executed on the specified local hardware, but they do not demonstrate comparative efficiency, suitability for real-time analysis, or scalability to substantially larger intelligence collections. Such claims would require additional workload characterization and comparative measurements.

Taken together, the answers to RQ1--RQ3 establish the feasibility and characteristic quality profile of the integrated workflow in the examined cases. The method produces structured, inspectable fields with strong atomicity and meaningfulness results, while its principal semantic challenges concern distinguishing prerequisites from consequences and keeping generated descriptions fully aligned with evidence. The local runtime analysis further identifies generation as the dominant computational cost. These findings provide concrete design priorities for subsequent enrichment systems: improve condition typing, strengthen claim-level support checking, and reduce generation overhead. Comparative effectiveness and analyst productivity remain separate questions for future evaluation.

\subsection{Threats to Validity}
\label{sec:threats_validity}
Following the taxonomy proposed by Wohlin\etal~\cite{wohlin_experimentation_2024}, we discuss threats to construct, internal, external, and conclusion validity.

\paragraph{Construct validity.}
The principal constructs evaluated in this study are the semantic quality and evidence support of the generated attack-step enrichments. However, the absence of a validated ground-truth dataset containing \precond and \postcond for real-world attacks prevents a fully automated assessment of their correctness. We therefore adopt a structured manual annotation protocol covering condition validity, evidence support, meaningfulness, atomicity, type validity, and non-redundancy, as well as faithfulness, evidence support, specificity, and unsupported claims for enriched descriptions. Although these criteria capture complementary dimensions of enrichment quality, they remain dependent on human judgment and do not directly measure operational usefulness, completeness, or the effectiveness of subsequent analyst decisions. Moreover, evidence support assesses whether generated statements are justified by the retrieved context, rather than whether the retriever identifies all relevant information available in the source documents. Consequently, evidence-support rates should not be interpreted as direct measurements of retrieval recall or overall factual completeness.

\paragraph{Internal validity.}
The pipeline operates on analyst-provided threat representations and document collections assembled for the selected case studies. These inputs provide a controlled evaluation setting but may reduce the ambiguity and noise encountered in operational \cti environments. Furthermore, the quality of the generated enrichments depends on several interacting components, including retrieval, prompting, local \slm generation, and deduplication. Although the models and retrieval configuration are kept fixed throughout the case-study evaluation, the absence of component-level ablations prevents us from isolating their individual contributions. In particular, the quality effect of similarity-based deduplication cannot be established without comparison with an otherwise identical pipeline that lacks this component. Model non-determinism represents an additional potential source of variability~\cite{perkovic_hallucinations_2024}, as repeated executions may produce different descriptions or conditions even with unchanged inputs. Retrieved evidence is retained for inspection, but no independent automated verification of generated claims is established by the present experiments.

\paragraph{External validity.}
The evaluation considers 10 public malware case studies that cover different malware families, platforms, and adversarial behaviors. Although this diversity provides an initial assessment across heterogeneous threat scenarios, the selected cases do not represent the full range of operational \cti, which may include incomplete reports, conflicting sources, multilingual documents, organization-specific intelligence, and substantially larger collections of evidence. Moreover, the pipeline assumes the availability of an analyst-guided intermediate representation containing candidate attack steps and their associated \mitre metadata. Its effectiveness may therefore depend on the completeness and quality of this initial representation. Another limitation concerns retrieval-configuration transferability: the hybrid retriever was configured using SQuAD, a general-domain dataset rather than a specialized \cti benchmark, \changed{due to the limitations of available benchmarks described in Section~\ref{sec:datasets_case_studies}}. Although the downstream manual evaluation provides preliminary evidence of its applicability to the examined cases, domain-specific retrieval performance and optimality remain unverified.

\paragraph{Conclusion validity.}
The evaluation combines automatic measurements with manual annotations for 860 generated conditions and 110 enriched descriptions, and is assessed by two annotators. Although predefined annotation criteria were provided, several metrics require subjective judgments, as reflected in the observed disagreements. In particular, intermediate scores introduce uncertainty when distinguishing fully valid or supported outputs from partially satisfactory or ambiguous ones. To make this uncertainty explicit, we report strict outcomes alongside optimistic and pessimistic interpretations of intermediate annotations. These alternative interpretations provide sensitivity bounds for the reported quality estimates but do not eliminate annotator subjectivity or establish ground-truth correctness. Accordingly, the reported results should be interpreted as preliminary evidence of feasibility, evidence grounding, and semantic enrichment quality, rather than as definitive proof of general performance or superiority over alternative approaches.

\section{Conclusion}
\label{sec:conclusion}

This paper presented and evaluated a methodology for evidence-linked enrichment of analyst-defined \cti attack steps. The local-first workflow combines hybrid lexical and semantic retrieval with a locally deployed \slm to generate contextual descriptions, candidate \precond and \postcond, and links to the source passages used during generation. Its enriched representation preserves the input attack-step scaffold and organizes the additional semantic fields within tactic-aligned milestones. The methodological contribution is a structured, inspectable transformation from existing threat descriptions and heterogeneous evidence into explicit step-level requirements and effects, accompanied by an empirical assessment of the quality of those enrichments.

Across ten public malware case studies, two annotators independently assessed 860 generated conditions and 110 enriched descriptions. Under strict joint-positive scoring, 765 conditions (88.95\%) were judged atomic, 667 (77.56\%) meaningful, and 602 (70.00\%) directly supported by the retrieved evidence. Among descriptions, 93 of 110 (84.55\%) were judged step-specific. These results show that the workflow frequently produces individually inspectable, context-specific enrichment fields. The evaluation also identifies where this capability is less reliable: 465 conditions (54.07\%) were jointly judged correctly typed as prerequisites or consequences, while 46 descriptions (41.82\%) were judged fully faithful and 41 (37.27\%) free from unsupported claims.

The central empirical finding is that statement form, operational semantics, and evidential adequacy must be evaluated separately. A condition can be concise and meaningful yet assigned to the wrong temporal role; similarly, a description can be specific to an attack step without being fully supported by its evidence. This distinction has direct implications for evidence-linked threat representations: provenance makes generated assertions inspectable, while type correctness and factual faithfulness remain separate validation tasks. The present results therefore support an analyst-guided use of the representation, with particular attention to precondition/postcondition typing and the justification of generated claims.

The study is bounded by its ten case studies, fixed local configuration, and subjective annotation criteria; it does not establish comparative superiority or measured gains in analyst productivity. Future work will target more reliable condition typing, claim-level evidence checking, and refinement of deduplication, alongside \cti-specific retrieval evaluation and comparative component studies. Analyst-centered assessments will be needed to measure the value of the enriched representation in concrete investigative tasks. Within these bounds, the findings establish both the feasibility of local, evidence-linked attack-step enrichment and specific reliability challenges that subsequent systems should address.

\section*{Data availability}
The source code, configuration files, case-study inputs, and annotation files are openly available on Zenodo at \url{https://doi.org/10.5281/zenodo.24463164}~\cite{deidda_paper-slm-attackgraphs_2026} and on GitHub at \url{https://github.com/MrDeiv/Paper-SLM-AttackGraphs}. The generated outputs and evaluation results supporting this study are openly available on Zenodo at \url{https://doi.org/10.5281/zenodo.24463589}~\cite{deidda_paper-slm-attackgraphs_data_2026}.

\section*{Declaration of competing interest}
The authors declare that they have no known competing financial interests or personal relationships that could have appeared to influence the work reported in this paper.

\section*{Funding}
This work was partially supported by the projects SINTESI (SIstemi cyber-fisici INTElligenti e SIcuri per la sanità, l’industria e la società), CUP B29H26000410007, and SUNS (Secure Ubiquitous Networks and Services), CUP B29H26000190007, under Action 1.1.2—“Support for a limited number of strategic research chains”, within the “Coesione Italia 21-27—Ricerca, Innovazione e Competitività per la transizione verde e digitale” Programme, co-funded by the European Union.

\section*{Declaration of generative AI and AI-assisted technologies in the manuscript preparation process}
During the preparation of this work, the authors used Claude (Anthropic) to revise and condense the text. After using this tool, the authors reviewed and edited the content as needed and took full responsibility for the published article.

\printcredits

\bibliographystyle{cas-model2-names}

\bibliography{cas-refs}

@misc{gao_retrieval-augmented_2024,
	title = {Retrieval-Augmented Generation for Large Language Models: A Survey},
	doi = {10.48550/arXiv.2312.10997},
	author = {Gao, Yunfan and Xiong, Yun and Gao, Xinyu and Jia, Kangxiang and Pan, Jinliu and Bi, Yuxi and Dai, Yi and Sun, Jiawei and Wang, Meng and Wang, Haofen},
	year = {2024}
}

@article{AINSLIE2023103352,
    title = {Cyber-threat intelligence for security decision-making: A review and research agenda for practice},
    journal = {Computers \& Security},
    volume = {132},
    year = {2023},
    doi = {https://doi.org/10.1016/j.cose.2023.103352},
    author = {Scott Ainslie and Dean Thompson and Sean Maynard and Atif Ahmad},
}

@article{bm25,
author = {Robertson, Stephen and Zaragoza, Hugo},
title = {The Probabilistic Relevance Framework: BM25 and Beyond},
year = {2009},
issue_date = {April 2009},
publisher = {Now Publishers Inc.},
address = {Hanover, MA, USA},
volume = {3},
number = {4},
issn = {1554-0669},
doi = {10.1561/1500000019},
journal = {Found. Trends Inf. Retr.},
month = apr,
pages = {333–389},
numpages = {57}
}

@inproceedings{tural_retrieval-augmented_2024,
	title = {Retrieval-Augmented Generation ({RAG}) and {LLM} Integration},
	doi = {10.1109/ISAS64331.2024.10845308},
	booktitle = {2024 8th International Symposium on Innovative Approaches in Smart Technologies ({ISAS})},
	author = {Tural, Büşra and Örpek, Zeynep and Destan, Zeynep},
	year = {2024},
}

@inproceedings{kucecka_selective_2013,
	title = {Selective chunking — Easy and effective way to estimate text similarity},
	doi = {10.1109/CINTI.2013.6705226},
	booktitle = {2013 {IEEE} 14th International Symposium on Computational Intelligence and Informatics ({CINTI})},
	author = {Kučečka, Tomas and Chudâ, Daniela and Samuhel, Patrik},
	year = {2013},
}

@inproceedings{bhagdev_hybrid_2008,
	title = {Hybrid Search: Effectively Combining Keywords and Semantic Searches},
	doi = {10.1007/978-3-540-68234-9_41},
	booktitle = {The Semantic Web: Research and Applications},
	author = {Bhagdev, Ravish and Chapman, Sam and Ciravegna, Fabio and Lanfranchi, Vitaveska and Petrelli, Daniela},
	year = {2008},
}

@inproceedings{chen_keyword-based_2011,
	title = {Keyword-based search and exploration on databases},
	doi = {10.1109/ICDE.2011.5767958},
	booktitle = {2011 {IEEE} 27th International Conference on Data Engineering},
	author = {Chen, Yi and Wang, Wei and Liu, Ziyang},
	year = {2011}
}

@inproceedings{huang_embedding-based_2020,
	location = {Virtual Event {CA} {USA}},
	title = {Embedding-based Retrieval in Facebook Search},
	doi = {10.1145/3394486.3403305},
	booktitle = {Proceedings of the 26th {ACM} {SIGKDD} International Conference on Knowledge Discovery \& Data Mining},
	author = {Huang, Jui-Ting and Sharma, Ashish and Sun, Shuying and Xia, Li and Zhang, David and Pronin, Philip and Padmanabhan, Janani and Ottaviano, Giuseppe and Yang, Linjun},
	year = {2020}
}

@misc{naveed_comprehensive_2024,
	title = {A Comprehensive Overview of Large Language Models},
	doi = {10.48550/arXiv.2307.06435},
	author = {Naveed, Humza and Khan, Asad Ullah and Qiu, Shi and Saqib, Muhammad and Anwar, Saeed and Usman, Muhammad and Akhtar, Naveed and Barnes, Nick and Mian, Ajmal},
	year = {2024}
}

@inproceedings{perkovic_hallucinations_2024,
	title = {Hallucinations in {LLMs}: Understanding and Addressing Challenges},
	doi = {10.1109/MIPRO60963.2024.10569238},
	pages = {2084--2088},
	booktitle = {2024 47th {MIPRO} {ICT} and Electronics Convention ({MIPRO})},
	author = {Perković, Gabrijela and Drobnjak, Antun and Botički, Ivica},
	year = {2024},
}

@misc{xu_intelex_2024,
	title = {{IntelEX}: A {LLM}-driven Attack-level Threat Intelligence Extraction Framework},
	doi = {10.48550/arXiv.2412.10872},
	author = {Xu, Ming and Wang, Hongtai and Liu, Jiahao and Lin, Yun and Liu, Chenyang Xu Yingshi and Lim, Hoon Wei and Dong, Jin Song},
	year = {2024},
}

@article{zhang_attackg_2025,
title = {AttacKG+: Boosting attack graph construction with Large Language Models},
journal = {Computers \& Security},
volume = {150},
pages = {104220},
year = {2025},
issn = {0167-4048},
doi = {10.1016/j.cose.2024.104220},
author = {Yongheng Zhang and Tingwen Du and Yunshan Ma and Xiang Wang and Yi Xie and Guozheng Yang and Yuliang Lu and Ee-Chien Chang}
}

@inproceedings{li_attackg_2022,
author = {Li, Zhenyuan and Zeng, Jun and Chen, Yan and Liang, Zhenkai},
title = {AttacKG: Constructing Technique Knowledge Graph from Cyber Threat Intelligence Reports},
year = {2022},
isbn = {978-3-031-17139-0},
publisher = {Springer-Verlag},
address = {Berlin, Heidelberg},
url = {https://doi.org/10.1007/978-3-031-17140-6_29},
doi = {10.1007/978-3-031-17140-6_29},
booktitle = {Computer Security – ESORICS 2022: 27th European Symposium on Research in Computer Security, Copenhagen, Denmark, September 26–30, 2022, Proceedings, Part I},
pages = {589–609},
numpages = {21},
location = {Copenhagen, Denmark}
}

@article{saeed_systematic_2023,
AUTHOR = {Saeed, Saqib and Suayyid, Sarah A. and Al-Ghamdi, Manal S. and Al-Muhaisen, Hayfa and Almuhaideb, Abdullah M.},
TITLE = {A Systematic Literature Review on Cyber Threat Intelligence for Organizational Cybersecurity Resilience},
JOURNAL = {Sensors},
VOLUME = {23},
YEAR = {2023},
NUMBER = {16},
ARTICLE-NUMBER = {7273},
PubMedID = {37631808},
ISSN = {1424-8220},
DOI = {10.3390/s23167273}
}

@inproceedings{krauss_analysis_2022,
	title = {Analysis of Threat Intelligence Information Exchange via the {STIX} Standard},
	doi = {10.1109/ICECCME55909.2022.9988073},
	booktitle = {2022 International Conference on Electrical, Computer, Communications and Mechatronics Engineering ({ICECCME})},
	author = {Krauss, Oliver and Papesh, Konstantin},
	year = {2022},
}

@inproceedings{phillips_graph-based_1998,
	title = {A graph-based system for network-vulnerability analysis},
	doi = {10.1145/310889.310919},
	booktitle = {Proceedings of the 1998 workshop on New security paradigms},
	author = {Phillips, Cynthia and Swiler, Laura Painton},
	year = {1998},
}

@inproceedings{dantu_risk_2004,
	title = {Risk management using behavior based attack graphs},
	volume = {1},
	doi = {10.1109/ITCC.2004.1286496},
	booktitle = {International Conference on Information Technology: Coding and Computing, 2004. Proceedings. {ITCC} 2004.},
	author = {Dantu, R. and Loper, K. and Kolan, P.},
	year = {2004}
}

@article{noel_measuring_nodate,
	title = {Measuring Security Risk of Networks Using Attack Graphs},
	volume = {1},
    url = {https://csis.gmu.edu/noel/pubs/2010_IJNGC.pdf},
	author = {Noel, Steven and Jajodia, Sushil},
    year={2010},
    journal={International Journal of Next-Generation Computing}
}

@inproceedings{noel_correlating_2004,
	title = {Correlating intrusion events and building attack scenarios through attack graph distances},
	doi = {10.1109/CSAC.2004.11},
	booktitle = {20th Annual Computer Security Applications Conference},
	author = {Noel, S. and Robertson, E. and Jajodia, S.},
	year = {2004},
}

@article{WAGNER2019101589,
	title = {Cyber threat intelligence sharing: Survey and research directions},
	volume = {87},
	issn = {0167-4048},
	doi = {https://doi.org/10.1016/j.cose.2019.101589},
	pages = {101589},
	journaltitle = {Computers \& Security},
	author = {Wagner, Thomas D. and Mahbub, Khaled and Palomar, Esther and Abdallah, Ali E.},
	date = {2019},
}

@article{barnum_standardizing_nodate,
  title={Standardizing cyber threat intelligence information with the structured threat information expression (stix)},
  author={Barnum, Sean},
  journal={Mitre Corporation},
  volume={11},
  number={2012},
  pages={1--22},
  year={2012}
}

@inproceedings{10.1145/2994539.2994542,
author = {Wagner, Cynthia and Dulaunoy, Alexandre and Wagener, G\'{e}rard and Iklody, Andras},
title = {MISP: The Design and Implementation of a Collaborative Threat Intelligence Sharing Platform},
year = {2016},
doi = {10.1145/2994539.2994542},
booktitle = {Proceedings of the 2016 ACM on Workshop on Information Sharing and Collaborative Security},
}

@article{arazzi_nlp-based_2025,
	title = {{NLP}-based techniques for Cyber Threat Intelligence},
        year = {2025},
	url = {https://linkinghub.elsevier.com/retrieve/pii/S1574013725000413},
	doi = {10.1016/j.cosrev.2025.100765},
	journaltitle = {Computer Science Review},
	author = {Arazzi, Marco and R. Arikkat, Dincy and Nicolazzo, Serena and Nocera, Antonino and Rehiman K.A., Rafidha and P., Vinod and Conti, Mauro},
}

@inproceedings{qiang_quality_2018,
	title = {A Quality Evaluation Method of Cyber Threat Intelligence in User Perspective},
	doi = {10.1109/TrustCom/BigDataSE.2018.00049},
	booktitle = {2018 17th {IEEE} International Conference On Trust, Security And Privacy In Computing And Communications/ 12th {IEEE} International Conference On Big Data Science And Engineering ({TrustCom}/{BigDataSE})},
    year={2018},
	author = {Qiang, Li and Zhengwei, Jiang and Zeming, Yang and Baoxu, Liu and Xin, Wang and Yunan, Zhang},
}

@article{yang_automated_2025,
title = {An automated dynamic quality assessment method for cyber threat intelligence},
journal = {Computers \& Security},
volume = {148},
pages = {104079},
year = {2025},
issn = {0167-4048},
doi = {10.1016/j.cose.2024.104079},
author = {Libin Yang and Menghan Wang and Wei Lou}
}

@inproceedings{sillaber_data_2016,
	title = {Data Quality Challenges and Future Research Directions in Threat Intelligence Sharing Practice},
	doi = {10.1145/2994539.2994546},
	booktitle = {Proceedings of the 2016 {ACM} on Workshop on Information Sharing and Collaborative Security},
	publisher = {{ACM}},
	author = {Sillaber, Christian and Sauerwein, Clemens and Mussmann, Andrea and Breu, Ruth},
        year={2016}
}

@techreport{smith2024evaluating,
  title = {Evaluating Chunking Strategies for Retrieval},
  author = {Smith, Brandon and Troynikov, Anton},
  year = {2024},
  institution = {Chroma},
  url = {https://research.trychroma.com/evaluating-chunking},
}

@inproceedings{rajpurkar-etal-2016-squad,
    title = "{SQ}u{AD}: 100,000+ Questions for Machine Comprehension of Text",
    author = "Rajpurkar, Pranav  and
      Zhang, Jian  and
      Lopyrev, Konstantin  and
      Liang, Percy",
    booktitle = "Proceedings of the 2016 Conference on Empirical Methods in Natural Language Processing",
    year = "2016",
    doi = "10.18653/v1/D16-1264",
}

@misc{tang2025lrctilargelanguagemodelbased,
      title={LRCTI: A Large Language Model-Based Framework for Multi-Step Evidence Retrieval and Reasoning in Cyber Threat Intelligence Credibility Verification}, 
      author={Fengxiao Tang and Huan Li and Ming Zhao and Zongzong Wu and Shisong Peng and Tao Yin},
      year={2025},
      url={https://arxiv.org/abs/2507.11310}, 
}

@misc{gu2025surveyllmasajudge,
      title={A Survey on LLM-as-a-Judge}, 
      author={Jiawei Gu and Xuhui Jiang and Zhichao Shi and Hexiang Tan and Xuehao Zhai and Chengjin Xu and Wei Li and Yinghan Shen and Shengjie Ma and Honghao Liu and Saizhuo Wang and Kun Zhang and Yuanzhuo Wang and Wen Gao and Lionel Ni and Jian Guo},
      year={2025},
      url={https://arxiv.org/abs/2411.15594}, 
}

@misc{cheng2024crucialgreconstructintegratedattack,
      title={CRUcialG: Reconstruct Integrated Attack Scenario Graphs by Cyber Threat Intelligence Reports}, 
      author={Wenrui Cheng and Tiantian Zhu and Tieming Chen and Qixuan Yuan and Jie Ying and Hongmei Li and Chunlin Xiong and Mingda Li and Mingqi Lv and Yan Chen},
      year={2024},
      url={https://arxiv.org/abs/2410.11209}, 
}

@misc{alam2023lookingiocsautomaticallyextracting,
      title={Looking Beyond IoCs: Automatically Extracting Attack Patterns from External CTI}, 
      author={Md Tanvirul Alam and Dipkamal Bhusal and Youngja Park and Nidhi Rastogi},
      year={2023},
      doi={https://doi.org/10.1145/3607199.3607208}
}

@INPROCEEDINGS {orbinato_automatic_2022,
author = { Orbinato, Vittorio and Barbaraci, Mariarosaria and Natella, Roberto and Cotroneo, Domenico },
booktitle = { 2022 IEEE 33rd International Symposium on Software Reliability Engineering (ISSRE) },
title = {{ Automatic Mapping of Unstructured Cyber Threat Intelligence: An Experimental Study: (Practical Experience Report) }},
year = {2022},
doi = {10.1109/ISSRE55969.2022.00027},
publisher = {IEEE Computer Society}}

@misc{siracusano_action_2023,
  title = {Time for a{CTI}on: Automated Analysis of Cyber Threat Intelligence in the Wild},
  author = {Siracusano, Giuseppe and Sanvito, Davide and Gonzalez, Roberto and Srinivasan, Manikantan and Kamatchi, Sivakaman and Takahashi, Wataru and Kawakita, Masaru and Kakumaru, Takahiro and Bifulco, Roberto},
  year = {2023},
  doi = {10.48550/arXiv.2307.10214},
}

@misc{ttprint2026,
    title={CTI2TTP: Evidence-Grounded TTP Extraction from Cyber Threat Intelligence via Diverge-then-Converge Verification}, 
    author={Yutong Cheng and Changze Li and Raihan Sultan Pasha Basuki and Qian Cui and Wei Ding and Peng Gao},
    year={2026},
    doi={https://doi.org/10.48550/arXiv.2605.25836}
}

@misc{hou_reachable_2026,
  title = {An Automated Framework for Extracting Reachable Attack Chains from Cyber Threat Intelligence Reports},
  author = {Hou, Wenbo and Hu, Ning and Wang, Xueping and Gu, Jiahao and Luo, Wenjian},
  year = {2026},
  doi = {10.48550/arXiv.2607.19742}
}

@misc{ctifoundry2026,
  title = {{CTIFoundry}: An Agent-Native Corpus Scaffold for Cyber Threat Intelligence},
  author = {Cheng, Yutong and Li, Changze and Cui, Qian and Ding, Wei and Wang, Lingzhi and Chen, Yan and Gao, Peng},
  year = {2026},
  doi = {10.48550/arXiv.2608.18613}
}

@misc{sawarkar_blended_2024,
  title = {Blended {RAG}: Improving {RAG} (Retriever-Augmented Generation) Accuracy with Semantic Search and Hybrid Query-Based Retrievers},
  author = {Sawarkar, Kunal and Mangal, Abhilasha and Solanki, Shivam Raj},
  year = {2024},
  doi = {10.48550/arXiv.2404.07220}
}

@article{hu_llm_tikg_2024,
title = {LLM-TIKG: Threat intelligence knowledge graph construction utilizing large language model},
journal = {Computers \& Security},
volume = {145},
pages = {103999},
year = {2024},
issn = {0167-4048},
doi = {10.1016/j.cose.2024.103999},
author = {Yuelin Hu and Futai Zou and Jiajia Han and Xin Sun and Yilei Wang}
}

@INPROCEEDINGS{10763668,
  author={Xia, Yuchen and Kim, Jiho and Chen, Yuhan and Ye, Haojie and Kundu, Souvik and Hao, Cong Callie and Talati, Nishil},
  booktitle={2024 IEEE International Symposium on Workload Characterization (IISWC)}, 
  title={Understanding the Performance and Estimating the Cost of LLM Fine-Tuning}, 
  year={2024},
  doi={10.1109/IISWC63097.2024.00027}}

@INPROCEEDINGS{10386611,
  author={Liu, Jiehui and Zhan, Jieyu},
  booktitle={2023 IEEE International Conference on Big Data (BigData)}, 
  title={Constructing Knowledge Graph from Cyber Threat Intelligence Using Large Language Model}, 
  year={2023},
  doi={10.1109/BigData59044.2023.10386611}}

@misc{belcak_small_2025,
	title = {Small Language Models are the Future of Agentic {AI}},
	doi = {10.48550/arXiv.2506.02153},
	author = {Belcak, Peter and Heinrich, Greg and Diao, Shizhe and Fu, Yonggan and Dong, Xin and Muralidharan, Saurav and Lin, Yingyan Celine and Molchanov, Pavlo},
	year = {2025},
}

@book{wohlin_experimentation_2024,
	title = {Experimentation in Software Engineering},
	doi = {10.1007/978-3-662-69306-3},
	publisher = {Springer},
	author = {Wohlin, Claes and Runeson, Per and Höst, Martin and Ohlsson, Magnus C. and Regnell, Björn and Wesslén, Anders},
	year = {2024},
}

@inproceedings{10.1145/3319535.3354239,
author = {Kokulu, Faris Bugra and Soneji, Ananta and Bao, Tiffany and Shoshitaishvili, Yan and Zhao, Ziming and Doup\'{e}, Adam and Ahn, Gail-Joon},
title = {Matched and Mismatched SOCs: A Qualitative Study on Security Operations Center Issues},
year = {2019},
doi = {10.1145/3319535.3354239},
booktitle = {Proceedings of the 2019 ACM SIGSAC Conference on Computer and Communications Security},
}

@inproceedings{thimmaraju_human_2025,
	title = {Human Performance in Security Operations: A Survey on Burnout, Well-Being and Flow State Among Practitioners},
	doi = {10.14722/wosoc.2025.23002},
	booktitle = {Proceedings 2025 Workshop on Security Operation Center Operations and Construction},
	author = {Thimmaraju, Kashyap and Rispens, Sybe Izaak and Ahn, Gail-Joon},
	year = {2025},
}

@online{strom_mitre_nodate,
	title = {{MITRE} {ATT}\&{CK}: Design and Philosophy},
	author = {Strom, Blake E and Applebaum, Andy and Miller, Doug P and Nickels, Kathryn C and Pennington, Adam G and Thomas, Cody B},
    year = {2018}
}

@Article{jcp5030063,
AUTHOR = {Joy, Anooja and Chandane, Madhav and Nagare, Yash and Kazi, Faruk},
TITLE = {Threat Intelligence Extraction Framework (TIEF) for TTP Extraction},
JOURNAL = {Journal of Cybersecurity and Privacy},
VOLUME = {5},
YEAR = {2025},
NUMBER = {3},
ARTICLE-NUMBER = {63},
ISSN = {2624-800X},
DOI = {10.3390/jcp5030063}
}

@misc{wagner_towards_2024,
	title = {Towards Evaluation Guidelines for Empirical Studies involving {LLMs}},
	doi = {10.48550/arXiv.2411.07668},
	number = {{arXiv}:2411.07668},
	publisher = {{arXiv}},
	author = {Wagner, Stefan and Barón, Marvin Muñoz and Falessi, Davide and Baltes, Sebastian},
}

@misc{ctihal,
      title={CTI-HAL: A Human-Annotated Dataset for Cyber Threat Intelligence Analysis}, 
      author={Sofia Della Penna and Roberto Natella and Vittorio Orbinato and Lorenzo Parracino and Luciano Pianese},
      year={2025},
      doi={https://doi.org/10.48550/arXiv.2504.05866}, 
}

@misc{ctisum,
      title={CTISum: A New Benchmark Dataset For Cyber Threat Intelligence Summarization}, 
      author={Wei Peng and Junmei Ding and Wei Wang and Lei Cui and Wei Cai and Zhiyu Hao and Xiaochun Yun},
      year={2025},
      doi={https://doi.org/10.48550/arXiv.2408.06576}, 
}

@misc{ctinexus,
      title={CTINexus: Automatic Cyber Threat Intelligence Knowledge Graph Construction Using Large Language Models}, 
      author={Yutong Cheng and Osama Bajaber and Saimon Amanuel Tsegai and Dawn Song and Peng Gao},
      year={2025},
      doi={https://doi.org/10.48550/arXiv.2410.21060}, 
}

@misc{cticonnect,
      title={CTIConnect: A Benchmark for Retrieval-Augmented LLMs over Heterogeneous Cyber Threat Intelligence}, 
      author={Yutong Cheng and Yang Liu and Changze Li and Dawn Song and Peng Gao},
      year={2026},
      doi={https://doi.org/10.48550/arXiv.2510.11974}, 
}

@article{deepseek-ai_deepseek-r1_2025,
	title = {{DeepSeek}-R1: Incentivizing Reasoning Capability in {LLMs} via Reinforcement Learning},
	volume = {645},
	issn = {0028-0836, 1476-4687},
    year= {2025},
	doi = {10.1038/s41586-025-09422-z},
	journaltitle = {Nature},
	author = {{DeepSeek}-{AI} and Guo, Daya and Yang, Dejian and Zhang, Haowei and Song, Junxiao and Wang, Peiyi and Zhu, Qihao and Xu, Runxin and Zhang, Ruoyu and Ma, Shirong and Bi, Xiao and Zhang, Xiaokang and Yu, Xingkai and Wu, Yu and Wu, Z. F. and Gou, Zhibin and Shao, Zhihong and Li, Zhuoshu and Gao, Ziyi and Liu, Aixin and Xue, Bing and Wang, Bingxuan and Wu, Bochao and Feng, Bei and Lu, Chengda and Zhao, Chenggang and Deng, Chengqi and Zhang, Chenyu and Ruan, Chong and Dai, Damai and Chen, Deli and Ji, Dongjie and Li, Erhang and Lin, Fangyun and Dai, Fucong and Luo, Fuli and Hao, Guangbo and Chen, Guanting and Li, Guowei and Zhang, H. and Bao, Han and Xu, Hanwei and Wang, Haocheng and Ding, Honghui and Xin, Huajian and Gao, Huazuo and Qu, Hui and Li, Hui and Guo, Jianzhong and Li, Jiashi and Wang, Jiawei and Chen, Jingchang and Yuan, Jingyang and Qiu, Junjie and Li, Junlong and Cai, J. L. and Ni, Jiaqi and Liang, Jian and Chen, Jin and Dong, Kai and Hu, Kai and Gao, Kaige and Guan, Kang and Huang, Kexin and Yu, Kuai and Wang, Lean and Zhang, Lecong and Zhao, Liang and Wang, Litong and Zhang, Liyue and Xu, Lei and Xia, Leyi and Zhang, Mingchuan and Zhang, Minghua and Tang, Minghui and Li, Meng and Wang, Miaojun and Li, Mingming and Tian, Ning and Huang, Panpan and Zhang, Peng and Wang, Qiancheng and Chen, Qinyu and Du, Qiushi and Ge, Ruiqi and Zhang, Ruisong and Pan, Ruizhe and Wang, Runji and Chen, R. J. and Jin, R. L. and Chen, Ruyi and Lu, Shanghao and Zhou, Shangyan and Chen, Shanhuang and Ye, Shengfeng and Wang, Shiyu and Yu, Shuiping and Zhou, Shunfeng and Pan, Shuting and Li, S. S. and Zhou, Shuang and Wu, Shaoqing and Ye, Shengfeng and Yun, Tao and Pei, Tian and Sun, Tianyu and Wang, T. and Zeng, Wangding and Zhao, Wanjia and Liu, Wen and Liang, Wenfeng and Gao, Wenjun and Yu, Wenqin and Zhang, Wentao and Xiao, W. L. and An, Wei and Liu, Xiaodong and Wang, Xiaohan and Chen, Xiaokang and Nie, Xiaotao and Cheng, Xin and Liu, Xin and Xie, Xin and Liu, Xingchao and Yang, Xinyu and Li, Xinyuan and Su, Xuecheng and Lin, Xuheng and Li, X. Q. and Jin, Xiangyue and Shen, Xiaojin and Chen, Xiaosha and Sun, Xiaowen and Wang, Xiaoxiang and Song, Xinnan and Zhou, Xinyi and Wang, Xianzu and Shan, Xinxia and Li, Y. K. and Wang, Y. Q. and Wei, Y. X. and Zhang, Yang and Xu, Yanhong and Li, Yao and Zhao, Yao and Sun, Yaofeng and Wang, Yaohui and Yu, Yi and Zhang, Yichao and Shi, Yifan and Xiong, Yiliang and He, Ying and Piao, Yishi and Wang, Yisong and Tan, Yixuan and Ma, Yiyang and Liu, Yiyuan and Guo, Yongqiang and Ou, Yuan and Wang, Yuduan and Gong, Yue and Zou, Yuheng and He, Yujia and Xiong, Yunfan and Luo, Yuxiang and You, Yuxiang and Liu, Yuxuan and Zhou, Yuyang and Zhu, Y. X. and Xu, Yanhong and Huang, Yanping and Li, Yaohui and Zheng, Yi and Zhu, Yuchen and Ma, Yunxian and Tang, Ying and Zha, Yukun and Yan, Yuting and Ren, Z. Z. and Ren, Zehui and Sha, Zhangli and Fu, Zhe and Xu, Zhean and Xie, Zhenda and Zhang, Zhengyan and Hao, Zhewen and Ma, Zhicheng and Yan, Zhigang and Wu, Zhiyu and Gu, Zihui and Zhu, Zijia and Liu, Zijun and Li, Zilin and Xie, Ziwei and Song, Ziyang and Pan, Zizheng and Huang, Zhen and Xu, Zhipeng and Zhang, Zhongyu and Zhang, Zhen},
}

@article{cohen,
author = {Jacob Cohen},
title ={A Coefficient of Agreement for Nominal Scales},
journal = {Educational and Psychological Measurement},
volume = {20},
number = {1},
pages = {37-46},
year = {1960},
doi = {10.1177/001316446002000104},
}

@misc{sankararaman2024provenance,
      title={Provenance: A Light-weight Fact-checker for Retrieval Augmented LLM Generation Output}, 
      author={Hithesh Sankararaman and Mohammed Nasheed Yasin and Tanner Sorensen and Alessandro Di Bari and Andreas Stolcke},
      year={2024},
      doi={https://doi.org/10.48550/arXiv.2411.01022}
}

@ARTICLE{kgv2024,
author={Wu, Zongzong and Zhao, Ming and Li, Yufeng and Li, Huan and Peng, Shisong and Kato, Nei and Tang, Fengxiao},
journal={ IEEE Transactions on Mobile Computing },
title={{ KGV: Integrating Large Language Models With Knowledge Graphs for Cyber Threat Intelligence Credibility Assessment }},
year={2026},
volume={25},
number={10},
ISSN={1558-0660},
doi={10.1109/TMC.2026.3692451},
}

@misc{trace_cti_2026,
  title = {{TRACE-CTI}: Auditable Post-Extraction Governance of {TTP} Claims with Knowledge Graphs},
  author = {Valletta, Federico and Longo, Giacomo and Russo, Enrico and Merlo, Alessio},
  year = {2026},
  doi = {10.48550/arXiv.2607.24563}
}

@article{gwet_computing_2008,
	title = {Computing inter-rater reliability and its variance in the presence of high agreement},
	volume = {61},
	issn = {0007-1102},
	doi = {10.1348/000711006X126600},
    journaltitle = {The British Journal of Mathematical and Statistical Psychology},
	shortjournal = {Br J Math Stat Psychol},
	author = {Gwet, Kilem Li},
    year={2008}
}

@inproceedings{cormack_reciprocal_2009,
	title = {Reciprocal rank fusion outperforms condorcet and individual rank learning methods},
	doi = {10.1145/1571941.1572114},
	booktitle = {Proceedings of the 32nd international {ACM} {SIGIR} conference on Research and development in information retrieval},
	publisher = {Association for Computing Machinery},
	author = {Cormack, Gordon V. and Clarke, Charles L A and Buettcher, Stefan},
	date = {2009}
}

@misc{yang2025qwen3technicalreport,
      title={Qwen3 Technical Report}, 
      author={An Yang and Anfeng Li and Baosong Yang and Beichen Zhang and Binyuan Hui and Bo Zheng and Bowen Yu and Chang Gao and Chengen Huang and Chenxu Lv and Chujie Zheng and Dayiheng Liu and Fan Zhou and Fei Huang and Feng Hu and Hao Ge and Haoran Wei and Huan Lin and Jialong Tang and Jian Yang and Jianhong Tu and Jianwei Zhang and Jianxin Yang and Jiaxi Yang and Jing Zhou and Jingren Zhou and Junyang Lin and Kai Dang and Keqin Bao and Kexin Yang and Le Yu and Lianghao Deng and Mei Li and Mingfeng Xue and Mingze Li and Pei Zhang and Peng Wang and Qin Zhu and Rui Men and Ruize Gao and Shixuan Liu and Shuang Luo and Tianhao Li and Tianyi Tang and Wenbiao Yin and Xingzhang Ren and Xinyu Wang and Xinyu Zhang and Xuancheng Ren and Yang Fan and Yang Su and Yichang Zhang and Yinger Zhang and Yu Wan and Yuqiong Liu and Zekun Wang and Zeyu Cui and Zhenru Zhang and Zhipeng Zhou and Zihan Qiu},
      year={2025},
      url={https://arxiv.org/abs/2505.09388}, 
}

@misc{deidda_paper-slm-attackgraphs_2026,
	title = {Paper-{SLM}-{AttackGraphs}: Source code, data, and annotations for evidence-grounded attack-step enrichment},
	author = {Deidda, Nicola and Regano, Leonardo and Sanna, Alessandro and Maiorca, Davide and Giacinto, Giorgio},
	howpublished = {GitHub repository [software]},
	url = {https://github.com/MrDeiv/Paper-SLM-AttackGraphs},
	note = {Accessed 23 November 2026},
	year = {2026}
}

@misc{deidda_paper-slm-attackgraphs_data_2026,
	title = {Paper-{SLM}-{AttackGraphs}: Generated outputs and evaluation data for evidence-grounded attack-step enrichment},
	author = {Deidda, Nicola and Regano, Leonardo and Sanna, Alessandro and Maiorca, Davide and Giacinto, Giorgio},
	howpublished = {Zenodo [dataset]},
	doi = {10.5281/zenodo.24463589},
	year = {2026}
}

\appendix
\section{\acrshort{rag} Hyperparameters Tuning}
\label{app:rag}
Although CTIConnect~\cite{cticonnect} has recently been proposed as a benchmark for retrieval-augmented \llm{s} over heterogeneous \cti sources, we do not use it for tuning the retrieval hyperparameters of our pipeline. CTIConnect is designed to evaluate retrieval and reasoning for \cti question-answering tasks, including entity linking, entity attribution, and multi-document synthesis over structured and unstructured sources such as CVE, CWE, CAPEC, MITRE ATT\&CK, and vendor threat reports. This setting is related to ours because both involve retrieval over heterogeneous \cti. However, the evaluation unit and the optimization objective differ. CTIConnect provides expert-verified question--answer pairs and evaluates whether a model can retrieve and integrate the information needed to answer a query. In contrast, our pipeline retrieves case-specific evidence chunks to enrich a user-provided attack-step skeleton with grounded descriptions, \precond, and \postcond.

\changed{We introduced SQuAD~\cite{rajpurkar-etal-2016-squad} as a generic and reproducible retrieval benchmark in Section~\ref{sec:evaluation_design}. Then, we evaluate the alignment of the evidence for the generated enrichments in the \ncsc case studies according to the manual annotation protocol described later (Section~\ref{sec:manual_evaluation_protocol}).}

The SQuAD training set contains approximately 87,600 question--answer entries. We segment unique contexts using the semantic chunking configuration described in Section~\ref{sec:preprocessing_evaluation} and retain chunk offsets. We mark a chunk as relevant if its character span overlaps the reference answer. This produces 45,295 chunks and 86,398 evaluable queries, each with at least one identified relevant chunk.

\changed{To determine the trade-off between the number of retrieved chunks and the quality, we evaluate the lexical and vector retrievers independently by varying $k_{bm25}$ and $k_{vdb}$ over a discrete set of $k$, \ie, $K=\{1,3,5,10,20,50,100,200\}$. These values have been arbitrarily chosen.}

We assess retrieved-context quality using metrics computed from retrieved chunk texts and the answer-bearing reference chunk texts. We separately report rank-based Hit@$k$ and MRR@$k$ against the chunk IDs: Hit@$k$ indicates whether any such chunk is found among the top $k$ results, and MRR@$k$ is the mean reciprocal rank of the first such match.

We report the \bm-related retrieval metrics in Figure~\ref{fig:bm25_main}. Figure~\ref{fig:bm25_prf} shows Context Precision, Context Recall\footnote{\url{https://docs.ragas.io/en/latest/concepts/metrics/available_metrics/}}, F1, and F2 as $k$ varies. Recall increases from $0.411$ at $k=1$ to $0.797$ at $k=200$. Precision stabilizes after the first retrieved chunks, reaching approximately $0.494$ from $k=20$ onward. F1 increases from $0.522$ at $k=50$ to $0.524$ at $k=200$, while F2 increases from $0.571$ to $0.576$ over the same interval. The ranking metrics in Figure~\ref{fig:bm25_ranking} show that Hit@$k$ continues to grow with $k$, whereas MRR@$k$ becomes almost stable after $k=20$.

The marginal-gain analysis in Figure~\ref{fig:bm25_marginal} supports selecting $k_{bm25}=100$ as a practical operating point. Moving from $k=50$ to $k=100$ increases Recall by $0.041$ and F2 by $0.003$, while moving from $k=100$ to $k=200$ yields an F2 gain of $0.002$ and a Recall gain of $0.035$. The mean number of retrieved context tokens increases from approximately $5{,}936$ at $k=100$ to $12{,}146$ at $k=200$. We therefore select $k_{bm25}=100$ as a recall-oriented setting that limits the growth of retrieved context on this benchmark.

\begin{figure*}[h!]
    \centering
    \begin{subfigure}[t]{0.48\textwidth}
        \centering
        \includegraphics[width=\linewidth]{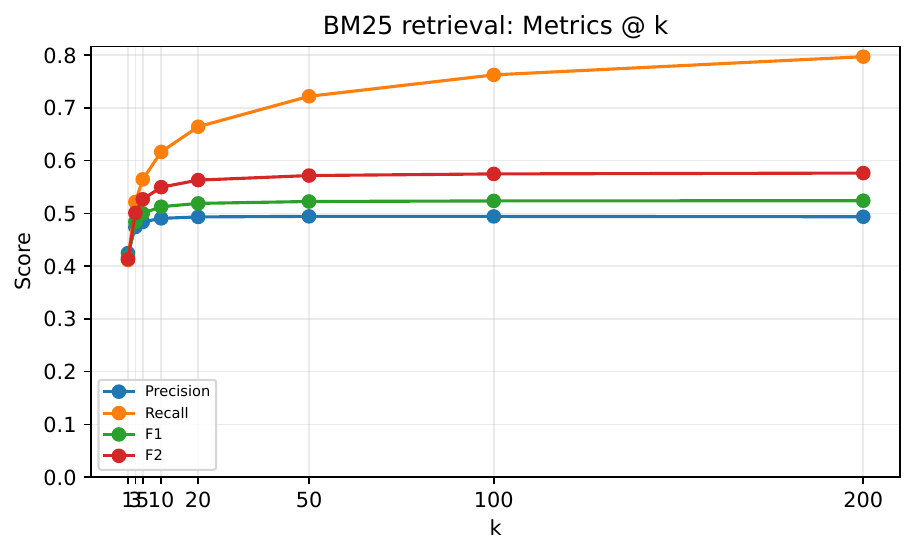}
        \caption{Precision, Recall, F1, and F2 at varying $k$.}
        \label{fig:bm25_prf}
    \end{subfigure}
    \hfill
    \begin{subfigure}[t]{0.48\textwidth}
        \centering
        \includegraphics[width=\linewidth]{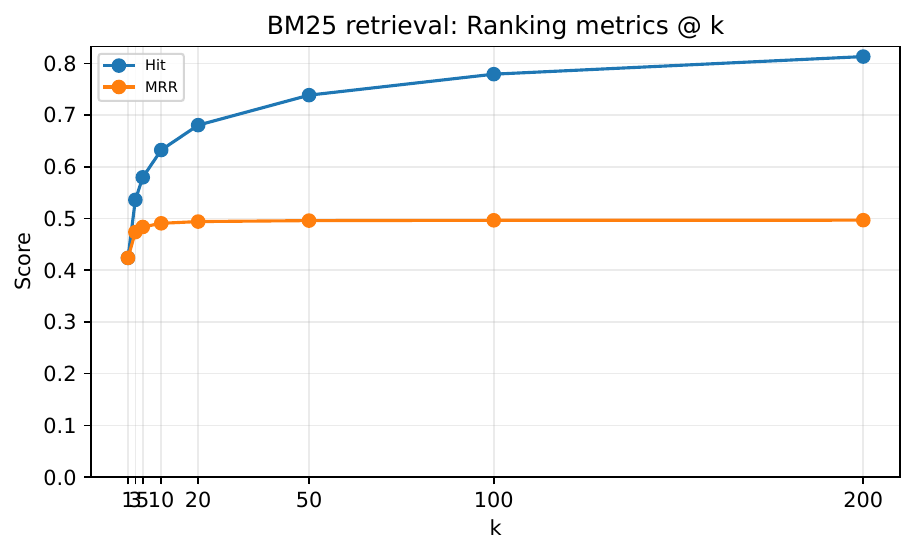}
        \caption{Ranking metrics at varying $k$.}
        \label{fig:bm25_ranking}
    \end{subfigure}

    \vspace{3ex}

    \begin{subfigure}[t]{0.48\textwidth}
        \centering
        \includegraphics[width=\linewidth]{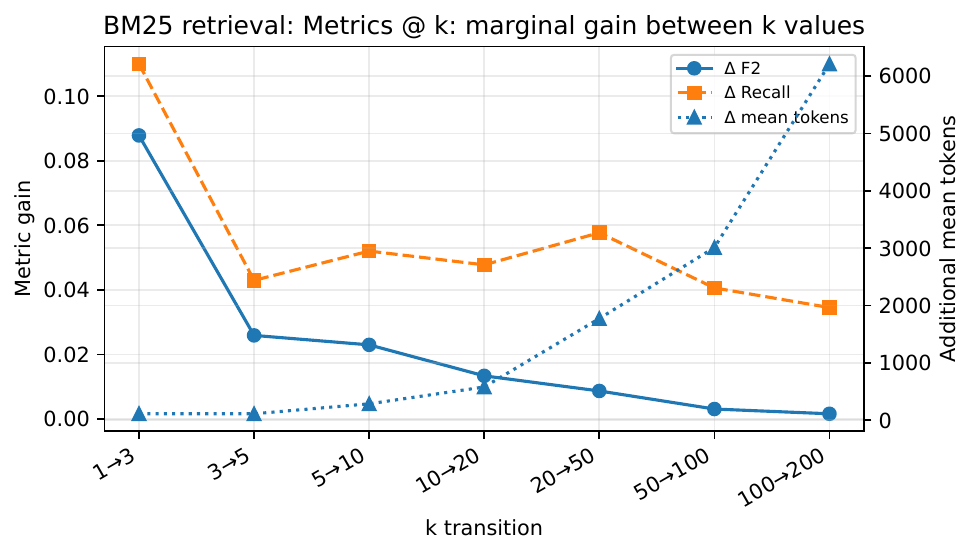}
        \caption{Marginal gains between retrieval depths.}
        \label{fig:bm25_marginal}
    \end{subfigure}

    \caption{BM25 retrieval-configuration analysis on SQuAD.}
    \label{fig:bm25_main}
\end{figure*}

Concerning the vector retriever, the metrics show a similar saturation pattern, as shown in Figure~\ref{fig:vec_main}. Recall increases from $0.527$ at $k=1$ to $0.961$ at $k=200$ (Figure~\ref{fig:vec_prf}), and Hit@$k$ increases from $0.541$ to $0.969$ (Figure~\ref{fig:vec_ranking}). F1 remains at approximately $0.688$ from $k=50$ to $k=200$, while F2 changes from $0.755$ to $0.759$. MRR@$k$ is also almost unchanged after $k=50$. Moving from $k=50$ to $k=100$ increases Recall by $0.024$ and F2 by $0.002$, while increasing the mean retrieved context from approximately $2{,}552$ to $5{,}022$ tokens. At $k=200$, the mean retrieved context reaches approximately $9{,}830$ tokens. We therefore select $k_{vdb}=50$ as the default value.

\begin{figure*}[h!]
    \centering
    \begin{subfigure}[t]{0.48\textwidth}
        \centering
        \includegraphics[width=\linewidth]{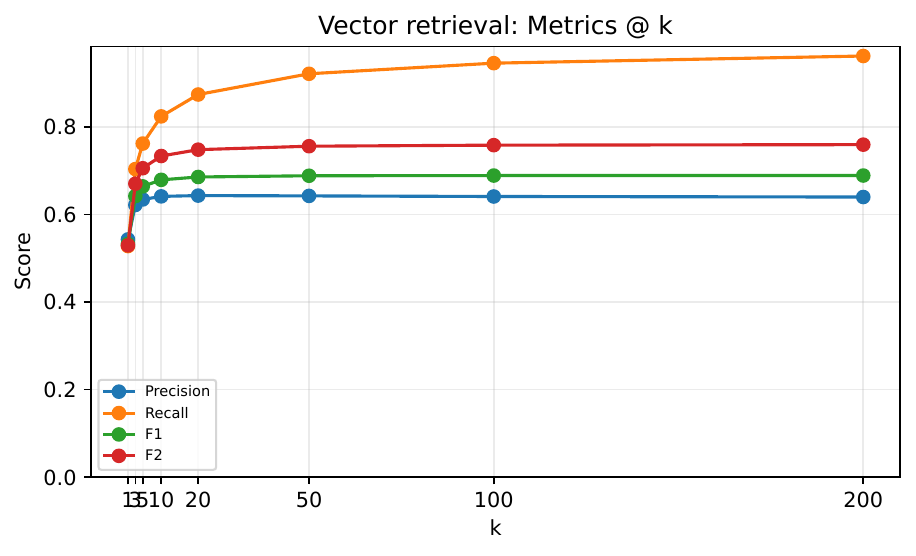}
        \caption{Precision, Recall, F1, and F2 at varying $k$.}
        \label{fig:vec_prf}
    \end{subfigure}
    \hfill
    \begin{subfigure}[t]{0.48\textwidth}
        \centering
        \includegraphics[width=\linewidth]{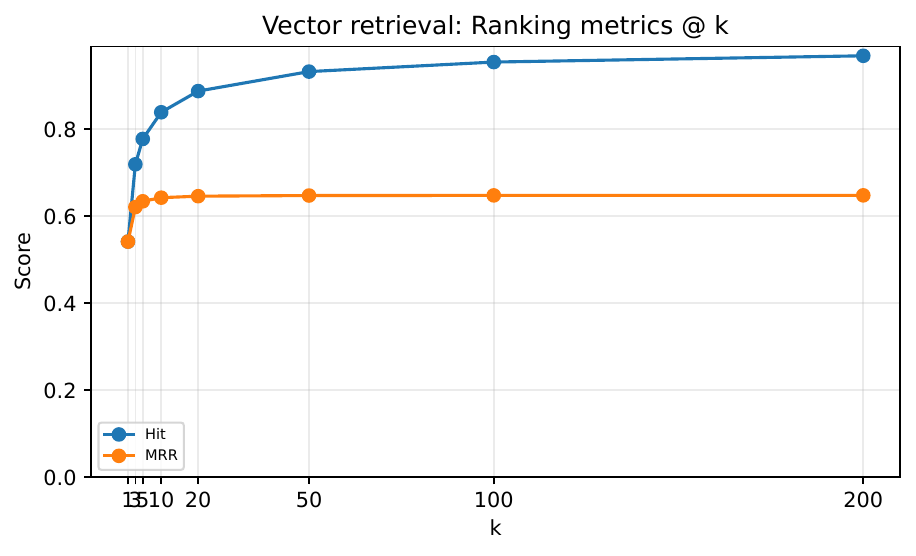}
        \caption{Ranking metrics at varying $k$.}
        \label{fig:vec_ranking}
    \end{subfigure}

    \vspace{3ex}

    \begin{subfigure}[t]{0.48\textwidth}
        \centering
        \includegraphics[width=\linewidth]{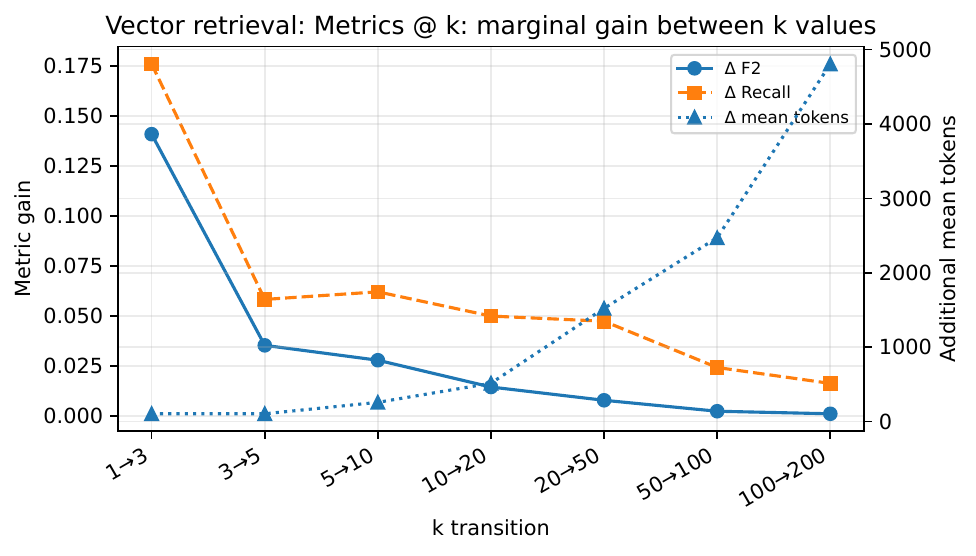}
        \caption{Marginal gains between retrieval depths.}
        \label{fig:vec_marginal}
    \end{subfigure}

    \caption{Vector retrieval-configuration analysis on SQuAD.}
    \label{fig:vec_main}
\end{figure*}

The hybrid retriever combines the candidate evidence units returned by the lexical and vector retrievers, configured with $k_{bm25}=100$ and $k_{vdb}=50$, respectively. Before accounting for overlapping results, these two candidate sets contain at most 150 entries, which motivates evaluating a final hybrid retrieval depth of up to $k_h=150$. However, the number of distinct evidence units may be lower when the same chunk is retrieved by both components.  

The hybrid retriever combines the top-$k_{bm25}$ and top-$k_{vdb}$ ranked lists through weighted reciprocal-rank fusion, as specified in Section~\ref{sec:retrieval_evidence_selection}. Its weights apply to reciprocal-rank contributions. Consequently, differences in the original score scales do not affect the fusion rule. The implementation uses the library\footnote{\url{https://www.langchain.com/}} defaults for the rank-fusion constant ($c=60$) and document identity; after fusion, the pipeline removes repeated chunk identifiers and truncates the list to $k_h$ results.

The results favor a vector-dominant configuration, as shown in Figure~\ref{fig:hybrid_weights}. The settings $w_{bm25}=0.25$ and $w_{vdb}=0.75$ provide the reported trade-off between retrieval quality and ranking metrics.

\begin{figure*}[h!]
    \centering
    \begin{subfigure}[t]{0.48\textwidth}
        \centering
        \includegraphics[width=\linewidth]{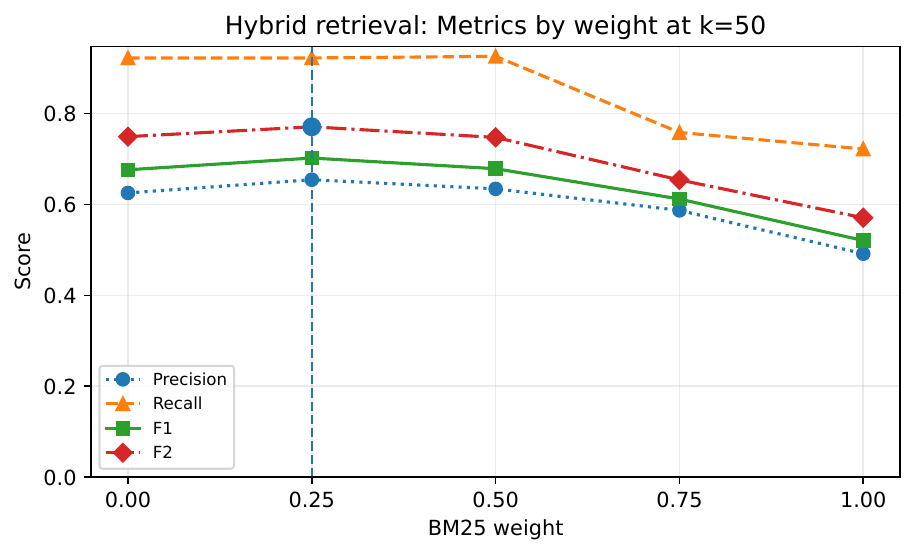}
        \caption{Precision, Recall, F1, and F2 at varying $w_{bm25}$. Hence, $w_{vdb} = 1-w_{bm25}$.}
        \label{fig:hybrid_prf}
    \end{subfigure}
    \hfill
    \begin{subfigure}[t]{0.48\textwidth}
        \centering
        \includegraphics[width=\linewidth]{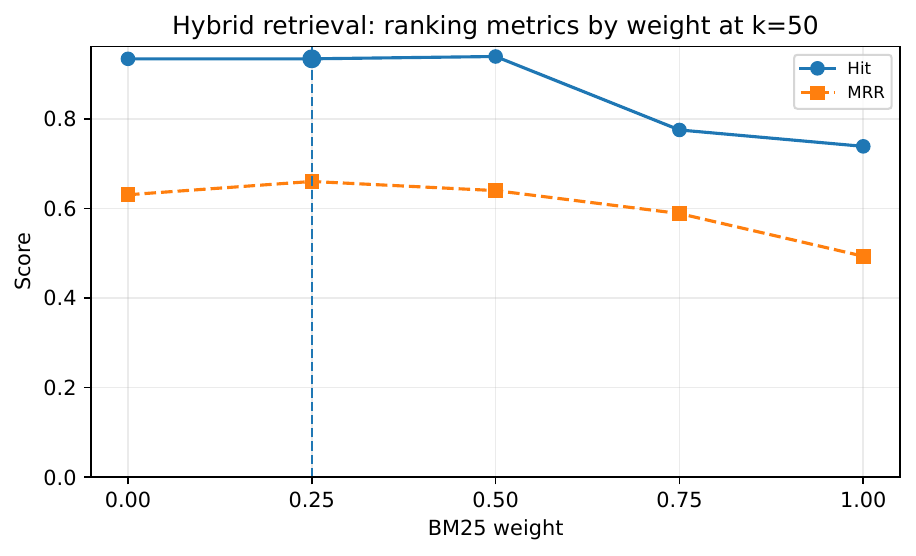}
        \caption{Ranking metrics at varying $w_{bm25}$.}
        \label{fig:hybrid_ranking}
    \end{subfigure}

    \caption{Weights analysis for the hybrid retrieval configuration.}
    \label{fig:hybrid_weights}
\end{figure*}

With this weighting, the highest reported F2 score is obtained at $k_h=150$, with F2 equal to $0.773$, Recall equal to $0.949$, and Hit@$k$ equal to $0.960$. This configuration requires an average retrieved context of approximately $7{,}887$ tokens (Figure~\ref{fig:hybrid_tokens}). In contrast, $k_h=50$ achieves F2 equal to $0.771$, Recall equal to $0.922$, Hit@$k$ equal to $0.934$, and MRR equal to $0.660$, with approximately $2{,}556$ retrieved context tokens. These results are reported in Figure~\ref{fig:hybrid_main}.

We consequently select $k_{bm25}=100$, $k_{vdb}=50$, $k_h=50$, $w_{bm25}=0.25$, and $w_{vdb}=0.75$ as the default hybrid-retrieval configuration. The SQuAD evaluation and the \cti case-study pipeline both call the shared production retrieval implementation, ensuring that the same rank-fusion procedure is used.

The selected configuration sacrifices a small amount of measured retrieval effectiveness on SQuAD in exchange for a substantially smaller retrieved context. It is used as a practical setting for the subsequent \cti enrichment experiments, not as evidence that the weighting or cut-off is optimal for \cti retrieval.

\begin{figure*}[h!]
    \centering
    \begin{subfigure}[t]{0.48\textwidth}
        \centering
        \includegraphics[width=\linewidth]{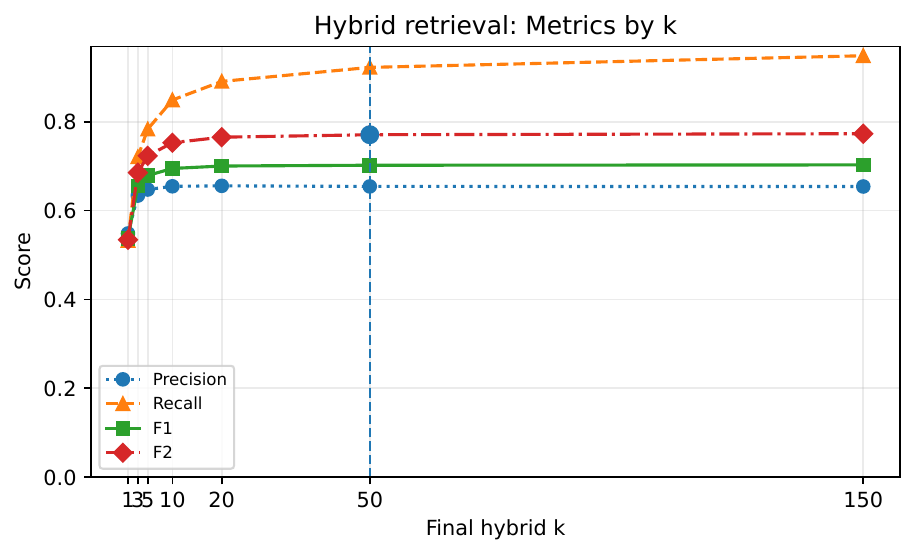}
        \caption{Precision, Recall, F1, and F2 at varying $k_h$.}
        \label{fig:hybrid_prf}
    \end{subfigure}
    \hfill
    \begin{subfigure}[t]{0.48\textwidth}
        \centering
        \includegraphics[width=\linewidth]{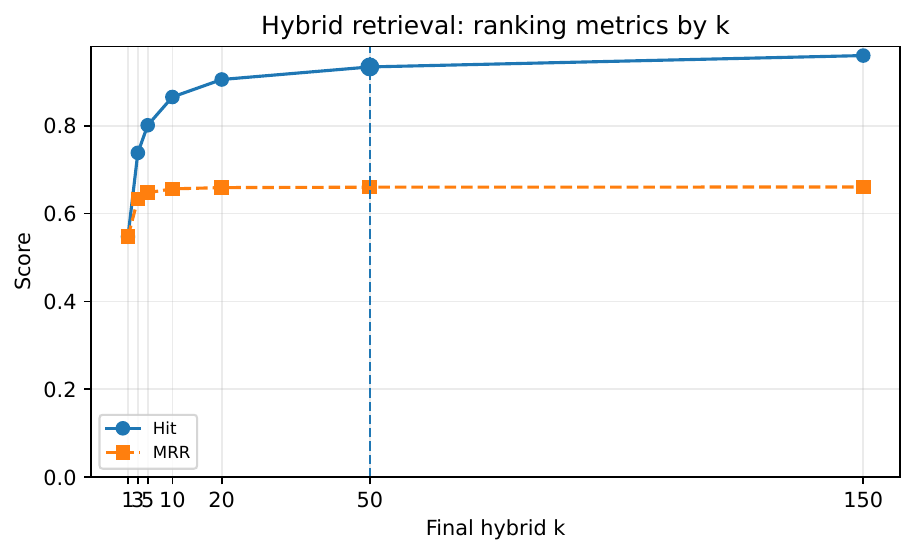}
        \caption{Ranking metrics at varying $k_h$.}
        \label{fig:hybrid_ranking}
    \end{subfigure}

    \vspace{3ex}

    \begin{subfigure}[t]{0.48\textwidth}
        \centering
        \includegraphics[width=\linewidth]{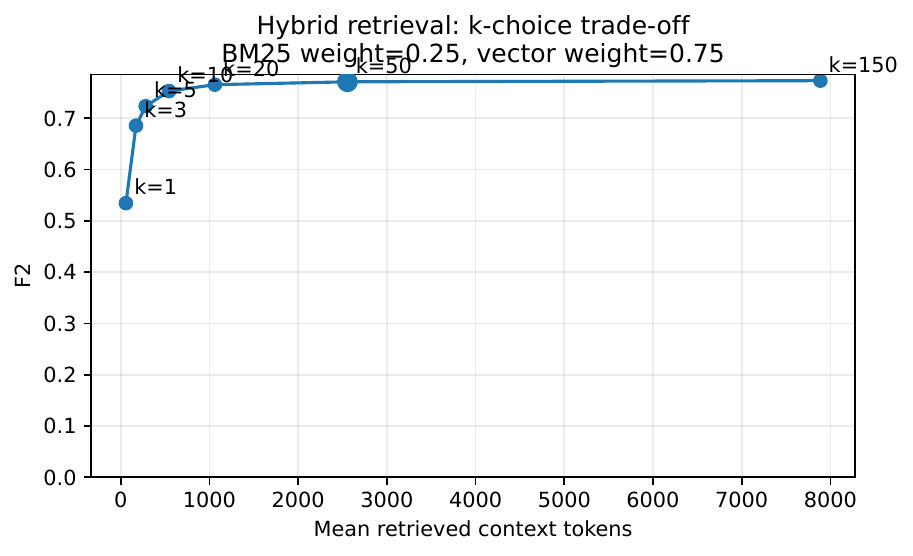}
        \caption{Mean retrieved context tokens at varying $k_h$.}
        \label{fig:hybrid_tokens}
    \end{subfigure}

    \caption{Hybrid retrieval-configuration analysis on SQuAD.}
    \label{fig:hybrid_main}
\end{figure*}


\clearpage

\bio{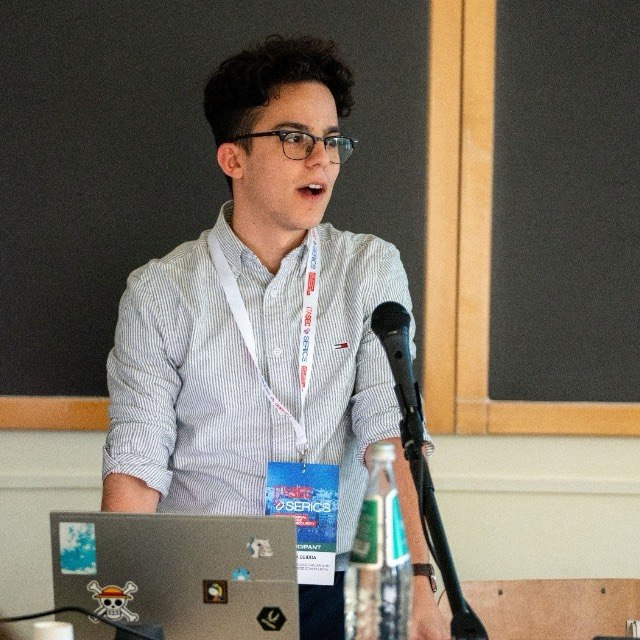}
Nicola Deidda received his Master's degree in Computer Engineering, Cybersecurity, and Artificial Intelligence from the University of Cagliari, Italy, in 2024. He is a PhD student in the National Ph.D. program in Cybersecurity, a joint program with IMT School for Advanced Studies of Lucca. His research focuses on AI applications in Cyber Threat Intelligence.
\endbio

\bio{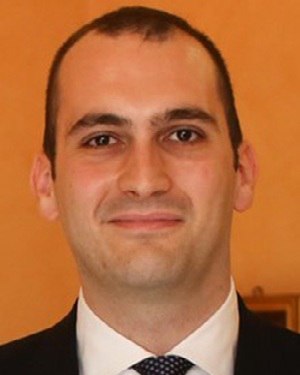}
Leonardo Regano obtained his Master's Degree in Computer Engineering (final grade: 110/110) from the Polytechnic University of Turin in 2015 and his Ph.D. in Computer Engineering and Systems in 2019. He currently holds the position of Assistant Professor (RTD-A) in Computer Engineering at the University of Cagliari. Previously, he worked as a Research Fellow from 2015 to 2020 and Senior Research Fellow from 2020 to 2023 at the Polytechnic University of Turin. His research interests primarily focus on the security of computer systems and networks, particularly the security and protection of software, applications of machine learning and deep learning for Cybersecurity, and the security of software networks.
\endbio

\bio{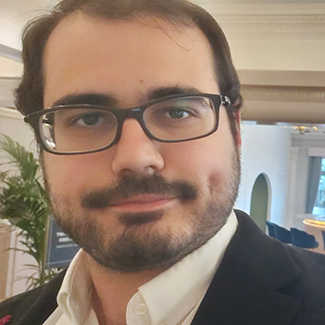}
Alessandro Sanna is a Postdoctoral Researcher at the University of Cagliari. His research focuses on malware detection and threat intelligence, primarily on Living-off-the-Land malware and the use of ML/DL for malware and threat detection. In particular, he has published works on Microsoft Office, Windows, and IoT malware, focusing on untangling obfuscated malicious software. He has also experience in industry, having worked with Abissi S.r.l. during his Ph.D., and he is certified as a Lead Auditor for ISO/IEC 27001:2022 (Information Security, Cybersecurity and Privacy Protection).
\endbio

\bio{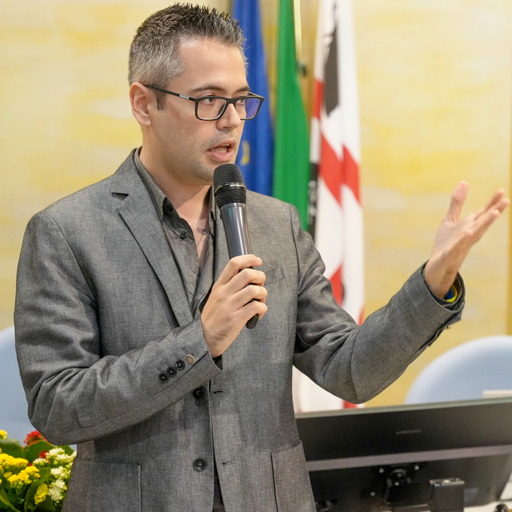}
Davide Maiorca received the Master of Science degree in Electronic Engineering (110/110 cum laude) from the University of Cagliari in 2012 and a Ph.D. in Electronic Engineering and Computer Science in 2016. He has been working at PRA Lab since 2012 and currently holds a position as an Associate Professor of Computer Engineering at the University of Cagliari. His research fields include analyzing and detecting X86 and Android malware, malicious documents and multimedia applications (e.g., PDF, Microsoft Office), and Adversarial Machine Learning. In 2021, he received the National Scientific Qualification (ASN) as Associate Professor (ING-INF/05). Dr. Maiorca authored more than 25 research papers and has served as a Program Committee member and reviewer for international conferences and journals. He is a member of IEEE, Computer Society, SMC, ACM, INSTICC, AIxIA and CVPL (ex-GIRPR – Gruppo Italiano Ricercatori Pattern Recognition).
\endbio

\newpage

\bio{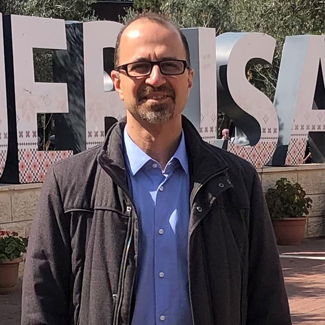}
Giorgio Giacinto is a Professor of Computer Engineering at the University of Cagliari, Italy, where he serves as the coordinator of the M.Sc. degree in Computer Engineering, Cybersecurity and Artificial Intelligence. His research interests are in machine learning for malware analysis and detection, and he published more than 180 papers in international conferences and journals. He is the Editor in Chief of the "Security Engineering \& Applications" section of the Journal of Cybersecurity and Privacy. He is a member of the managing committees of the Cybersecurity National Lab and the Artificial Intelligence \& Intelligent Systems Lab within the CINI consortium, Italy. He also represents the Cybersecurity National Lab within the European Cybersecurity Organization (ECSO). He is a Fellow of the IAPR (International Association for Pattern Recognition) and a Senior Member of the IEEE Computer Society and ACM.
\endbio

\end{document}